\documentclass[aps,prd,11pt,a4paper,nofootinbib,oneside,superscriptaddress,eqsecnum]{revtex4-1}
\pdfoutput = 1

\usepackage{amsmath,amssymb,amsfonts,color}
\usepackage{tensor,slashed,paralist,cases,mathrsfs}
\usepackage{float,cancel,xcolor}
\usepackage{graphicx}
\usepackage{dcolumn}
\usepackage{bm}
\usepackage[colorlinks=true,urlcolor=blue,linkcolor=blue,citecolor=blue,linktocpage=true]{hyperref}

\newcommand{\beq}{\begin{eqnarray}}
\newcommand{\eeq}{\end{eqnarray}}
\newcommand{\bpmatrix}{\begin{pmatrix}}
\newcommand{\epmatrix}{\end{pmatrix}}
\newcommand{\ba}{\begin{array}}
\newcommand{\ea}{\end{array}}

\renewcommand{\eqref}[1]{Eq.~(\ref{#1})}

\newcommand{\bc}{\begin{center}}
\newcommand{\ec}{\end{center}}

\begin{document}

\vspace*{1.5em}

\title{ A Potentially Detectable Gamma-Ray Line in the Fermi Galactic Center Excess ---
In Light of One-Step Cascade Annihilations of Secluded (Vector) Dark Matter via the  Higgs Portal}

\author{Kwei-Chou Yang}
\email{kcyang@cycu.edu.tw}

\affiliation{Department of Physics and Center for High Energy Physics, Chung Yuan Christian University, 
200 Chung Pei Road, Taoyuan 32023, Taiwan}


\begin{abstract}

We show the presence of a potentially detectable gamma-ray line in the Fermi Galactic center excess in light of the secluded (vector) dark matter (DM) model in which the hidden scalar, nearly degenerate with DM in mass, mediates the  interaction of the secluded DM with the Standard Model (SM) due to its mixing with the SM Higgs.
 We find that the parameter region $m_X\in[60, 132]~\text{GeV}$ can provide a good fit to the Fermi Galactic center gamma-ray excess spectrum, appearing a prominent gamma-ray line with the energy $\in [30, 66]$~GeV.  
 The best fit gives $m_X\simeq m_S \simeq 86$~GeV with a $p$-value$\, =0.42$, so that the resultant gamma-ray line, arising from the decay of the scalar mediator into $\gamma\gamma$, peaks at 43~GeV.   We derive constraints on the annihilation cross section from the Fermi-LAT gamma-ray line search,  gamma-ray observations of the Fermi-LAT dwarf spheroidal galaxies, and Planck cosmic microwave background measurement.  For the secluded vector DM model, the parameter space constrained by the current XENON1T and future LUX-ZEPLIN is shown. 
  Finally,  for the mixing angle between the Higgs sectors, we discuss its lower bound, which is required by the big bang nucleosynthesis constraint and relevant to the hidden sector decoupling temperature.

\end{abstract}
\maketitle
\newpage

\section{Introduction}

The existence of non-baryonic dark matter (DM) is evident from various cosmological observations and measurements \cite{Adam:2015rua,Ade:2015xua}. 
Moreover, the majority of the matter density in our Universe is dominated by the DM. Currently, one of the favorable DM candidates is the so-called  weakly interacting massive particles (WIMPs). For this scenario, dark matter, having mass of order GeV $-$ TeV and interacting with the Standard Model (SM) particles at the electroweak scale, can give the correct relic abundance today. Meanwhile, the nonrelativistic WIMPs,  following Boltzmann suppression, remains thermal equilibrium with the bath until freeze-out. However,  the DM models built based on the WIMPs scenario are increasingly constrained due to the null results from the direct detection and collider experiments. 

Instead, a paradigm of DM was proposed to suggest that (WIMP) dark matter is secluded within one of the hidden sectors, and is very weakly coupled to  the visible sector via a metastable mediator which is lighter than the DM \cite{Pospelov:2007mp,Hambye:2008bq,Lebedev:2011iq,Ko:2014gha,Berlin:2014pya,Escudero:2017yia,Ko:2014loa,Abdullah:2014lla,Martin:2014sxa,Kim:2016csm,Acharya:2016fge,Yang:2017zor,Profumo:2017obk,Yang:2018fje,Yang:2019bvg}. As such, the secluded DM may become undetectable or hard to detect in colliders and underground direct searches, but  can still produce viable signals in the indirect experiments \cite{Yang:2017zor,Yang:2018fje,Yang:2019bvg}. 

For the indirect DM searches, a number of studies have confirmed an excess of few-GeV gamma-rays from the region around the Galactic center (GC)  and suggested that the excess emission could arise from the DM annihilation  \cite{Goodenough:2009gk, Hooper:2010mq,Hooper:2011ti, Abazajian:2012pn, Gordon:2013vta, Huang:2013pda, Daylan:2014rsa, Calore:2014xka, Calore:2014nla,Karwin:2016tsw,TheFermi-LAT:2017vmf,Leane:2019xiy}.  The signal origin of GC excess is not conclusive yet.  Several interpretations,  recently proposed from the astrophysical point of view,  suggested that the excess can be better correlated with stellar over-density in the Galactic bulge and the nuclear stellar bulge \cite{Macias:2016nev,Macias:2019omb}, or described by point sources \cite{Lee:2015fea,Buschmann:2020adf}.
In this paper, we will focus on the secluded DM scenario for explaining the GC gamma-ray excess. In such a model,
compared with the WIMP case of direct annihilations to the SM, a multi-step cascade DM annihilation can accommodate a higher DM mass, allow a larger cross section in the fit, and broadens the spectrum of the secondary particles.

Not only for a conventional WIMP model but also for a secluded DM model,  gamma-ray lines are very likely to be expected at the loop level.
Thus, the gamma-ray line signal directly/indirectly reveals the particle nature of the underlying theory of dark matter. Moreover, it provides a striking signature which could be clearly distinguished from astrophysical backgrounds. It is interesting to note that the direct DM annihilation to the SM Higgs pair, $h\, h$,  gives a moderately good fit to the Fermi GC excess spectrum but with a $p$-value $=0.13$ at best, as long as the produced $h$ is approximately at rest \cite{Calore:2014nla} (c.f. $p$-value $=0.17$ obtained in Ref.~\cite{Agrawal:2014oha}). In this case, a detectable width of the gamma line with energy $\simeq m_h/2 \simeq 62.6$~GeV, about half of mass of DM,  is very sensitive to the Lorentz-boost from the Higgs rest frame to the DM center of mass (com) frame.
A  distinguishable line signal also depends on the energy resolution of the detector.

Motivated by the above gamma line results  \cite{Calore:2014nla,Agrawal:2014oha}, in this paper, we consider a secluded DM model, where the DM mainly annihilates into a pair of  lighter scalar mediators, $S$. 
For this secluded DM model, a generic case of $m_{\rm DM} \gtrsim m_S$ can have a good fit to the GC spectrum.  Instead, here we focus on the study of the GC gamma-ray spectrum with prominent lines which could be detectable.
 To have a clear gamma-line signal, we take into account the case that the scalar mediator is nearly degenerate with DM in mass.
As will be shown in Sec.~\ref{sec:results}, when the both masses of the DM and scalar mediator are about 86~GeV,  the $p$-value of the best GC spectrum fit can be as large as 0.42.

\begin{figure}[t!]
\begin{center}
\includegraphics[width=0.85\textwidth]{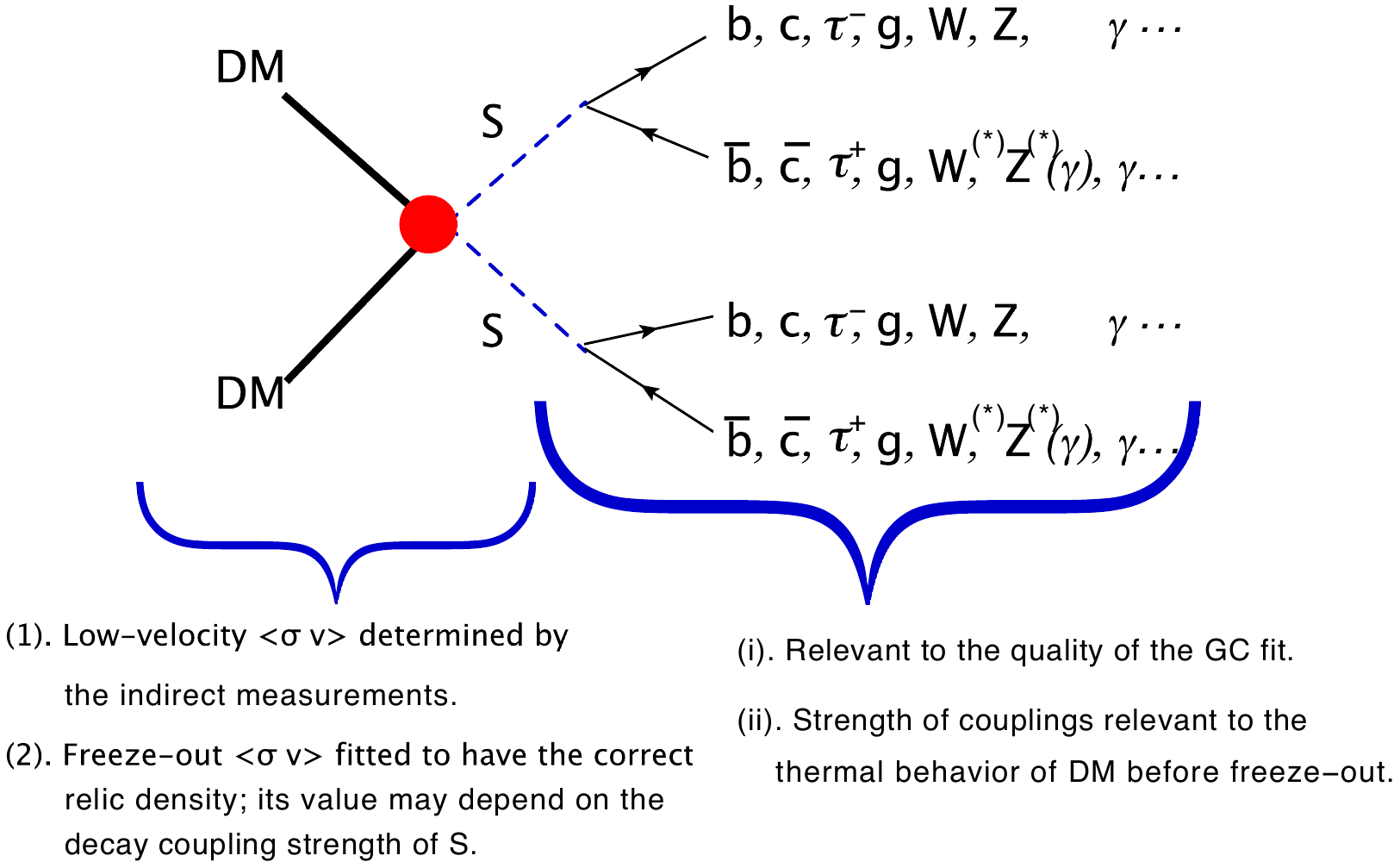}
\caption{ Gamma-ray spectrum generated from one-step cascade annihilations of dark matter via the Higgs portal, where the hidden Higgs mediator is denoted as ``$S$".
}
\label{fig:cascadesDM}
\end{center}
\end{figure}

The mediator, a mixture of the hidden sector Higgs and the SM Higgs, has a mass lighter than the 125~GeV Higgs observed at the LHC, so that 
 the resulting cascade decays can soften the gamma-ray spectrum to have a better fit to the peak at $1-3$~GeV.
For illustration, in Fig.~\ref{fig:cascadesDM}, we show one-step cascade annihilations of the secluded dark matter into a pair of scalar mediators which subsequently decay to the SM final states. 
In indirect searches, the qualitative fit of the gamma-ray spectrum is relevant to the decay channels of the mediator as well as the DM mass, which determines the initially kinetic energy of the mediator and thus the boost factor for the spectrum, while the DM annihilation cross section plays as an overall factor in the fit.  For determination of the DM relic abundance today, the thermodynamic evolution of the hidden sector before freeze-out depends on the strength of couplings between the mediator and SM \cite{Yang:2019bvg}. If the couplings are small enough, the hidden sector can be kinetically decoupled from the bath before it becomes nonrelativistic.  As such the freeze-out DM annihilation cross section required to give a correct relic density could be boosted above the conventional WIMP value \cite{Yang:2019bvg,Farina:2016llk,Pappadopulo:2016pkp,Dror:2016rxc}.  

In the analysis, we will use the Fermi GC gamma-ray excess spectrum obtained by Calore, Cholis, and Weniger (CCW)  \cite{Calore:2014xka}\footnote{
It should be note that the nature of the GC excess is under active debate. Besides the scenario that GC excess might arise from the DM annihilation, some newly proposed diffuse models could provide an even improved fit to the data by including various astrophysical phenomenologies, e.g., models correlating the excess with stellar over-density of the Galactic bulge  \cite{Macias:2016nev,Macias:2019omb}, or with point sources \cite{Lee:2015fea,Buschmann:2020adf}.}.
The result of CCW is based on the  Fermi Pass 7 dataset\footnote{
The extracted GC spectra do not have obvious difference among Fermi Pass 7 and  Pass 8 datasets \cite{Linden:2016rcf}. However, their results at low energies  can have appreciable difference, depending on event selections of the point sources in various datasets.
}, of which the energy resolution is about 10\% \cite{Ackermann:2012kna}. In the parameter plane of the DM annihilation cross section and DM mass, that is relevant to the spectral line(s) generated from the  Higgs portal one-step cascade annihilation of DM, we will further show the current bounds imposed by the Fermi-LAT observations of dwarf spheroidal galaxies (dSphs)  \cite{Fermi-LAT:2016uux}, by Fermi-LAT gamma-ray line search in the region around the GC \cite{Ackermann:2015lka}, by the Planck cosmic microwave background (CMB)  \cite{Ade:2015xua}, and by direct detections  \cite{Aprile:2017iyp,Akerib:2018lyp}.   The Fermi-LAT projected sensitivity with as much as 15 years of data \cite{Charles:2016pgz} as well as the high energy resolution detectors from forthcoming experiments \cite{Bernardini:2017han,Topchiev:2017xfp,Topchiev:2017gku} is capable of exploring the considered parameter space.
Thus, the Higgs portal scenario is very likely to be testable in the near future. See the details in Sec.~\ref{sec:results}.

To be more specific, we will consider a simplest secluded vector dark matter model in which the vector DM interacts with the SM mainly through the scalar mediator, which is a hidden physical Higgs state resulting from an extremely small mixing angle between the dark sector scalar singlet and the SM Higgs. 
However, one should note that the determination of the gamma line is nothing to do with the fundamental property of DM, but is related to the Higgs portal.

The layout of this paper is as follows.  In Sec.~\ref{sec:model}, we present a renormalizable vector DM model in which the dark sector described by the $U(1)_X$ gauge symmetry contains a gauge vector boson and a complex scalar.  Compared with the SM, four additional parameters, including the DM and mediator masses ($m_X, m_S$), DM-mediator coupling constant ($g_{\rm dm}$), and Higgs mixing angle ($\alpha$), are introduced.
In Sec.~\ref{sec:one-step}, we outline the formulation of the gamma-ray spectrum  with prominent lines, arising from a one-step cascade annihilation of DM to scalar mediators, which subsequently decay into SM particles through very small couplings, owing to the tiny Higgs mixing angle.
 In order to have a correct spectrum fit, for the mediator mass range $m_V \lesssim m_S \lesssim  2 m_V$, not only the usual two-body decay modes but also the three-body  decay modes, $S \to V V^* \to V f_1 \bar{f_2} $ with $V\equiv W, Z$, need to be taken into account. Furthermore, we calculate the expected the gamma lines originating from $S \to \gamma\gamma, \gamma Z$, where the continuum spectrum resulting from the $Z$ decay is also considered in the $S\to \gamma Z$ decay.
 In Sec.~\ref{sec:results}, we present the main analysis. In Sec.~\ref{sec:discussions}, we discuss the constraint on the mixing angle of the two scalar sectors from the thermodynamic point of view, and the scale-dependence of vacuum stability for the secluded vector DM model.
We conclude in Sec.~\ref{sec:conclusions}.

\section{The Model}\label{sec:model}

We consider the simplest abelian vector dark matter model, which is renormalizable. In this model, the vector dark matter, $X$, associated with a dark $U(1)_X$ gauge symmetry, interacts with the SM particles vis the Higgs portal, which originates from the mixture of the SM Higgs and the hidden complex scalar ($\Phi_S$).
In addition to the usual SM part, the relevant  Lagrangian, involving the dark kinetic terms and scalar potentials, are described by
\begin{align}
 {\cal L}_\text{hidden} = & -\frac{1}{4} X_{\mu\nu} X^{\mu\nu} + (D_\mu\Phi_S)^\dagger (D^\mu \Phi_S) 
    - \mu_{H}^2 |\Phi_H|^2 - \mu_{S}^2 |\Phi_S|^2 \nonumber \\  
 &- \frac{\lambda_H}{2} (\Phi_H^\dagger \Phi_H)^2 
 - \frac{\lambda_S}{2} (\Phi_S^\dagger \Phi_S)^2 
 - \lambda_{HS} (\Phi_H^\dagger \Phi_H) (\Phi_S^\dagger \Phi_S)  \;,
\label{eq:lagrangian}
\end{align}
where  $X_{\mu\nu} =\partial_\mu X_\nu -\partial_\nu X_\mu$,  $ D_\mu \Phi_S \equiv (\partial_\mu + i g_{\rm dm} Q_{\Phi_S} X_\mu )\Phi_S$, $\Phi_H= (H^+, H^0)^{\rm T}$ is the SM Higgs doublet, and $\Phi_S$ is the hidden complex scalar  with a $U_X(1)$ charge assignment $Q_{\Phi_S}$.  In the following, we will simply use  $Q_{\Phi_S}=1$. 
After spontaneous symmetry breaking, 
\begin{equation}
\Phi_H=\frac{1}{\sqrt{2}} (v_H + \phi_h + i \sigma_h), \quad
\Phi_S=\frac{1}{\sqrt{2}} (v_S + \phi_s + i \sigma_s),
\label{eq:vev}
\end{equation}
DM gets its mass, $m_X=g_{\rm dm} Q_{\Phi_S} v_S$, and the CP-odd state,  $\sigma_s$, is absorbed to be the longitudinal component of $X$, where the $Z_2$ symmetry, $X_\mu \to -X_\mu$ and $\Phi_S \to \Phi_S^*$, is preserved, so that DM is stabilized.  Under this $Z_2$ symmetry, all other fields are even.
The scalar fields $(\phi_h, \phi_s)$ can be rewritten in terms of mass eigenstates of physical Higgses $(h, S)$ as
\begin{align}
\phi_h = c_\alpha \, h  -  s_\alpha \, S \,, \\
\phi_s = s_\alpha \, h  + c_\alpha \, S \,, 
\end{align}
and the mass term in the Lagrangian is given by $-1/2\, (m_h^2 h^2 +m_S^2 S^2) = -1/2\, (\phi_h, \phi_s)\,  M_{\rm Higgs}^2 \,  (\phi_h, \phi_s)^\dagger $, where
\begin{equation}
M_{\rm Higgs}^2= 
\left(
\begin{array}{cc}
 \lambda_H v_H^2 & \lambda_{HS} v_S v_H \\ 
\lambda_{HS} v_S v_H &  \lambda_S v_S^2
\end{array}
\right) =
\left(
\begin{array}{cc}
 m_h^2 c_\alpha^2  + m_S^2 s_\alpha^2  & (m_h^2 -m_S^2 )s_\alpha c_\alpha \\ 
(m_h^2 -m_S^2 )s_\alpha c_\alpha & m_S^2 c_\alpha^2 + m_h^2 s_\alpha^2
\end{array}
\right) \,,
\label{eq:mass_matrix}
\end{equation}
and the abbreviations, $s_\alpha \equiv \sin\alpha$ and $c_\alpha \equiv \cos\alpha$, are used here and in the following.

In the analysis, we will use $v_H\simeq 246$~GeV and $m_h=125.18$~GeV \cite{pdg2018} as inputs, and take $m_X , m_S$, $g_{\rm dm}$ and $\alpha$ as the four independent parameters, i.e., the remaining $\lambda_S, \lambda_H, \lambda_{HS}$, and $v_S$ can be parametrized in terms of these parameters. 

\begin{figure}[t!]
\begin{center}
\includegraphics[width=0.45\textwidth]{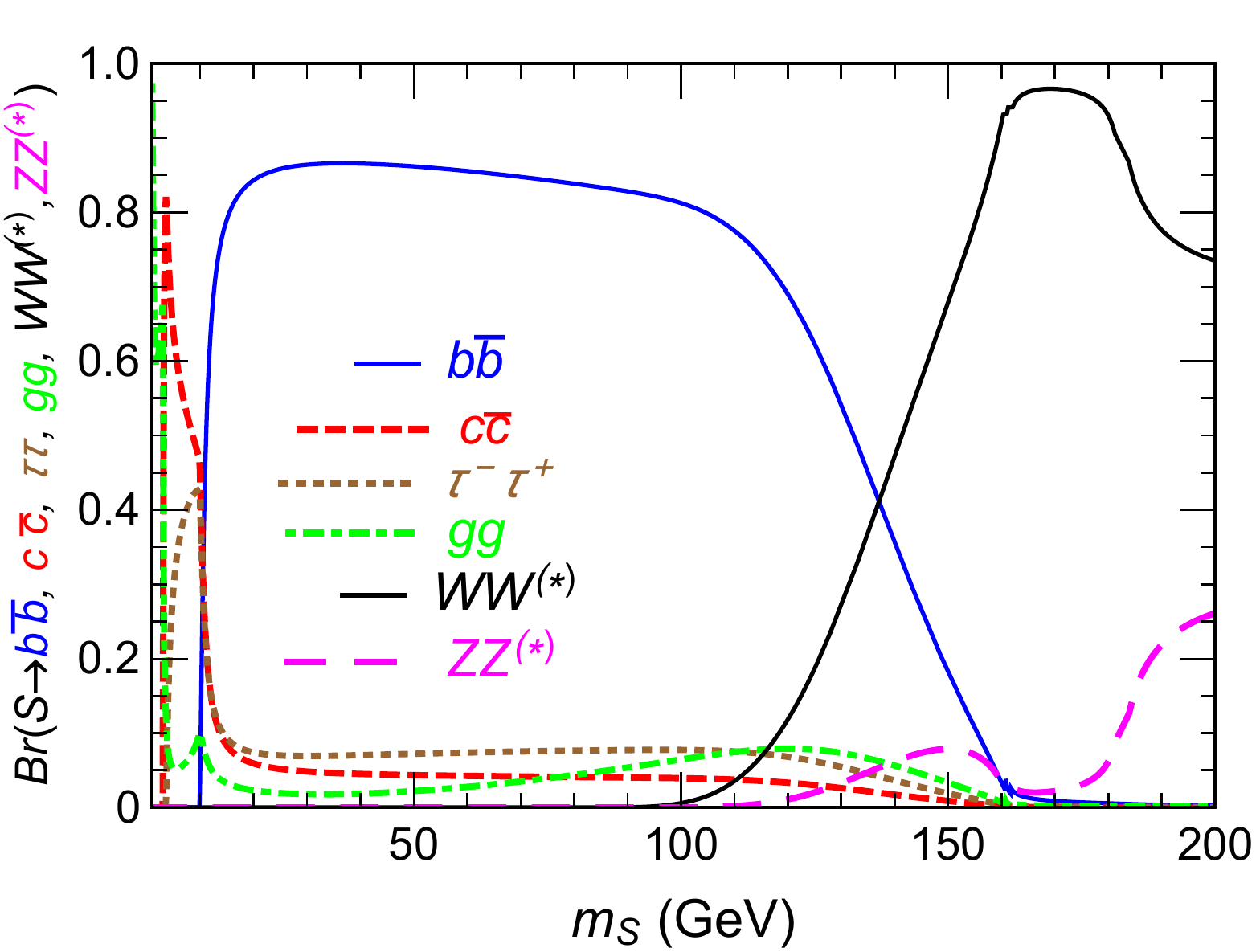}\hskip0.6cm
\includegraphics[width=0.468\textwidth]{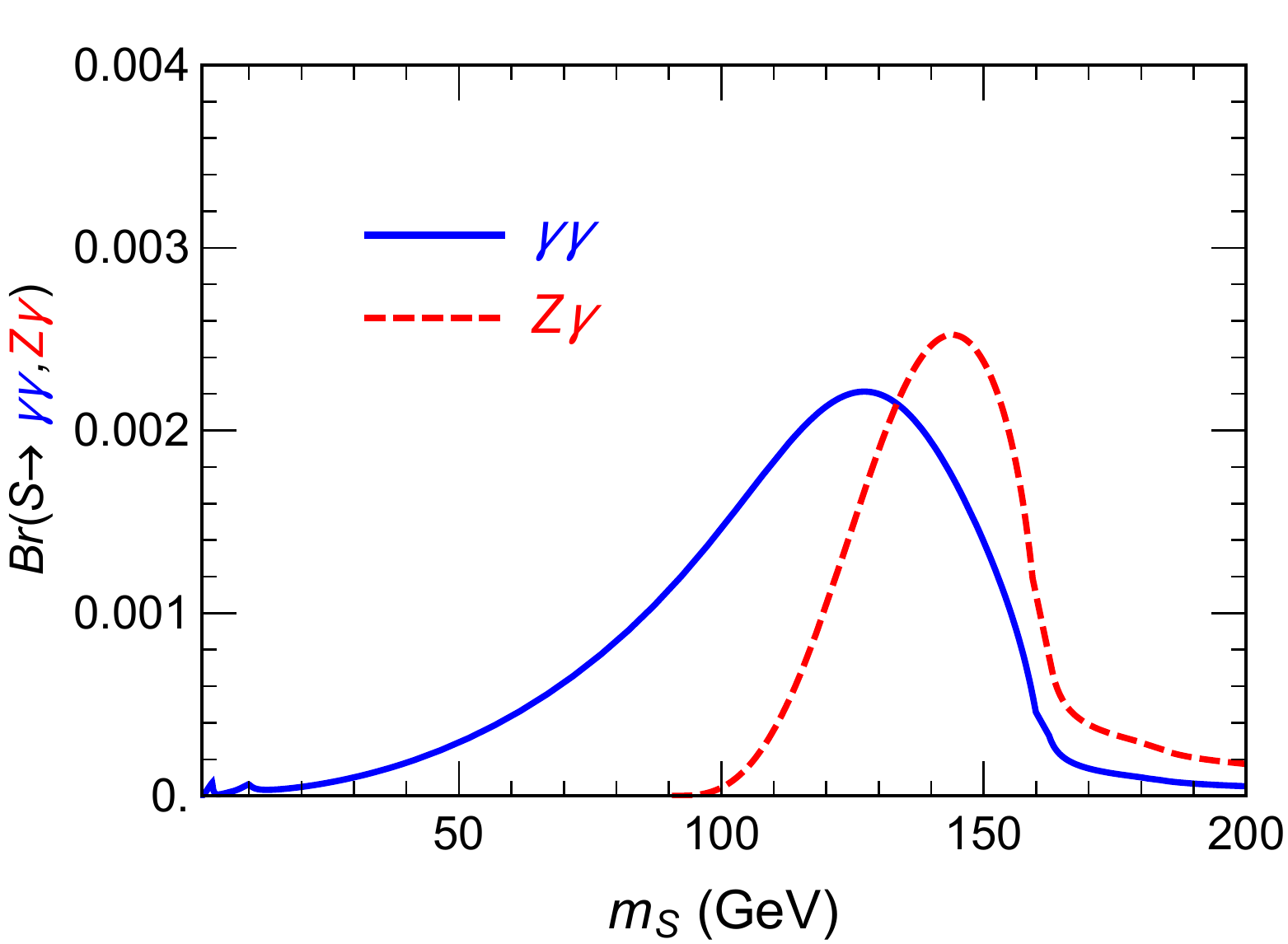}
\caption{The branching ratios (Brs) of the hidden scalar $S$ with $m_S <200$~GeV. The dominant modes are shown in the left panel \cite{Yang:2019bvg}, while $\gamma \gamma$ and $Z\gamma$ modes are given in the right panel.
The Brs are essentially independent of the value of $\alpha$. 
}
\label{fig:BrS}
\end{center}
\end{figure}

In  Fig.~\ref{fig:BrS}, we show  the branching  ratios of the hidden scalar, $S$, in the range $m_S <200$~GeV, where  a good fit of a photon spectrum showing gamma-line to the GC gamma-excess data can be obtained and will be discussed in the following analysis. For $m_S< 2m_h$, because all the decay widths of $S$  are proportional to $\sin^2\alpha$, its decay branching ratios are thus independent of the value of $\alpha$. The relevant formulas for decay widths of the hidden scalar $S$ are collected in Appendix~\ref{app:s-width}.

We consider the small $\alpha$ region, where the DM annihilation is dominated by $XX\to SS$, while $XX\to hh$ is negligible. More detailed discussions can be found in Appendix~\ref{app:XX2SS}. Moreover, we consider only a sliver region of the masses, where $(m_X -m_S) \ll m_X$, resulting in the produced $S$ to be close to rest, can account for the gamma-ray line phenomenon, and the value of $m_X (\approx m_S)$ is thus determined from the Fermi GC fit. The observed spectral line width, which depends on the energy resolution of the instrument, is very sensitive to the Lorentz-boost from the $S$ rest frame to the $XX$ com frame; the result is relevant to the mass difference of $m_X$ and $m_S$.

Using the low-velocity DM annihilation cross section obtained from the fit to the GC gamma-ray excess data, we can get the corresponding value of $g_{\rm dm}$ in this secluded vector DM model.

As for a small mixing angle $\alpha \lesssim 2 \times 10^{-6}$, the hidden sector has been thermally decoupled from the bath before it becomes nonrelativistic, such that the resulting  DM annihilation cross section that accounts for the correct relic density could be significantly boosted above the conventionally thermal WIMP value \cite{Yang:2019bvg}.

\section{Formulation of the Gamma spectrum with Prominent Lines Arising from  One-Step Cascade DM Annihilations}\label{sec:one-step}

\subsection{Basic formula of the differential gamma-ray flux originating from the DM annihilation}

The differential gamma-ray flux originating from the DM annihilation is given by
\begin{eqnarray} 
\label{eq:gammaflux}
\frac{d \Phi_\gamma}{dE} = \frac{\langle \sigma v\rangle_{\rm LV}}{8\pi m_X^2} 
\Bigg(\frac{dN_\gamma}{dE}\Bigg)_X \,  \frac{1}{\Delta\Omega}
 \underbrace{ \int_{\Delta\Omega}  \int_{\rm l.o.s.} ds \rho^2(r(s,\ell, b)) d\Omega }_{\text{J-factor}} \,.
\end{eqnarray}
Here, for the terms related to the particle physics, $\langle \sigma v\rangle_{\rm LV}$ is the DM annihilation cross section into two hidden Higgs scalars in the low-velocity limit\footnote{ 
For the thermally averaged annihilation cross section at the present day, the corresponding temperature is about $m_X (v_p/c)^2 /2$ with $v_p\sim 220~{\rm km}/s$ the most probable speed of the dark matter distribution (see Appendix B in Ref.~\cite{Yang:2017zor} for the relevant discussions). Thus, this s-wave cross section in the low-velocity limit, i.e. $T\to 0$, can be approximated as 
$\langle \sigma v \rangle_{\rm LV} = \sigma v_{\text{lab}}$ (see e.g. Ref.~\cite{Yang:2019bvg}). Here $v_{\rm lab}$ is the relative velocity measured in the laboratory frame, in which one of the incoming DM particles is at rest.}  (consistent with $T\to 0$), 
and  $(dN_\gamma / dE)_X$ is the resulting photon spectrum produced per DM annihilation in the com frame of DM.  On the other hand, the J-factor, related to the astrophysics, is the integral along the line of sight (l.o.s.) 
over the region of interest (ROI), which covers a rectangular solid angle $\Delta\Omega$ with galactic latitude and longitude  denoted by $b$ and $\ell$, respectively. For the GC data analysis, the  l.o.s. described by the coordinate $s$ is related to the distance to the GC by $r=(s^2 + r_\odot^2 - 2 r_\odot s \cos\ell \cos b)^{1/2}$ with $r_\odot$ being the distance from the Sun to the GC. 

For the GC gamma-ray excess study, we adopt the generalized Navarro-Frenk-White (gNFW) profile  \cite{Navarro:1995iw,Navarro:1996gj} as a canonical DM density distribution in our Galaxy,
 \begin{equation}
 \label{eq:gNFW}
 \rho(r)=\displaystyle \rho_{\odot} \left(\frac{r}{r_\odot}\right)^{-\gamma} \left(\frac{1+r/r_s}{1+r_\odot /r_s}\right)^{\gamma-3} \,,
 \end{equation}
where $\rho_\odot $ is  the local DM density corresponding to  $r = r_\odot$.

Below, we outline the calculation of the gamma-ray spectra generated from various $S$ decays following the DM annihilation $XX\to SS$.

\subsection{ Formulations of  the spectra arising from various channels of one-step cascade annihilations in the DM rest frame}\label{sec:gamma-spectrum}

For the process that DM annihilates into two on-shell mediators which subsequently decay through a small coupling into SM final states,  the resulting photon spectrum $(d{N}_\gamma/dE)_X$ defined in the com frame of DM can be written in terms of the spectrum $(dN'_\gamma/dE')_S$ described in the $S$ rest frame by considering a photon emitted at the angle $\theta'$ measured from the boost axis along which we can boost the $S$ rest frame by  a relative velocity $v = c \sqrt{1- m_S^2 / m_X^2}$ to obtain the result in the $X X$ com frame \cite{Elor:2015tva}. The result is given by
\begin{align}
\left( \frac{d {N_\gamma}}{dE} \right)_X
& = \frac{2}{m_X}\int_{-1}^{1} d\cos\theta'  \int^{1}_{0}  d x' 
\sum_f   {\rm Br}(S\to f)
  \left( \frac{d N'_\gamma}{dx'} \right)_S^f 
        \delta(2x -x' - \cos\theta' x' \sqrt{1-\epsilon^2} )  \nonumber \\
&=  \frac{2}{m_X}
  \sum_f   {\rm Br}(S\to f)  \int^{t_{\rm max}}_{t_{\rm min}} \frac{d x'}{x' \sqrt{1-\epsilon^2}} \Big( \frac{d N'_\gamma}{dx'} \Big)_S^f \,,
\label{eq:one-step}
\end{align}
where $x=E/m_X$, $x'=2E'/m_S$, and $\epsilon= m_S/m_X$, the factor ``2" on the right hand side (RHS) results from the increased multiplicity due to the fact that each $S$ decays to the SM final state $f$ with a branching fraction ${\rm Br}(S\to f)$, and, after performing the angular integration, the second line is the convolution integral with the bounds for $x'$, 
 \begin{equation}
 t_{\rm max} \equiv {\rm min} \Big[1, \frac{2 x}{\epsilon^2} (1+\sqrt{1-\epsilon^2}) \Big], \quad
  t_{\rm min} \equiv \frac{2 x}{\epsilon^2} (1-\sqrt{1-\epsilon^2}) \,.
   \end{equation}
  The kinematical range of the gamma-ray energy in the $X X$ com frame satisfies
\begin{align}
0 \leqslant E  \leqslant  \frac{m_X}{2} \left( 1+ \sqrt{1-\frac{m_S^2}{m_X^2}} \right)  \,.
\end{align}

\subsubsection{Gamma-ray spectrum generated from  $S \to \text{SM}~ \overline{\text{SM}} $ }

Considering the gamma spectrum arises from the $S$ decay into  a on-shell SM particle pair,  we employ the PPPC4DMID package\footnote{The package is also available from the website: ``http://www.marcocirelli.net/PPPC4DMID.html".}  \cite{Cirelli:2010xx,Ciafaloni:2010ti} with the replacement of the DM mass there by $m_S/2$ to generate  the direct spectra  $(d N'_\gamma/dx')_S$. This package,  including the electroweak corrections, was obtained by using PYTHIA 8.135 \cite{Sjostrand:2007gs}.  Below, we consistently use the results generated by PPPC4DMID as the essential inputs to obtain spectra of the remaining channels.

\subsubsection{Gamma-ray spectrum generated from the three-body decay $S \to V V^* \to V f_1 \bar{f_2} $ }

Below the $V V$ kinematical threshold but $m_S>m_V$, with $V \equiv W$ or $Z$,  the hidden scalar decays into a gauge boson pair, of which one ($V^*$) is off-shell,  resulting in $S \to V V^* \to V f_1 \bar{f_2} $.   In the $S$ rest frame, the gamma-ray spectrum generated from $S \to V V^*$ can be expressed as 
\begin{align}
(d N'_\gamma/dx')_S^{VV^*} = (d N'_\gamma/dx')_S^{V \to \gamma} + (d N'_\gamma/dx')_S^{V^* \to \gamma} \,, 
\end{align}
where, the former and latter terms on the RHS describe gamma-ray spectra that are produced by the cascade decays of $V$ and $V^*$, respectively. Here, as before, we define $x'= 2E' /m_S$ with $E'$ the photon energy measured in the $S$ rest frame.   The gamma-ray spectrum generated from $S\to V V^*$ can be obtained by convoluting the 3-body space with the spectrum arising from the cascade decay of $V$ and $V^*$. 
The relevant 3-body phase-space integral for the decay  $S \to V V^* \to V f_1 \bar{f_2} $ is given by  \cite{Kersevan:2004yh}
\begin{align}
\Phi_3 
=
\pi \int_{(m_1 + m_2)^2}^{(m_S - m_V)^2 }dM_2^2 
\frac{ \sqrt{\lambda(m_S^2, M_2^2, m_3^2)}}{ 2 m_S^2}  \,
\frac{ \sqrt{\lambda(M_2^2, m_1^2, m_2^2)}}{ 8M_2^2} d\Omega_2  \,,
\end{align}
where $\lambda(x,y,z) \equiv [x - (y+z)^2] (x-(y-z)^2]$, $M_2^2 \equiv p_{V^*}^{2}= (p_S - p_V)^2= m_S^2 - 2m_S E_V + m_V^2$, and the angle in $d\Omega_2$ is calculated in the com frame of $f_1$ and $\bar{f}_2$ with the invariant mass $M_2$.

The results will be briefly sketched as follows.
\begin{description}
\item[ (i)  $ (d N'_\gamma/dx')_S^{V \to \gamma}$] 

For a photon emitted from the $V$ cascade decay, the spectrum simply  satisfies the relation,
\begin{align}
 \Big( \frac{d N'_\gamma}{d E'} \Big)_S^{V\to\gamma}
  \propto &  \int_{(m_1 + m_2)^2}^{(m_S - m_V)^2 } \frac{d \Phi_3}{ dM_2^2} \,
       d M_2^2 \,  
        \Big( \frac{d N'_\gamma}{d E'} \Big)_{V(E_V)}^{ V\to \gamma}
     \,,
 \label{eq:vgamma-1}
\end{align}
where   $(d N^\prime_\gamma/d {E^\prime})_{V(E_V)}^{V \to \gamma}$ is the photon spectrum generated by the cascade decay of $V$ which has energy $E_V$ with respect to the $S$ rest frame. Changing variables,  
\begin{align}
&\xi   \equiv \frac{E_V}{m_S} \,, \quad
 x^\prime  \equiv \frac{2 E^\prime}{m_S} \,, \quad
\bar{x}  \equiv \frac{ E^\prime}{E_V} \equiv \frac{x^\prime}{2\xi} \,,    
 \nonumber \\
& \epsilon_1  \equiv \frac{m_1}{m_S}  \,, \quad
\epsilon_2  \equiv \frac{m_2}{m_S} \,, \quad
\epsilon_3  \equiv \frac{m_V}{m_S} \,,
\label{eq:parameter-3body}
\end{align}
with $m_1$ and $m_2$ being the masses of $f_1$  and $\bar{f}_2$, respectively,
we can recast the spectrum in the following form,
\begin{align}
    & \Big( \frac{d N'_\gamma}{dx'} \Big)_S^{V\to\gamma}  
  = \int_{\epsilon_3}^{\frac{ 1+\epsilon_3^2 - (\epsilon_1 + \epsilon_2)^2 }{2}}  d\xi  \,
   \frac{C_V}{2\xi} \nonumber\\
   &\quad \times
    \frac{ \sqrt{ (\xi^2 -\epsilon_3^2) \big(1-2\xi + \epsilon_3^2 - \big(\epsilon_1 + \epsilon_2)^2 \big)  \big(1-2\xi + \epsilon_3^2 - \big(\epsilon_1 - \epsilon_2)^2 \big) }}
                 {1-2\xi + \epsilon_3^2}    
      \Big( \frac{d N_\gamma^\prime }{d \bar{x}} \Big)_{V(E_V)}^{V\to\gamma}
    \,,
\end{align}
where $C_V$, which normalizes the spectrum, is given by
\begin{align}
C_V  
\equiv \left[
       \int_{\epsilon_3}^{\frac{ 1+\epsilon_3^2 - (\epsilon_1 + \epsilon_2)^2 }{2}} d\xi
              \frac{ \sqrt{ (\xi^2 -\epsilon_3^2) \big(1-2\xi + \epsilon_3^2 - \big(\epsilon_1 + \epsilon_2)^2 \big)  \big(1-2\xi + \epsilon_3^2 - \big(\epsilon_1 - \epsilon_2)^2 \big) }}
                 {1-2\xi + \epsilon_3^2}
    \right]^{-1} \,.
\label{eq:cv}
\end{align}
We will simply take $\epsilon_1=\epsilon_2=0$ and use $(1/2) (d N_\gamma^\prime/d{\bar x})_{\rm PPPC}^{V \to \gamma}$ to produce  
the direct spectrum  $ (d N_\gamma^\prime/d{\bar x})_{V(E_V)}^{V \to \gamma}$, where the subscript ``PPPC" denotes the result generated by the PPPC4DMID package but with the DM mass replaced by $E_V \equiv \xi m_S$. Here, the factor of ``1/2" accounts for the fact that in PPPC4DMID the spectrum is generated by two gauge bosons, $V$.  Note that the kinematic ranges of $\bar{x}$ and $x^\prime$ are given by $0 \leq \bar{x} \leq 1/2$ and $0\leq x^\prime \leq 1$.

\item [(ii) $ (d N_\gamma^\prime /dx')_S^{V^{*} \to \gamma}$]

 For a photon emitted from the $V^*$ cascade decays,  we have
\begin{align}
& \Big( \frac{d N'_\gamma}{dE'} \Big)_S^{{V^{*}}\to\gamma}
  \propto   \int_{(m_1 + m_2)^2}^{(m_S - m_V)^2 } \frac{d \Phi_3}{ dM_2^2} \, d M_2^2  
  \,
    \Big( \frac{d N^\prime_\gamma}{d {E'}} \Big)_{V^{*}(E_{V^*})}^{V^{*}\to\gamma}   \,,
\end{align}
where   $(d N^\prime_\gamma/d {E^\prime})_{V(E_{V^*})}^{V^* \to \gamma}$ is the photon spectrum generated by the cascade decay of $V^*$ which has energy $E_{V^*}=m_S-E_V$ with respect to the $S$ rest frame.

In order to compute $(d N^\prime_\gamma/d {E^\prime})_{V(E_{V^*})}^{V^* \to \gamma}$, we first consider the case with $V^*\equiv W^*$.
Note that the charges of $W$ have been summed in the width of $S\to W W^*$ given in Eq.~(\ref{HVV-3-2body}). Above the thresholds of the following channels, the ratio of the $W^{+*}$ decays approximately follows $\ell^+ \nu_\ell: U\bar{D} = 1 : N_c |V_{UD}|^2$, where $N_c\equiv 3$ is the number of colors, $V_{UD}$ is the Cabibbo-Kobayashi-Maskawa matrix, $\ell \in (e, \mu, \tau)$, $U\in (u,d)$ and $D\in (d,s,b)$.
As for $V^*\equiv Z^*$, its partial width satisfies 
\begin{align}
\Gamma(Z^* \to f \bar{f}) \propto 
\Bigg[ (g_V^2 +g_A^2) + 2(g_V^2 - 2 g_A^2) \frac{m_f^2}{m_{Z^*}} \Bigg]
\Bigg( 1- \frac{4m_f}{m_{Z^*}} \Bigg)^{1/2} \,,
\end{align}
where $g_V  = T_3 /2 -Q_f \sin^2\theta_W$ and $g_A = T_3 /2$ with $T_3$ and $Q_f$ being the weak isospin and electric charge of $f$, respectively.
For simplicity, we generically use $V^*\to f_{1,m}~ f_{2,m}$ to denote the two-body decay of the virtual vector boson.  We use the PPPC4DMID package to  obtain the spectrum,
\begin{align}
 \Big( \frac{d N^\prime_\gamma}{d {E'}} \Big)_{V^{*}(E_{V^*})}^{V^{*}\to\gamma}  
 \simeq &
  \frac{2}{m_S -E_V}
 \sum_m F_m \frac{1}{2} \Bigg[
      \left(\frac{d N^\prime_\gamma}{d \tilde{x}} \right)_{\rm PPPC}^{f_{1,m} \to \gamma}  
  + \left(\frac{d N^\prime_\gamma}{d \tilde{x}} \right)_{\rm PPPC}^{f_{2,m} \to \gamma}  
  \Bigg] 
   \,,      
  \label{eq:vstar-gamma}
\end{align}
where 
\begin{align}
\tilde{x} = \frac{2E^\prime}{m_S -E_V} \equiv \frac{x^\prime}{1-\xi} \,,
\end{align}
$(d N^\prime_\gamma / d \tilde{x} )_{\rm PPPC}^{f_{i,m}\to \gamma}$ (with $i\equiv 1,2$) is the gamma-ray spectrum arising from the cascade decays of the $f_{i,m}$ and $\bar{f}_{i,m}$ pair in the PPPC4DMID package with the DM mass replaced by $E_{V^*}/2 =(m_S-E_V)/2$.
Here, $F_m$, depending on parameters such as $N_c, g_V, g_A$ and CKM matrix elements as shown above, is the relative fraction for each channel, $m$, which is above the threshold.
Using the same notations as in Eq.~(\ref{eq:parameter-3body}), we can rewrite this spectrum in the following form,
\begin{align}
   \Big( \frac{d N'_\gamma}{dx'} \Big)_S^{V^*\to\gamma}  &  = \int_{\epsilon_3}^{\frac{ 1+\epsilon_3^2 - (\epsilon_1 + \epsilon_2)^2 }{2}}  d\xi 
   \,    \frac{C_V}{2(1-\xi)}
     \sum_m W_m \Bigg[
      \left(\frac{d N^\prime_\gamma}{d \tilde{x}} \right)_{\rm PPPC}^{f_{1,m} \to \gamma}  
  + \left(\frac{d N^\prime_\gamma}{d \tilde{x}} \right)_{\rm PPPC}^{f_{2,m} \to \gamma}  
  \Bigg]    \nonumber\\ 
&  \times 
  \frac{ \sqrt{ (\xi^2 -\epsilon_3^2) \big(1-2\xi + \epsilon_3^2 - \big(\epsilon_1 + \epsilon_2)^2 \big)  \big(1-2\xi + \epsilon_3^2 - \big(\epsilon_1 - \epsilon_2)^2 \big) }}
     {1-2\xi + \epsilon_3^2} 
    \,,
\end{align}
where $C_V$ is the normalization factor of the spectrum as given in Eq.~(\ref{eq:cv}).

\end{description}

\subsubsection{Gamma-ray spectrum generated from  $S \to \gamma \gamma $ }

We take the gamma line spectrum arising from the $S \to \gamma\gamma$ decay as a simple $\delta$-function form, 
\begin{align}
 \left( \frac{d N'_\gamma}{dx'} \right)_S^{\gamma \gamma}  = 2 \delta(x^\prime -1) \,,
\end{align}
in the rest frame of its parent $S$ particle. Therefore, for the line spectrum in the $XX$ com frame where the DM annihilates into two on-shell hidden scalars, each of which subsequently decays into two photons, the result can be written as 
\begin{align}
\left( \frac{d {N_\gamma}}{dx} \right)_X^{\gamma\gamma}
&=    {\rm Br}(S\to \gamma\gamma)  \frac{4}{  \sqrt{1-\epsilon^2}}  \,,
\label{eq:gammagamma-XX-frame}
\end{align}
where $x= E/m_X, \epsilon = m_S/ m_X$ (as defined previously), and 
\begin{equation}
\frac{1}{2} \Big( 1- \sqrt{1-\epsilon^2} \Big) \leq x  \leq  \frac{1}{2} \Big( 1 +  \sqrt{1-\epsilon^2} \Big) \,.
\end{equation}

When fitting the monochromatic line(s), which is likely much narrower than the experimental energy resolution, we account for the finite resolution of the instrument. The observed line spectrum measured by the detector at energy $E(=x\, m_X)$ can be modeled by convolving the signal with a Gaussian energy dispersion, 
\begin{align}
\left( \frac{d {N_\gamma}}{dx} \right)_X^{\gamma\gamma}  = \int_0^1 dx_0  \frac{1}{\sqrt{2\pi}  \sigma \, x} \,
  \exp\Big[ -\frac{ (x_0 -x)^2}{2 \sigma^2 x^2}  \Big]  
 S(x_0) \,,
\label{eq:gaussian}
\end{align}
where $\sigma$ is related to the detector energy resolution $\xi$ as $\sigma = \xi/ (2\sqrt{2\ln 2}) \simeq \xi/ 2.35$, which is the ratio of the full peak width at half maximum to mean energy  \cite{Lewin:1995rx},  and
\begin{equation}
S(x_0) \equiv   \left( \frac{d {N_\gamma}}{dx} \right)_X^{\gamma\gamma} \Bigg|_{x\to x_0} \,,
\end{equation}
is the result given by Eq.~(\ref{eq:gammagamma-XX-frame}) but with $x$ replaced by $x_0$.

\subsubsection{Gamma-ray spectrum generated from the decay $S \to Z \gamma $ }

 The decay $S \to Z \gamma $ exhibits a continuum spectrum plus a gamma line with a finite width, for which 
 at the $S$ rest frame, it has a central energy,
 \begin{align}
  E^\prime = \frac{m_S}{2} \left( 1- \frac{m_Z^2}{m_S^2} \right) \,,
 \end{align}
 depending on the mass of  $Z$.
 The  gamma line spectrum for this channel $S\to Z\gamma$ at the $S$ rest frame can be expressed in terms of the decay width,
 \begin{align}
\left(\frac{d N_\gamma^\prime}{d E^\prime} \right)_{S, {\rm line}}^{S\to Z\gamma} &=   \frac{1}{\Gamma_{Z\gamma}} \frac{d \Gamma_{Z\gamma}}{d E^\prime}  \,.
 \end{align}
Using the narrow width approximation for the resonance $Z$, the differential width can be written as 
\begin{align}
& \frac{d \Gamma_{Z\gamma}}{d M^2}   
 =  \frac{\Gamma_{Z\gamma} \cdot \Gamma_Z m_Z}{ {\bf B} }
    \left|\frac{1}{M^2 -m_Z^2 + i\Gamma_Z m_Z }\right|^2 \,,
\end{align}
where $\Gamma_Z$ is the total width of $Z$, $M^2 = m_S^2 - 2m_S E^\prime$, and ${\bf B}=\tan^{-1}\delta +\tan^{-1} \left(\frac{m_S^2-m _Z^2}{\Gamma_{Z} m_Z } \right)$ which is a normalization factor corresponding to 
\begin{align}
 (m_Z^2 - \delta\,  \Gamma_{Z} m_Z)  \leq M^2 \leq m_S^2  \,.
\end{align}
In the limit $\delta\to m_Z/\Gamma_Z$ and $\Gamma_Z\to 0$, one has ${\bf B} =\pi$. If taking $\delta=2 m_Z/m_S$, our narrow width approximation is consistent with that used in Refs.~\cite{Bertone:2009cb,Jackson:2009kg}. 
Numerically, we will use $\delta=3$. The result is insensitive to the value of $\delta\gtrsim 2$, especially when $\delta \gtrsim 3$. Changing the variable from $M^2$  to $E^\prime$, we obtain
 \begin{align}
\left(\frac{d N_\gamma^\prime}{d E^\prime} \right)_{S, {\rm line}}^{S\to Z\gamma} 
 & = \frac{\Gamma_{Z\gamma}}{2m_S} \frac{1}{\tan^{-1}\delta +\tan^{-1} \left(\frac{m_S^2-m _Z^2}{\Gamma_{Z} m_Z } \right) }
    \frac{1}{ \left[ E^\prime - \frac{m_S}{2} \left( 1-\frac{m_Z^2}{m_S^2} \right) \right]^2  + \frac{\Gamma_{Z}^2 m_Z^2}{4m_S^2} } \,,
 \end{align}
where 
\begin{align}
0 \leq E^\prime \leq \frac{m_S}{2} \left(1- \frac{m_Z^2}{m_S^2} \right) + \delta \frac{\Gamma_{Z}}{2m_S} \,.
\end{align}

For $S\to Z\gamma$, the continuum spectrum results from the cascade decay of $Z$. The energy of $Z$ emitted from $S\to Z\gamma$ in the $S$ rest frame are
\begin{align}
E_Z = \frac{m_S}{2}  \left(  1 + \frac{m_Z^2}{m_S^2} \right) \,.
\end{align}
Thus, the continuum spectrum in the $S$ rest frame can be written as
 \begin{align}
\left(\frac{d N_\gamma^\prime}{d E^\prime} \right)_{S, {\rm cont}}^{S\to Z\gamma} 
&= \frac{1}{2} \left( \frac{d N_\gamma^\prime}{d{\hat x}} \right)_{\rm PPPC}^{Z \to \gamma}
 \frac{2}{m_S + m_Z^2/m_S} \,,
 \label{eq:szgamma-con}
 \end{align}
 where $\hat{x} =E_\gamma^\prime/E_Z$ and we have used $(d N_\gamma^\prime/d{\hat x})_{\rm PPPC}^{Z \to \gamma}$ from PPPC4DMID in which the DM mass is replaced by $E_Z \equiv  (m_S +m_Z^2/m_S)/2$. In Eq.~(\ref{eq:szgamma-con}), the factor of ``1/2" is due to the fact that the PPPC4DMID spectrum is given by two $Z$ bosons.  
 
In summary,  the photon spectrum of $S\to Z\gamma$ in the $S$ rest frame is given by
 \begin{align}
\frac{d N_\gamma^\prime}{d x^\prime} (S\to Z\gamma)   
& =  
\frac{m_S}{2} \left[
\left(\frac{d N_\gamma^\prime}{d E^\prime} \right)_{S, {\rm line}}^{S\to Z\gamma}   + 
\left(\frac{d N_\gamma^\prime}{d E^\prime} \right)_{S, {\rm cont}}^{S\to Z\gamma} 
\right]  \nonumber\\
&= 
\left(\frac{d N_\gamma^\prime}{d x^\prime} \right)_{S, {\rm line}}^{S\to Z\gamma}   + 
\left(\frac{d N_\gamma^\prime}{d x^\prime} \right)_{S, {\rm cont}}^{S\to Z\gamma} \,,
 \label{eq:szgamma-t}
 \end{align}
where $x^\prime \equiv 2E^\prime /m_S$.  As shown in Eq.~(\ref{eq:gaussian}),  we  will further consider the energy resolution of the instrument for the gamma line signal by convolving the spectrum with a Gaussian kernel.

\subsubsection{Gamma-ray spectrum for ${\rm DM}~{\rm DM} \to SS$ with $m_S=m_h$, in comparison with  ${\rm DM}~{\rm DM} \to h h$}

\begin{figure}[htb!]
  \begin{center}
\includegraphics[width=0.32\textwidth]{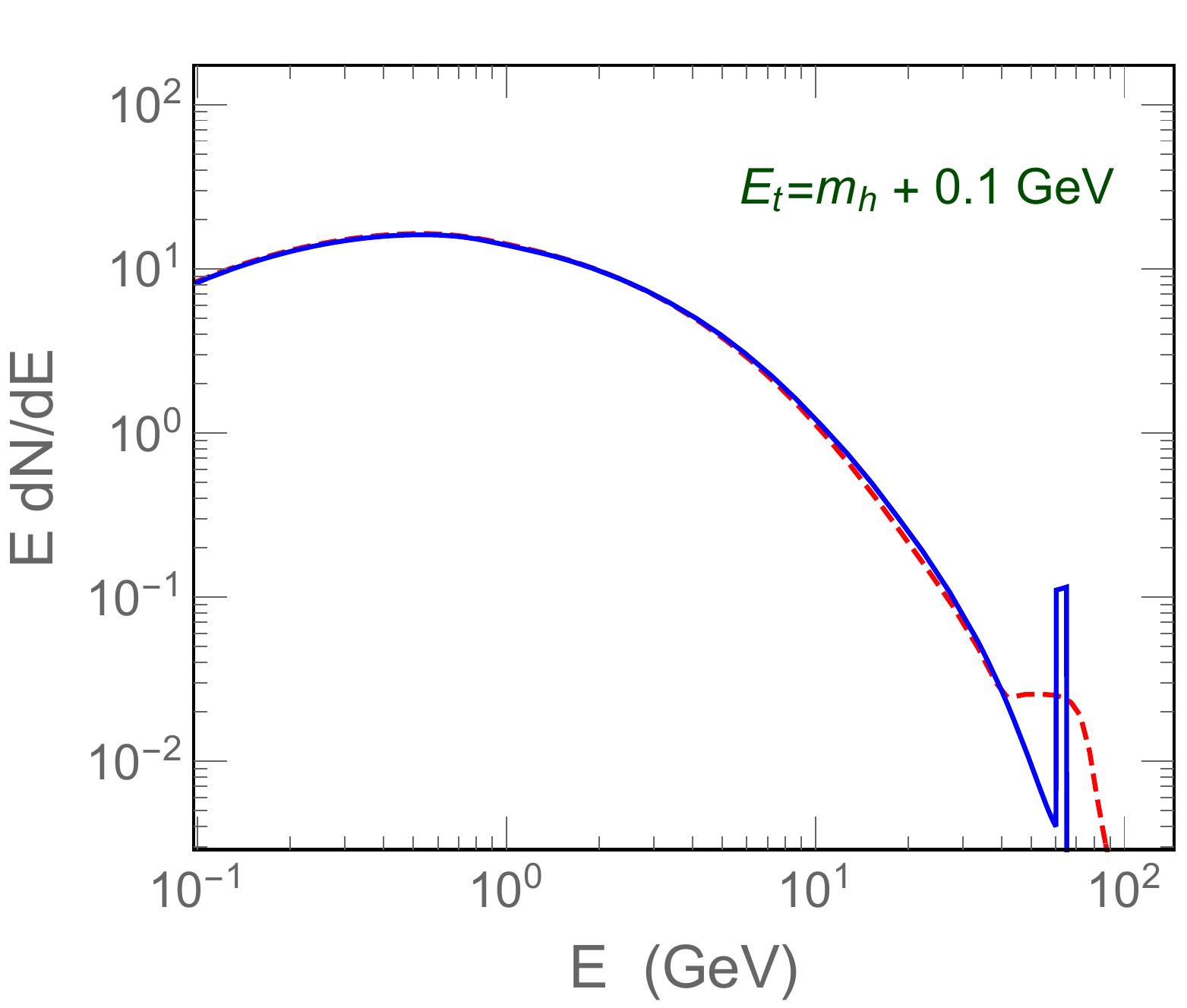}\hskip0.15cm
   \includegraphics[width=0.32\textwidth]{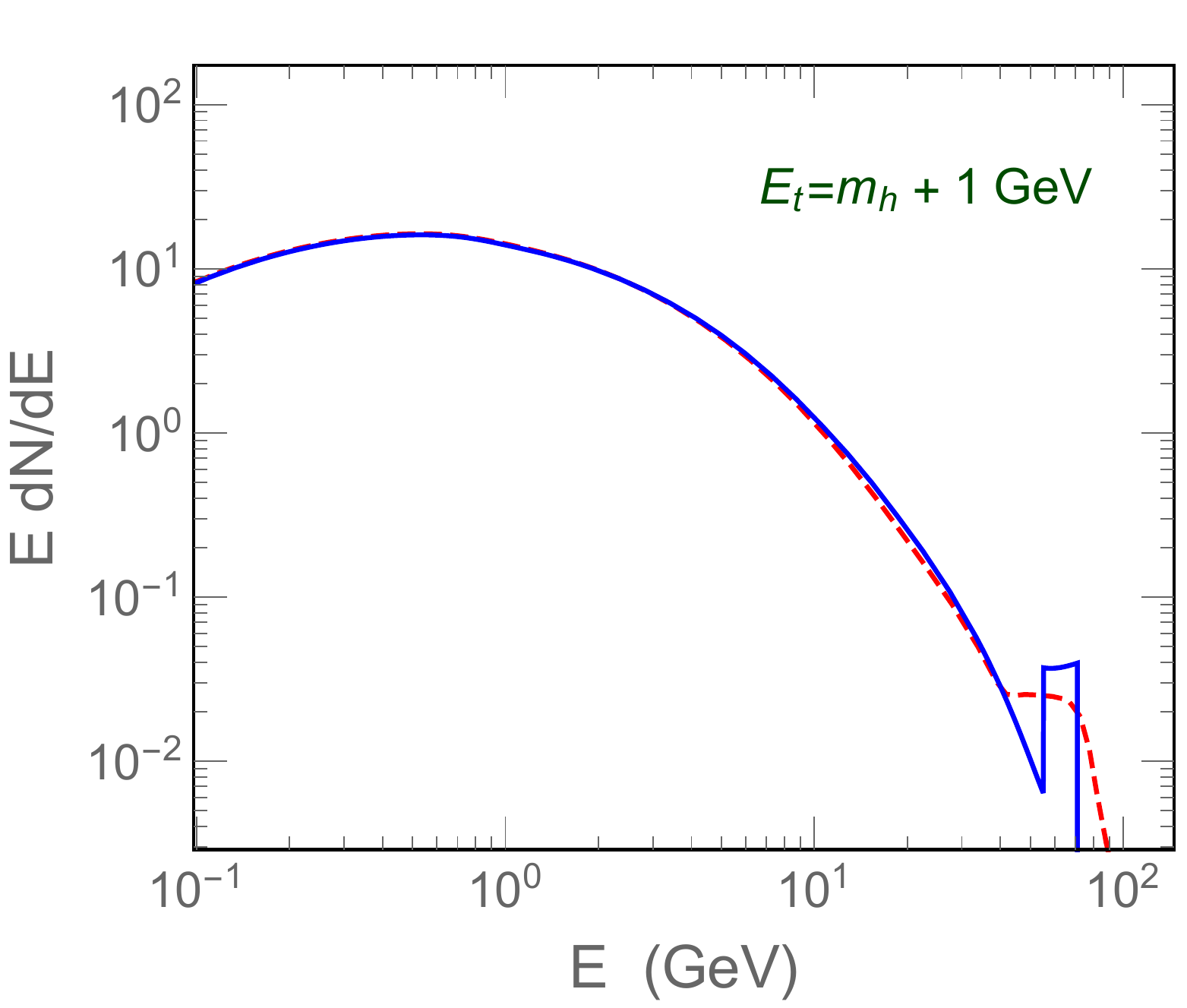} \hskip0.15cm
   \includegraphics[width=0.32\textwidth]{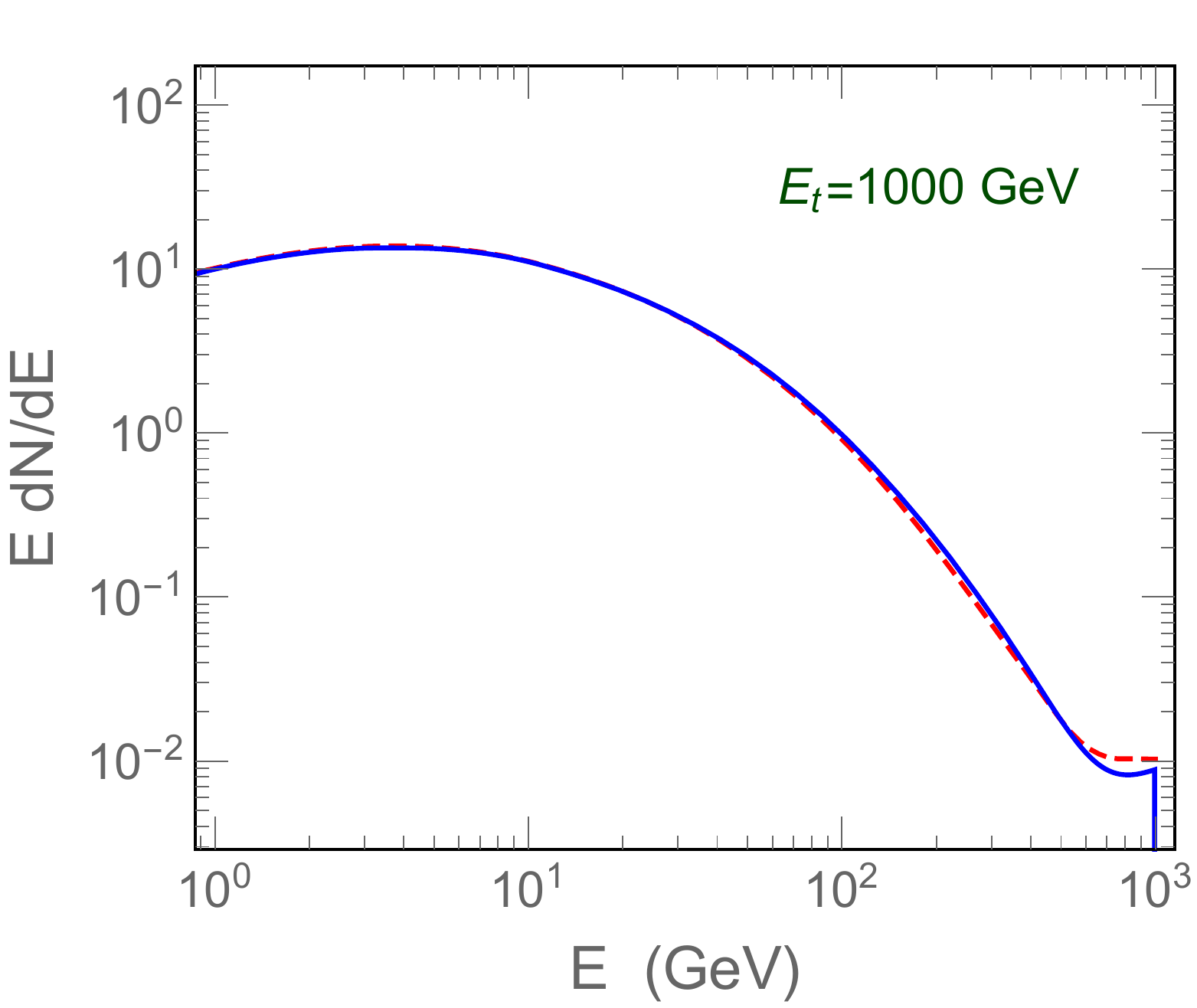} 
\caption{Gamma-ray spectra for  ${\rm DM}~{\rm DM} \to SS$ with $m_S=m_h$. Here $E_t (=m_{\rm DM})$,  is the total energy of a final state $S$. The solid blue curve is our result, in comparison with the case of  ${\rm DM}~{\rm DM} \to h h$ (dashed red curve) obtained  directly from the PPPC4DMID package  \cite{Cirelli:2010xx,Ciafaloni:2010ti}, which was generated from Pythia 8.135  \cite{Sjostrand:2007gs}.}
\label{fig:spectrumcheck}
\end{center}
\end{figure}

In Fig.~\ref{fig:spectrumcheck}, using the above formulas, we show the gamma-ray spectra for  ${\rm DM}~{\rm DM} \to SS$ (blue curve) with $m_S=m_h$, in comparison with the case of  ${\rm DM}~{\rm DM} \to h h$ (red curve) obtained by the PPPC4DMID package  \cite{Cirelli:2010xx,Ciafaloni:2010ti}, which was generated from Pythia 8.135  \cite{Sjostrand:2007gs}. Physically, in the limit $m_S=m_h$, the produced spectrum, independent of the mixing angle $\alpha$, should be the same for these two annihilation modes.  
For the case generating energetic $S$ particles, our result is in good agreement with PPPC4DMID, while for the case of the final states $S$ having a low kinetic energy,  our result has a better resolution for the spectrum at energies about the $S \to \gamma \gamma$ production (see Eq.~(\ref{eq:gammagamma-XX-frame})).

\section{Results}\label{sec:results}

\subsection{Fits to the Galactic center excess spectrum}

In order to satisfy the purpose of having a  good fit to the Fermi GC excess spectrum and to show the spectral line structure, we take into account three cases: (i) $m_S=0.99\, m_X$, (ii) $m_S=0.999\, m_X$, and (iii) $m_S=m_h$  (=125.18~GeV), for which the first two cases can figure out the boost dependence of the observed spectral line width due to the small mass difference of $m_X$ and $m_S$, and the third case is expected to be consistent with the WIMP case dominated by $XX\to h h$ as it should be. We can use the third case to evaluate the validity of our calculation. 

We fit the DM mass $m_X$ and low-velocity annihilation cross section $\langle\sigma v\rangle$ to the Fermi GC gamma-ray excess spectrum carried out by CCW \cite{Calore:2014xka}. CCW result covers the photon energy range 300 MeV$-$500 GeV, within ROI extended to $| \ell| \leq 20^\circ$ and $2^\circ \leq |b| \leq 20^\circ$.
We perform a $\chi^2$ fit, given by \cite{Calore:2014xka}
\begin{equation}
\chi^2
=
\sum_{ij \in \text{bins}}
\left(
  \frac{d\Phi_\gamma}{dE_i}(m_X,\langle\sigma v\rangle ) -
     \frac{d\Phi^{\text{obs}}_\gamma}{dE_i}
\right)
\cdot \Sigma_{ij}^{-1} \cdot
\left(
  \frac{d\Phi_\gamma}{dE_j}(m_X,\langle\sigma v\rangle ) -
     \frac{d\Phi^{\text{obs}}_\gamma}{dE_j}
\right)\,,
\end{equation}
where the covariance matrix $\Sigma$ contains  statistical error, correlated empirical model systematics and correlated residual systematics, for which the latter two are non-diagonal. Here  $ d\Phi_\gamma/dE_i $ and $ d\Phi_\gamma^{\rm obs}/dE_i$ respectively denote the model prediction and (CCW) central value of the observed flux in the $i{\rm th}$ energy bin with $i\in [1,24]$ in the energy range.

For the gNFW halo profile, we use the scale radius $r_s=20$~kpc,  $r_\odot=8.5$~kpc, $\gamma=1.2$ and $\rho_\odot=0.4$~GeV/cm$^3$ as canonical inputs in the analysis. 
Because the CCW analysis was performed on the  Fermi Pass 7 data, of which the energy resolution is about 10\%\footnote{The energy resolution of Pass 8 (P8R3\_SOURCE\_V2) has been improved to be about 6\%$-$8.5\% from 10~GeV to 200~GeV;  see ``http://www.slac.stanford.edu/exp/glast/groups/canda/lat\_Performance.htm". } \cite{Ackermann:2012kna}, we thus use $\xi=0.1$ for the line spectra, which are generated from $S \to \gamma\gamma$ and  $Z\gamma$, in the numerical fit.

 \begin{figure}[t!]
  \begin{center}
\includegraphics[width=0.31\textwidth]{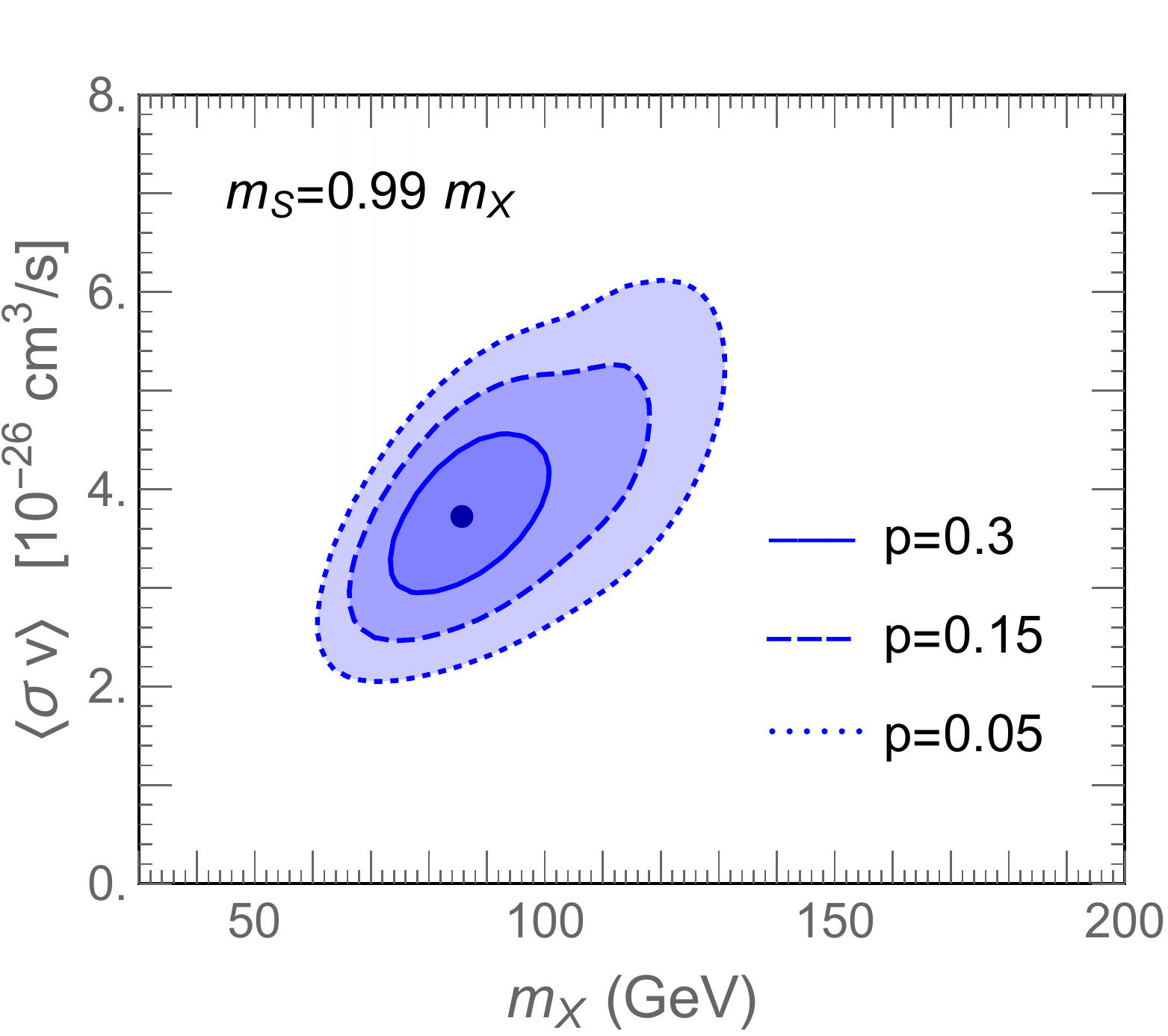}\hskip0.1cm
   \includegraphics[width=0.325\textwidth]{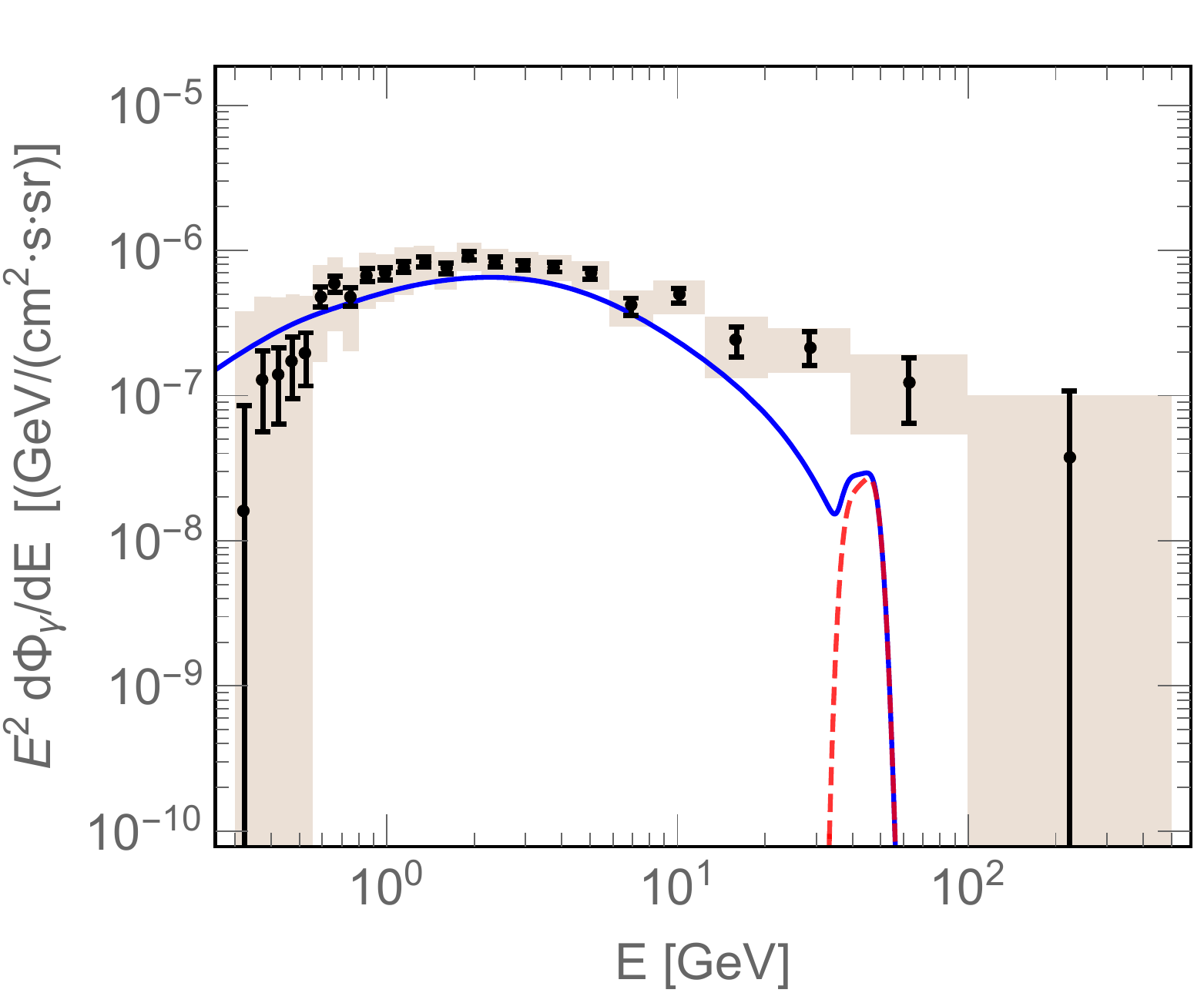} \hskip0.1cm
   \includegraphics[width=0.325\textwidth]{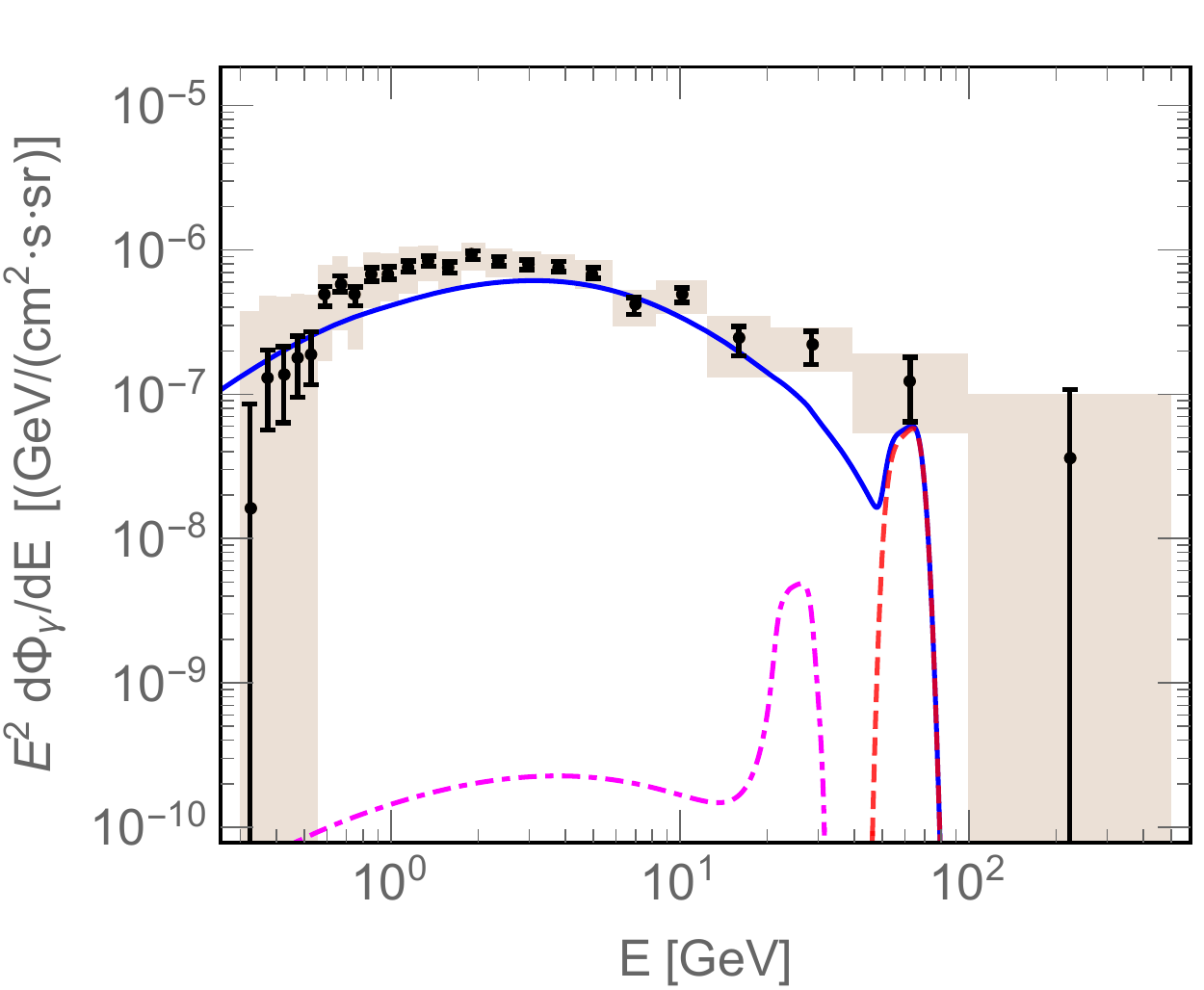} 
\\
\includegraphics[width=0.31\textwidth]{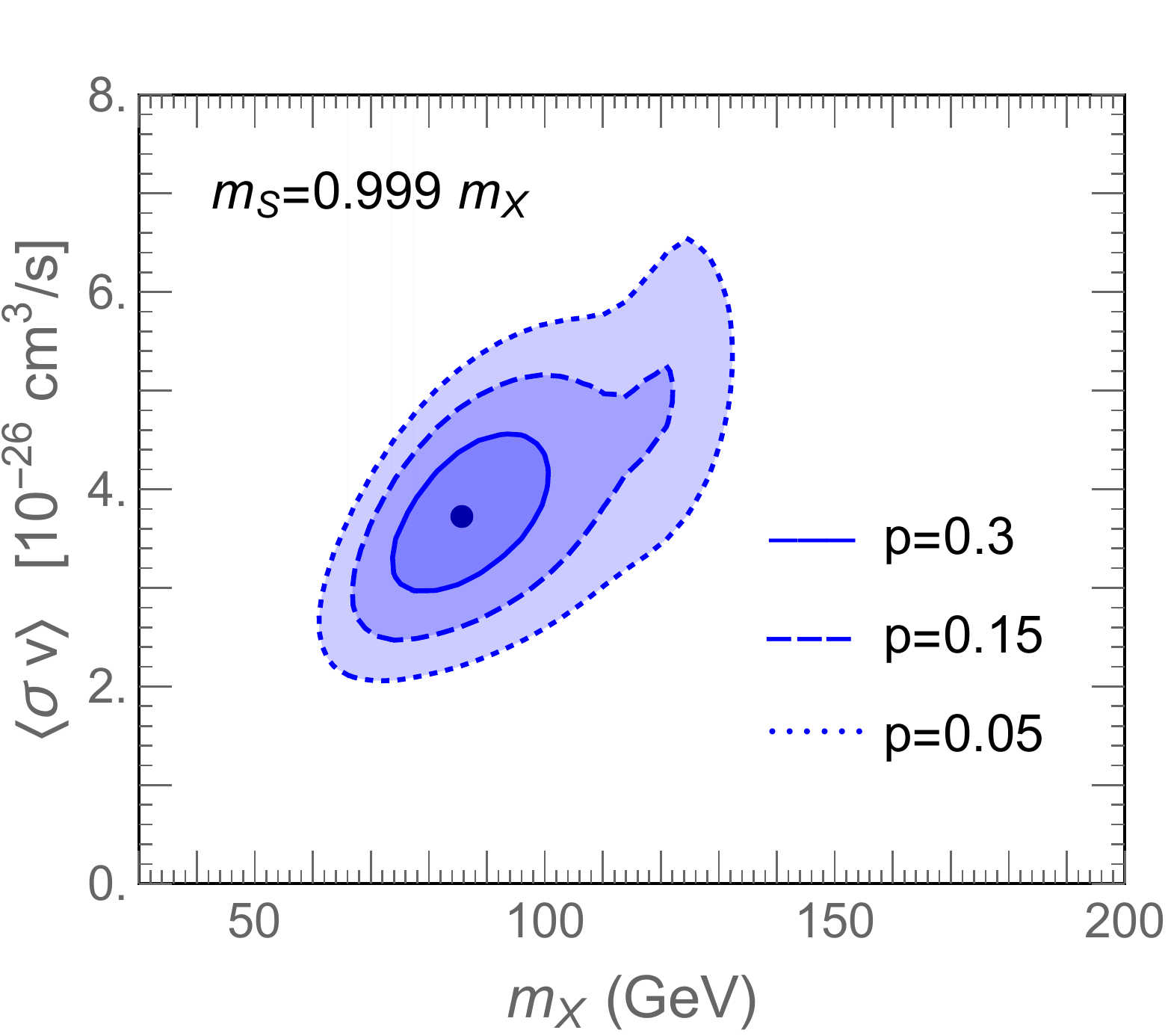}\hskip0.1cm
   \includegraphics[width=0.325\textwidth]{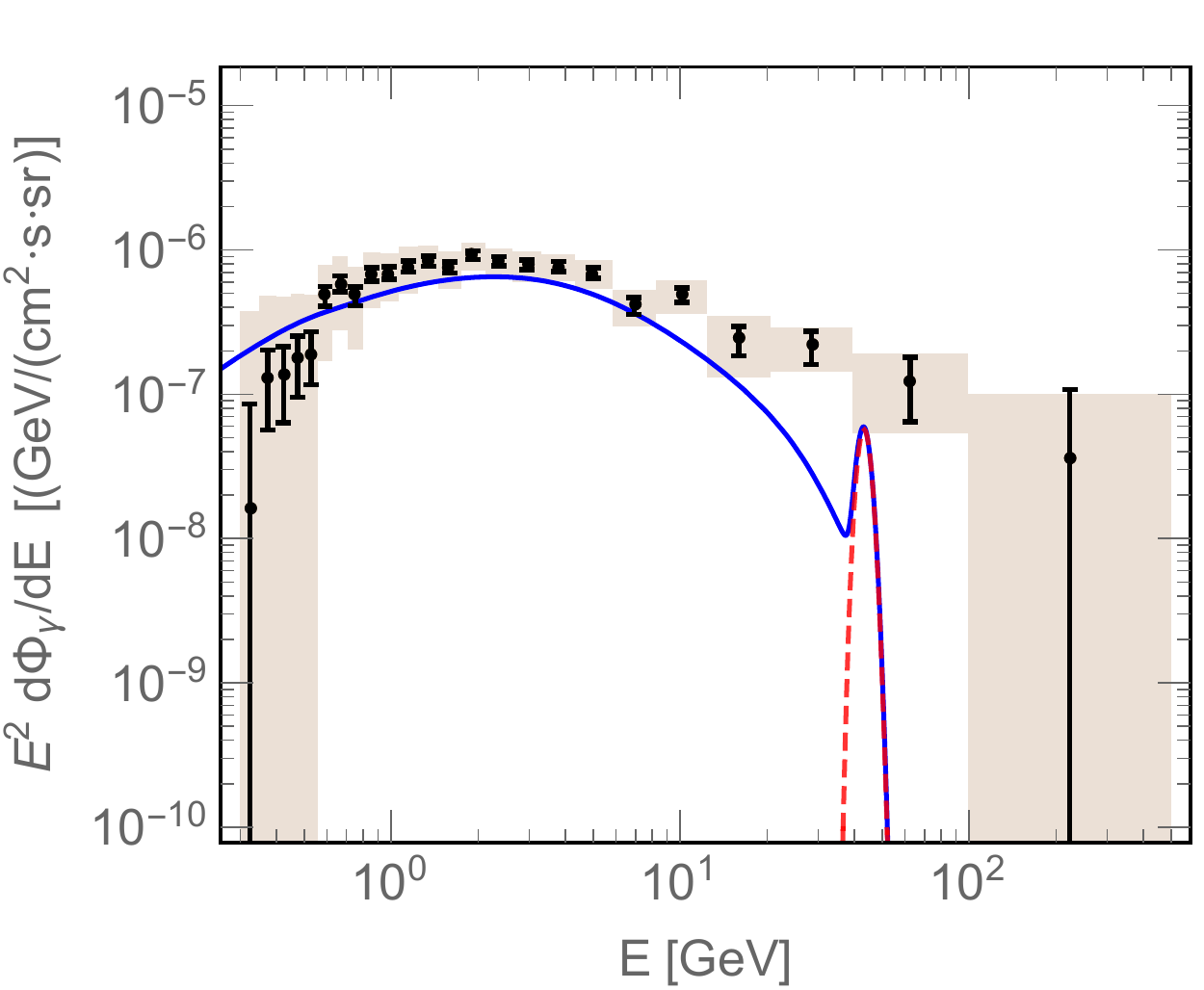} \hskip0.1cm
   \includegraphics[width=0.325\textwidth]{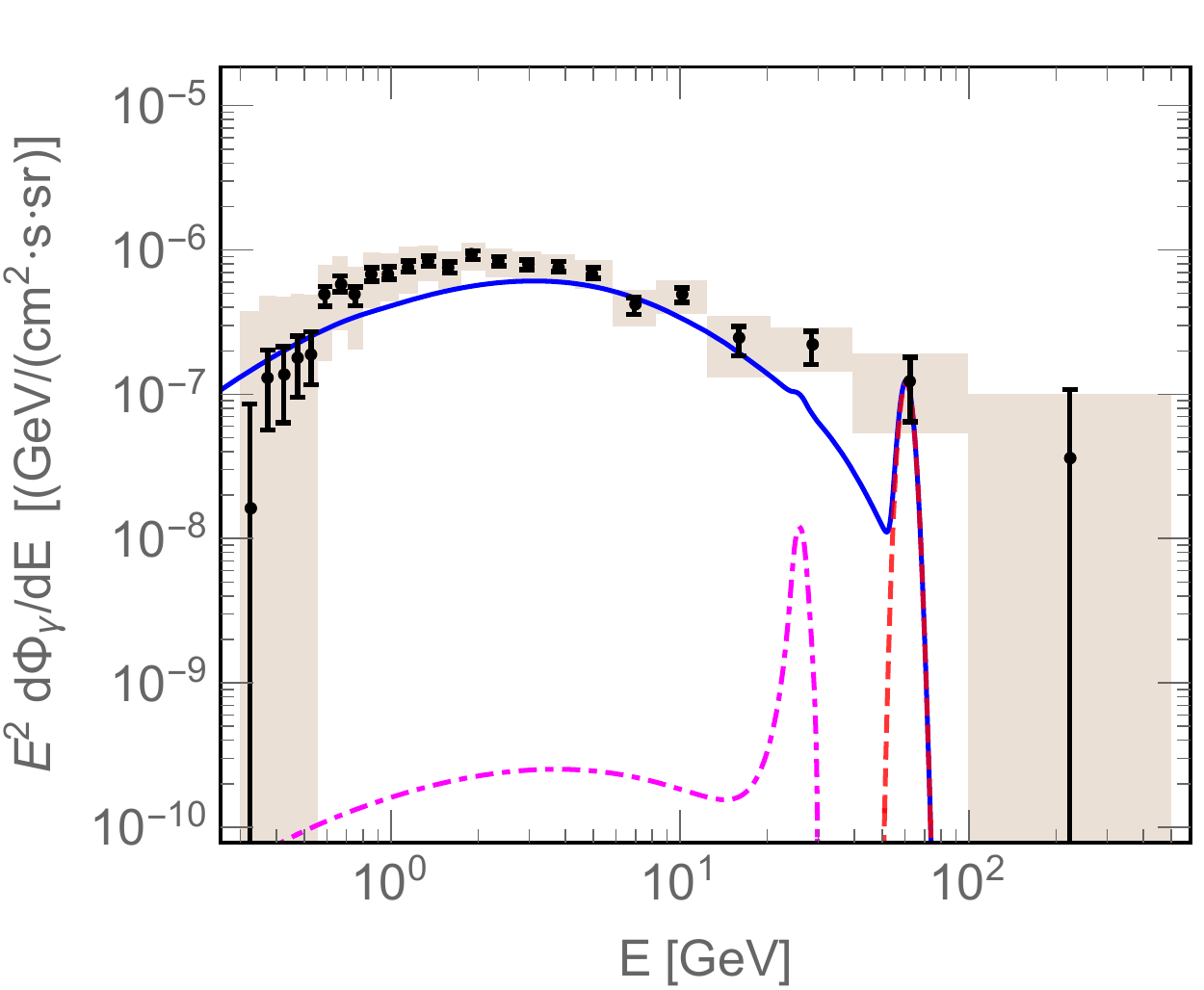} 
\\
\includegraphics[width=0.31\textwidth]{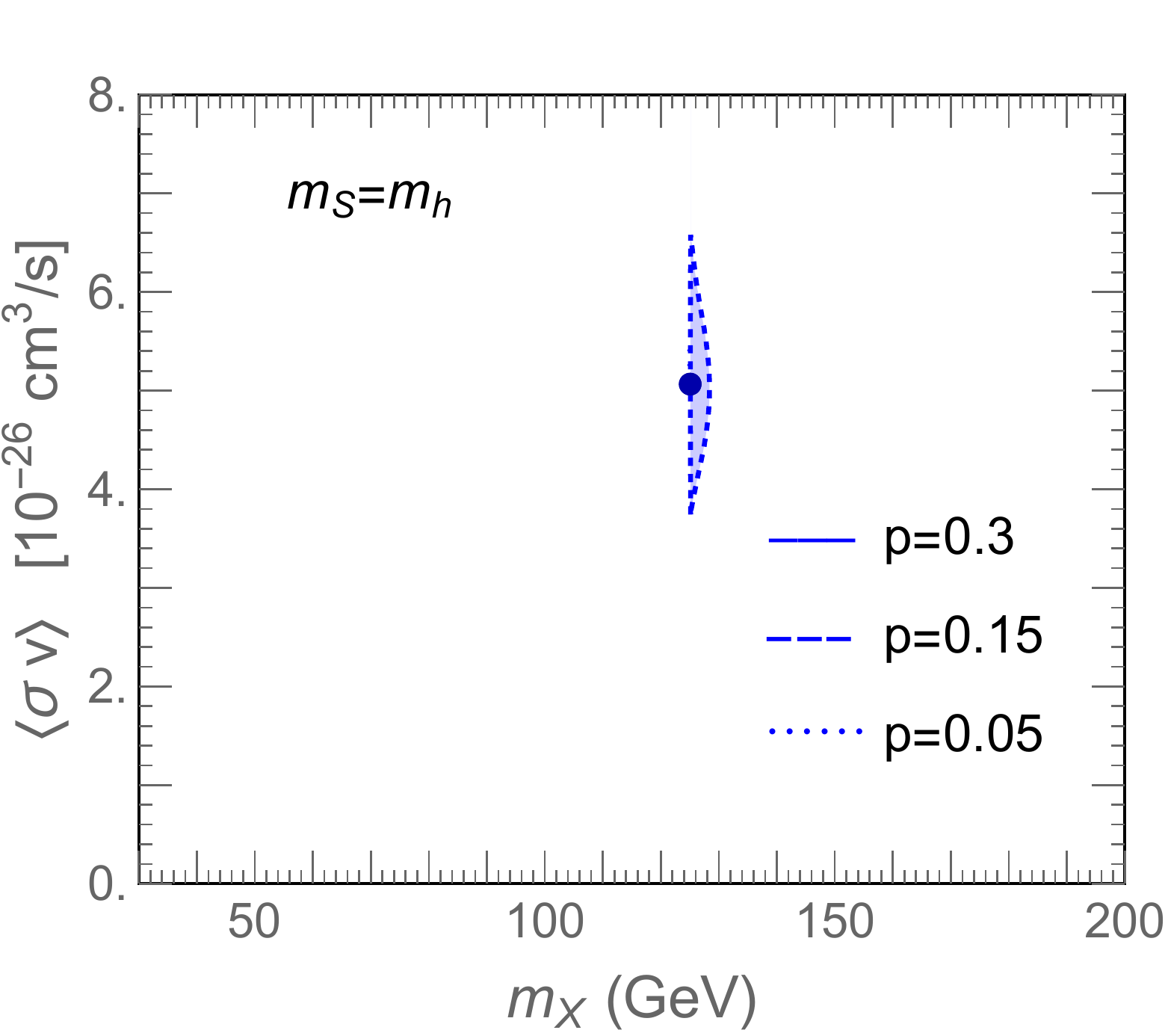}\hskip0.1cm
   \includegraphics[width=0.325\textwidth]{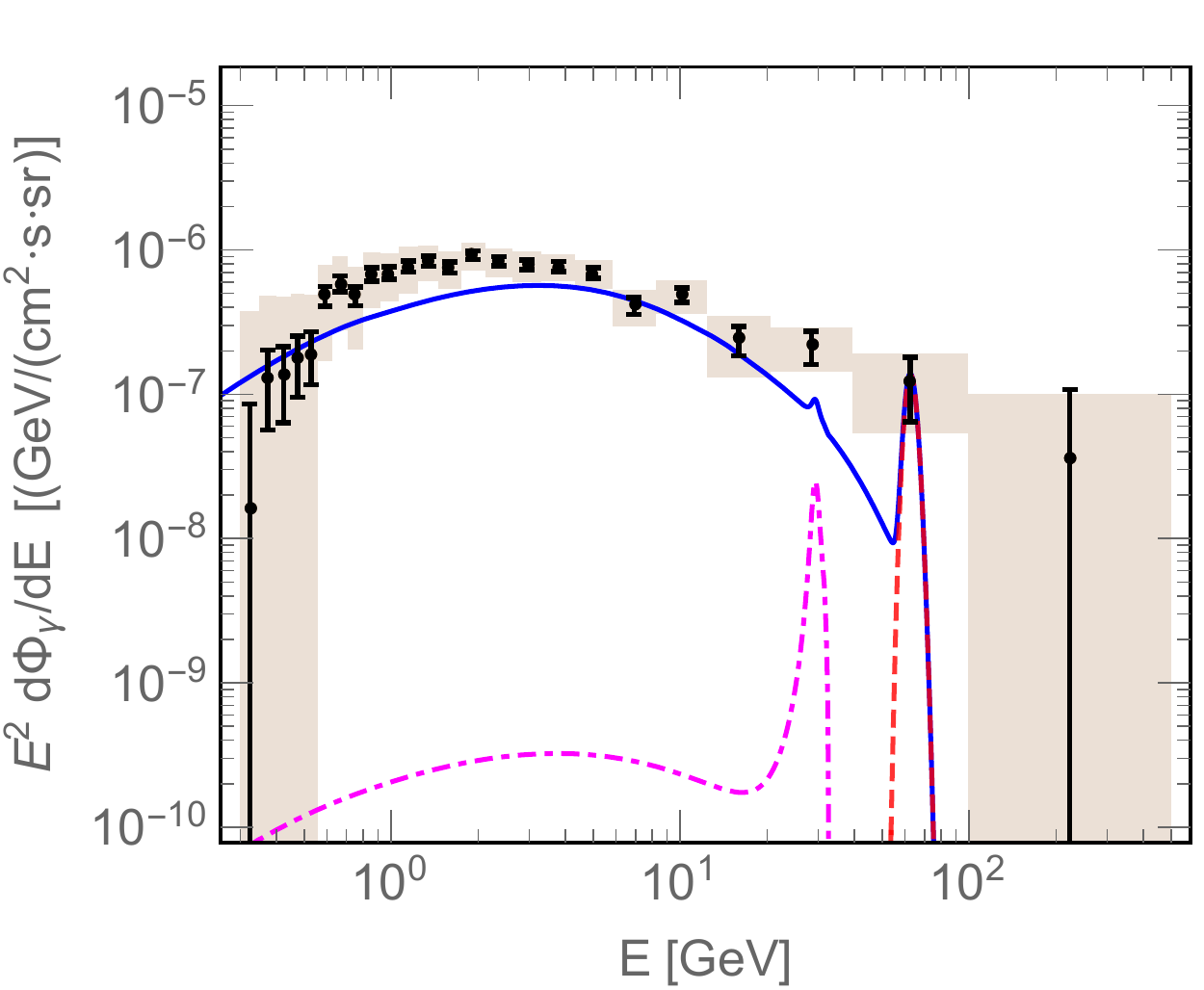} \hskip0.1cm
   \includegraphics[width=0.325\textwidth]{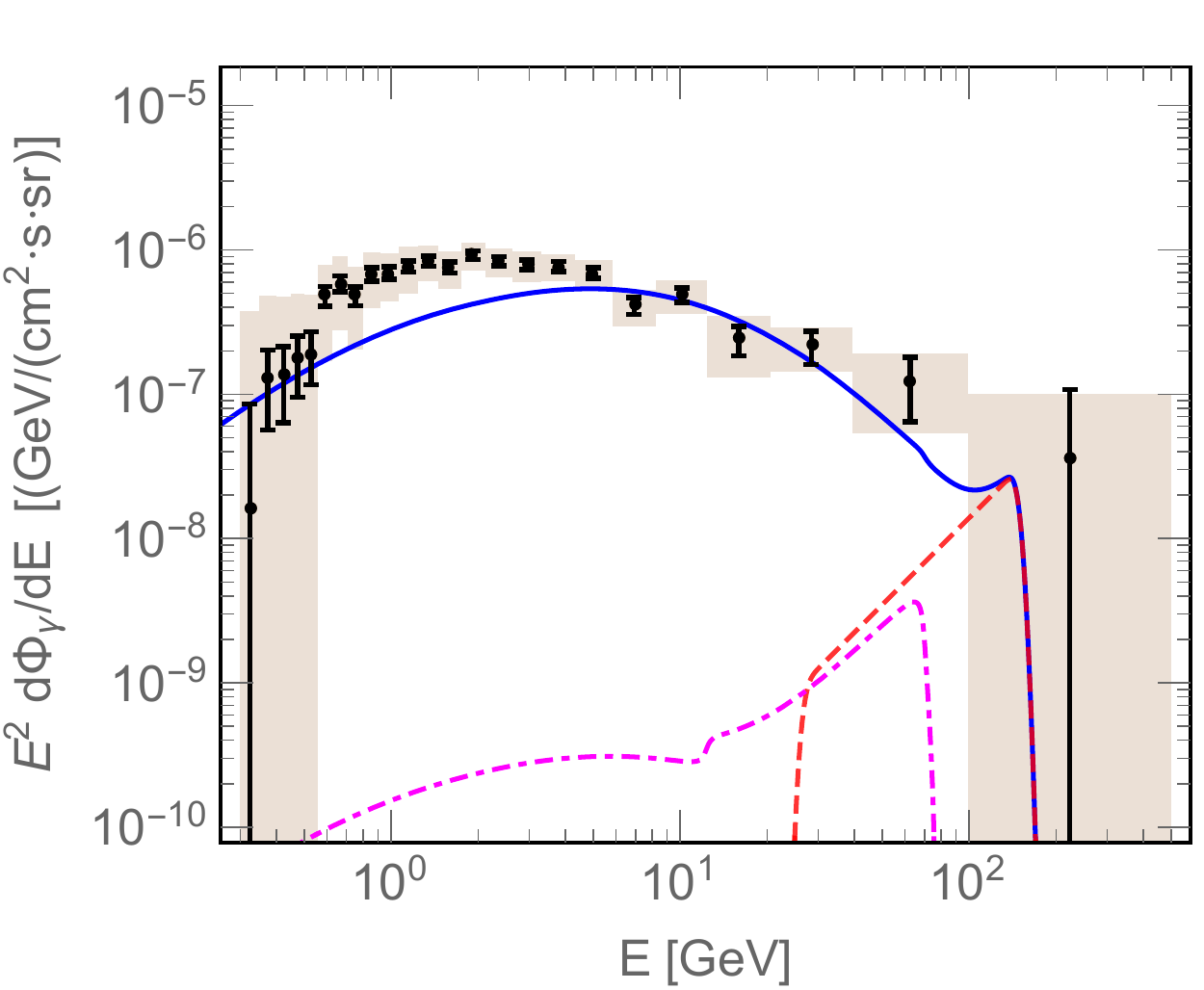} 
\caption{ Panels from up to down are, respectively, the results using $m_S=0.99\, m_X, 0.999\, m_X$ and $ m_h\, (=125.18~\text{GeV})$ as the input.  All analyses refer to $\rho_\odot = \text{0.4 GeV/cm}^3$ and $\gamma=1.2$ in the gNFW profile, and the detector energy resolution $\xi=0.1$.
 Left panel: GC excess data preferred regions, where the corresponding $p$-values of boundaries are given, and the best-fit point is denoted as the black dot.
Middle panel: The best fit (blue curve, using the best fit values of  $m_X$ and $\langle \sigma v\rangle$ as inputs) vs. GC excess spectrum \cite{Calore:2014xka} for which the error bars represent the statistical errors, while brown rectangles represent the diagonal part of the covariance matrix from systematical errors, including empirical model systematics and residual systematics.  The corresponding spectra resulting from $S \to \gamma \gamma$ and $Z \gamma$ are depicted by the dashed (red) and dot-dashed (magenta) curves. Right panel: same as the middle panel but the values for $m_X$ and $\langle \sigma v\rangle$ are rescaled by 1.4 times to draw all the corresponding curves; this is for illustrative purposes. See Table~\ref{tab:GC} for the best-fit values (for the middle panel) and the corresponding $p$-values, while for the right panel from up to down the theoretical curves  correspond to $p=0.12, 0.15$, and $2\times 10^{-6}$.}
\label{fig:gc-mxsv-resolution-s}
\end{center}
\end{figure}

\begin{table}[t]
\setlength{\tabcolsep}{5pt}
\renewcommand{\arraystretch}{1.3}
\center
\begin{tabular}{cccccc} 
\hline \hline 
$m_S/m_X$ & $m_S$ & $\langle\sigma v\rangle$   & $m_X$ &  $\chi^2_\text{min}/dof$  & $p$-\text{value} 
\\[-8pt]
input & input & $[10^{-26}$ cm$^{3}$ s$^{-1}]$   &  [GeV]  &   & \\
\hline
0.99   & --- & $3.74$ & $86.0$ & $22.66/22$  & $0.42$
\\
0.999 & --- & $3.74$ & $86.2$ & $22.69/22$  & $0.42$
\\
--- & $m_h$ & $5.17$ & $m_h$+0.019 & $29.40/22$  & $0.13$
 \\
\hline \hline
\end{tabular}
\caption{  \small  Values of the best fits to the GC gamma-ray excess spectrum for three values of $m_S/m_X= 0.99, 0.999$, and $m_S=m_h$ (=125.18~GeV).    The corresponding $p$-value of $\chi^2_{\rm{min}}$ is given, and $dof\equiv$ degrees of freedom. Here $\rho_\odot = \text{0.4 GeV/cm}^3$ and $\gamma=1.2$ are used as the canonical inputs.}
\label{tab:GC}
\end{table}

 \begin{figure}[t!]
  \begin{center}
   \includegraphics[width=0.33\textwidth]{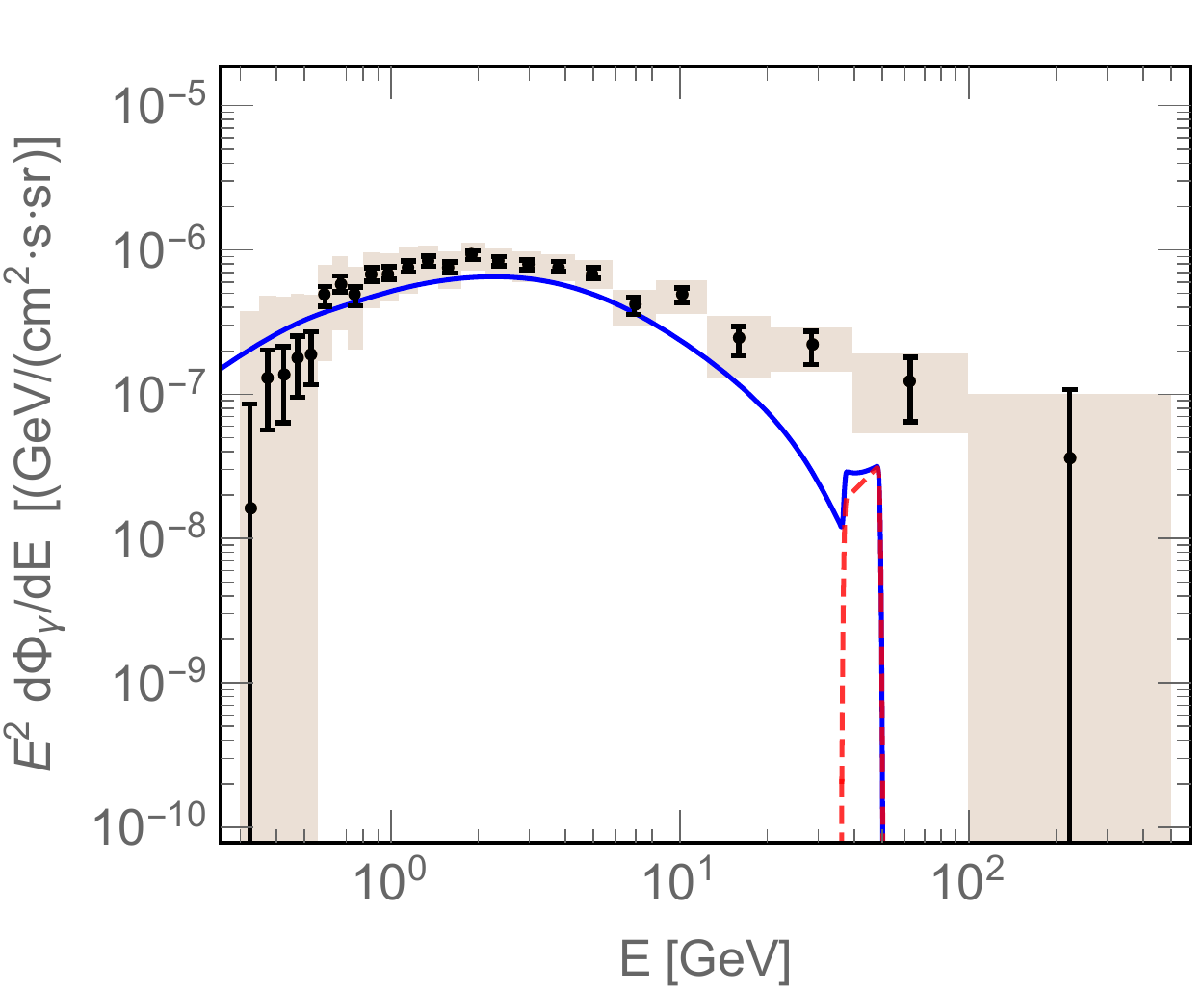} \hskip0.1cm
   \includegraphics[width=0.33\textwidth]{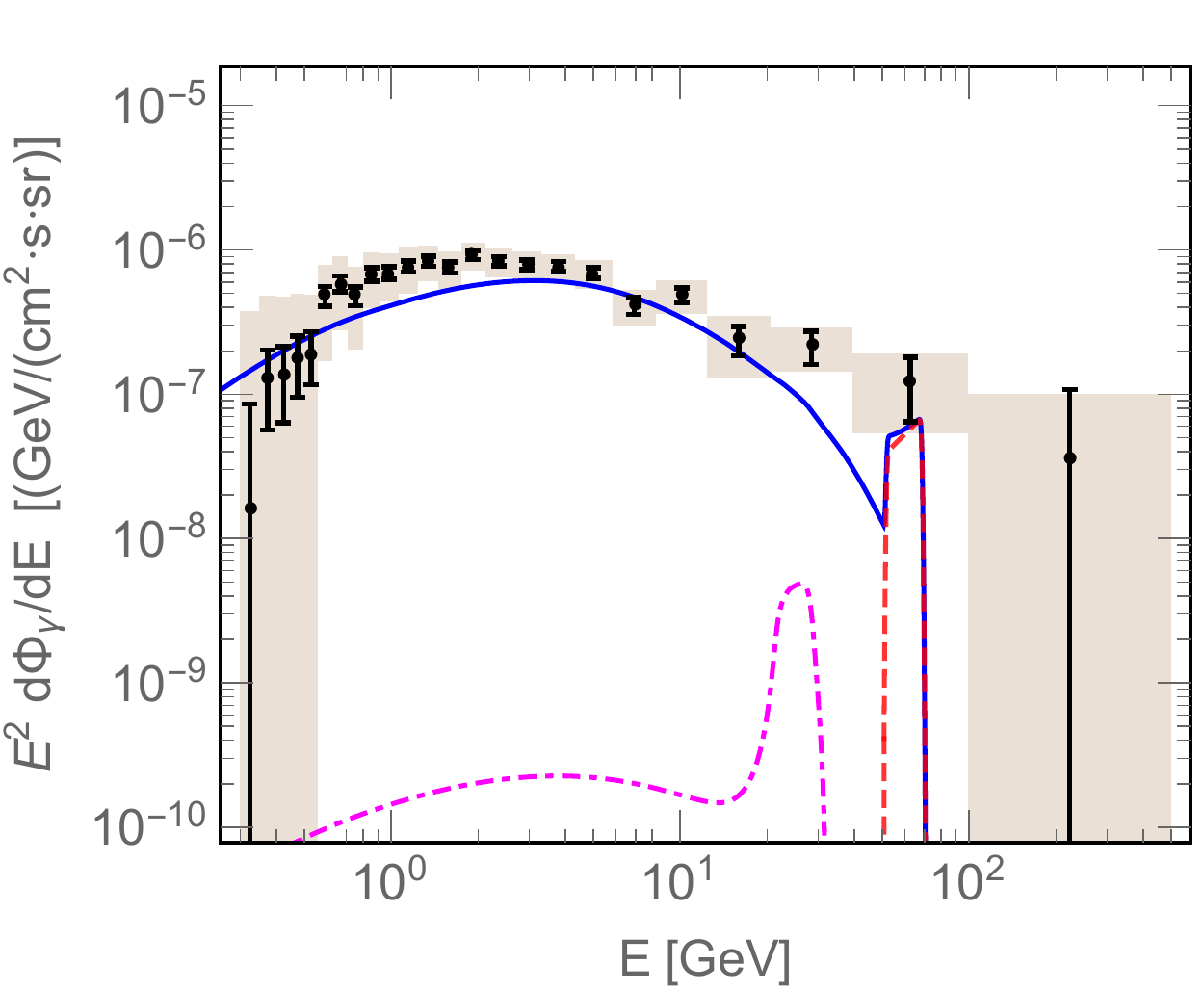} 
\\
   \includegraphics[width=0.33\textwidth]{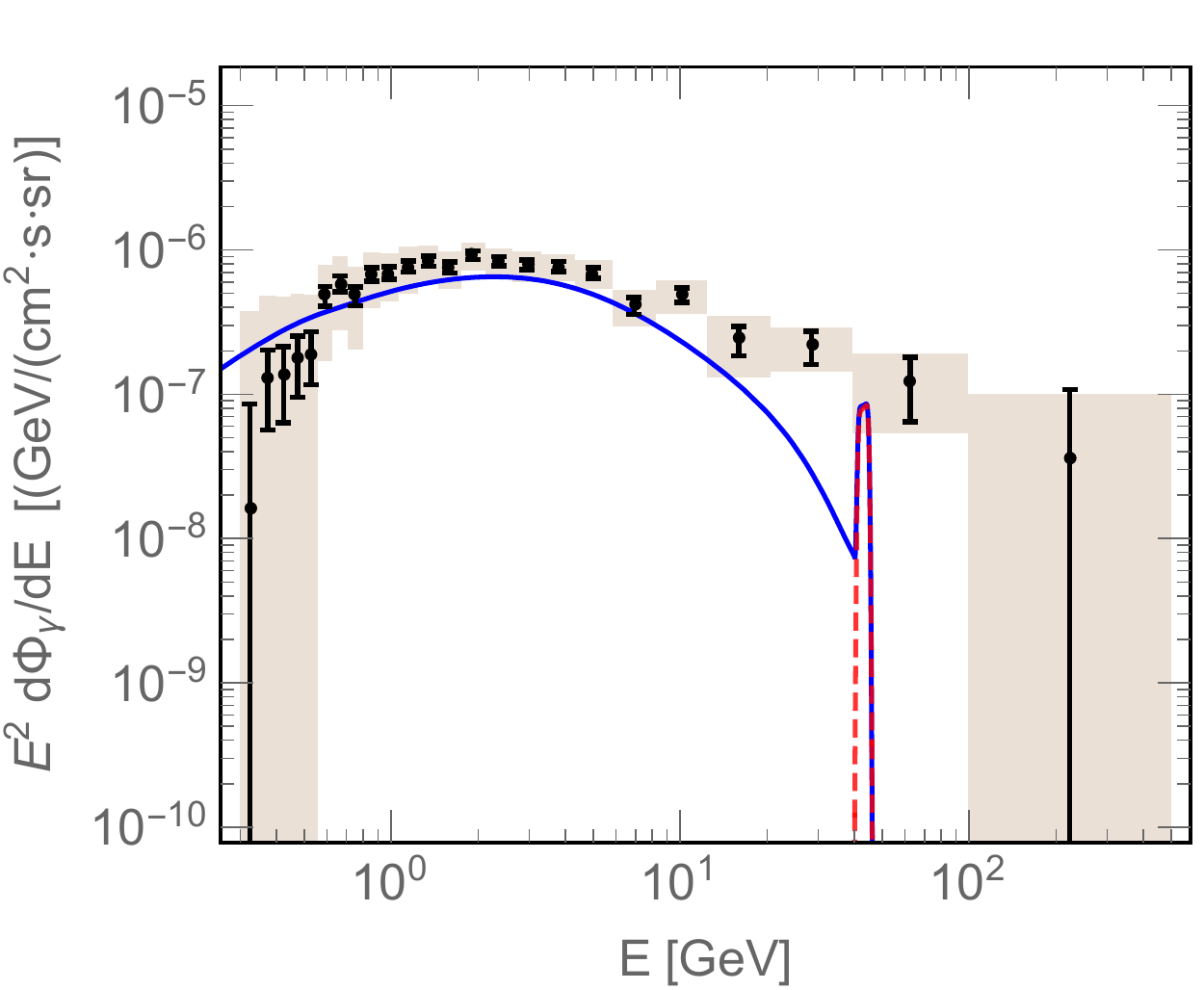} \hskip0.1cm
   \includegraphics[width=0.33\textwidth]{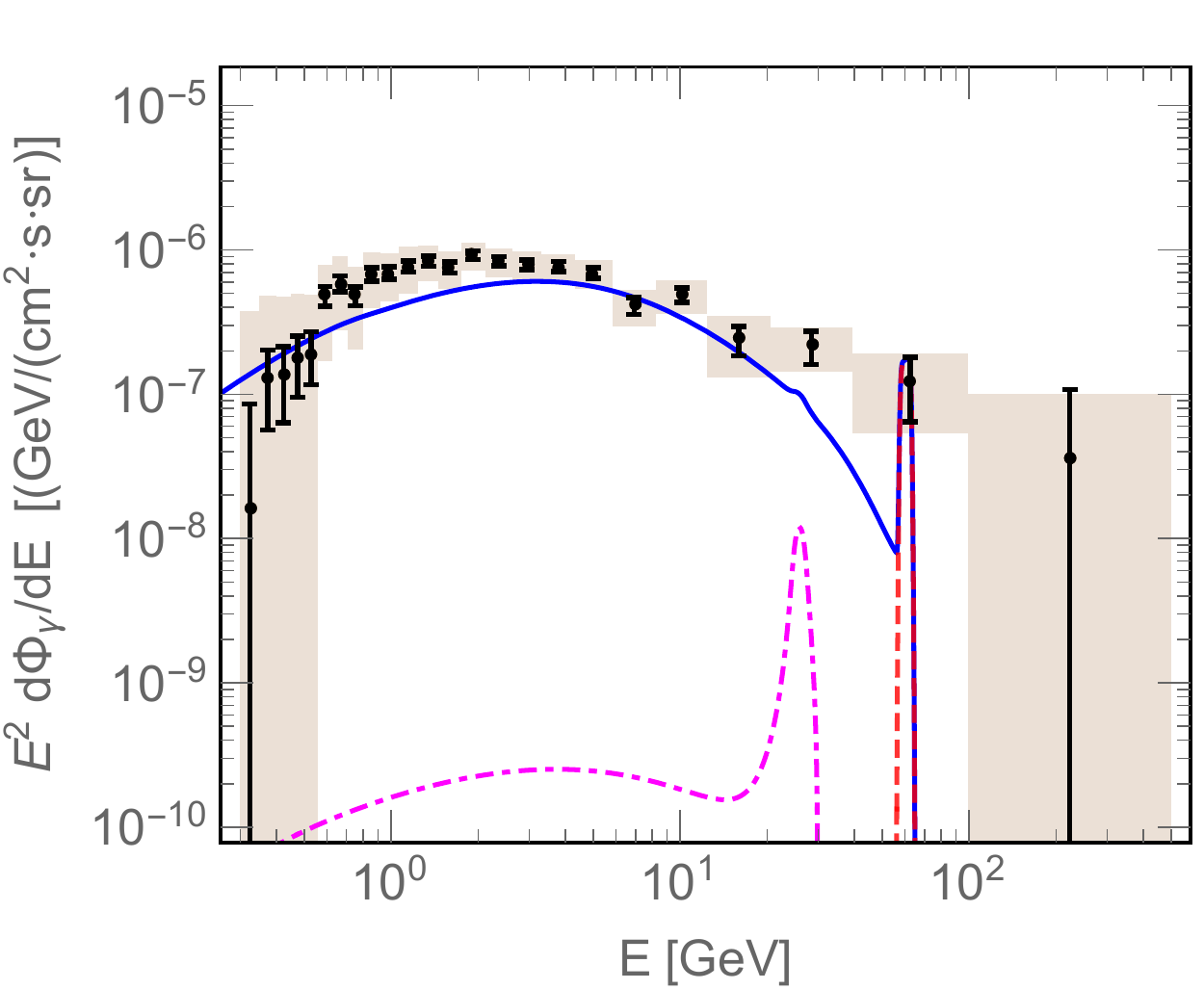} 
\\
   \includegraphics[width=0.33\textwidth]{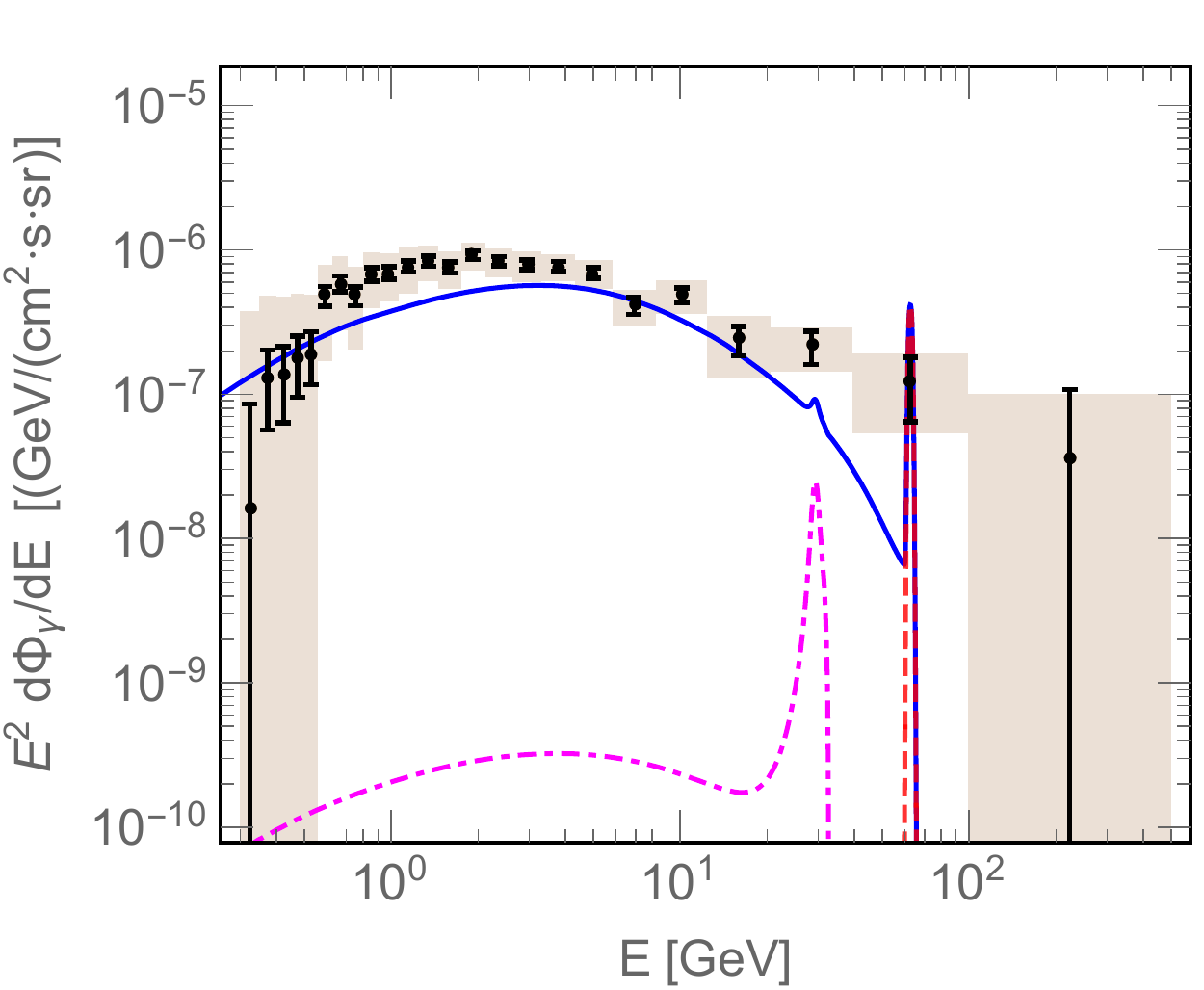} \hskip0.1cm
   \includegraphics[width=0.33\textwidth]{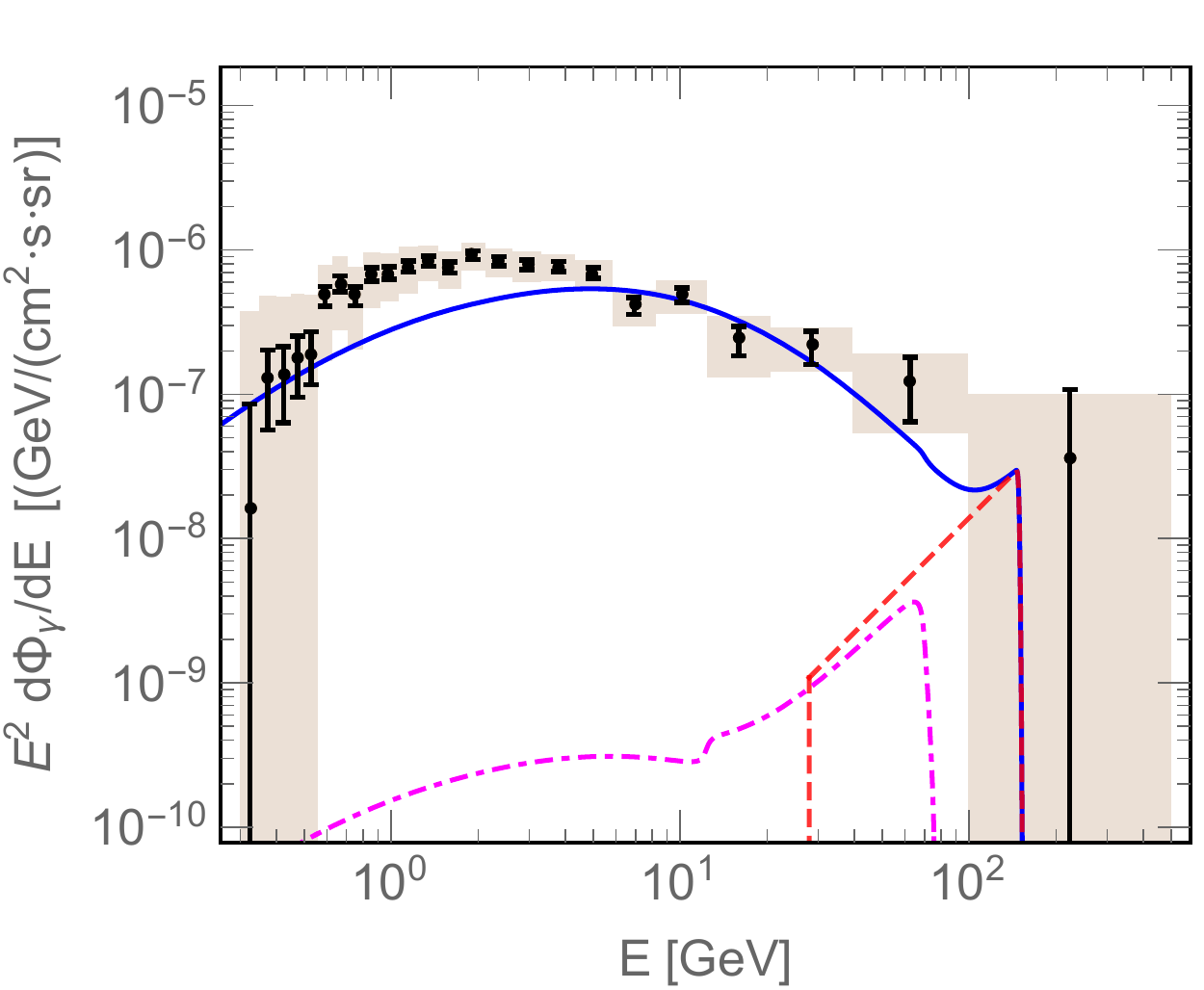} 
\caption{Same as Fig.~\ref{fig:gc-mxsv-resolution-s}, but the detector energy resolution $\xi=0.02$ is taken. The best-fit values of $m_X$ and $\langle \sigma v\rangle$ given in Table~\ref{tab:GC} are used in the left panel, while the best-fit values are further rescaled by 1.4 times in the right panel.}
\label{fig:gc-mxsv-resolution0.02}
\end{center}
\end{figure}

In Fig.~\ref{fig:gc-mxsv-resolution-s}, results of three cases, $m_S=0.99\, m_X$,  $m_S=0.999\, m_X$ and $m_S=m_h$,  are given. 
In the left panel of this figure, the GC fitted regions, providing a good fit, feature the $p$-values of 0.3, 0.15 and 0.05 denoted as the solid, dashed and dotted contours, respectively, on the plane of $m_X$ and $\langle\sigma v\rangle$. The corresponding best-fit values, together with their $p$-values and $\chi_{\rm min}$, are given in Table~\ref{tab:GC}.
 
Comparing with the GC gamma-ray excess data obtained by CCW, we show the spectrum in the middle panel of Fig.~\ref{fig:gc-mxsv-resolution-s} using the best fit values of $m_X$ and $\langle \sigma v\rangle$. For the illustrative purposes,  in the right panel, by multiplying the best fit values of $m_X$ and $\langle \sigma v\rangle$ by 1.4, we draw the theoretical spectrum (the blue curve), where  from up to down the $p$-values are respectively 0.12, 0.15, and $2\times10^{-6}$, for which the last one is poor in the fit.

Using the canonical parameter set, for a nearly degenerate case with $m_X \simeq m_S$,  we show the parameter space that provides a good-fit result ($p$-value $\geq 0.05$) located in the range of $m_X\in[60, 132]~\text{GeV}$ and $\langle \sigma v \rangle \in [ 2.0, 6.8]\times 10^{-26}\, \text{cm}^3/s$, corresponding to the energy  of the gamma-ray line $\in [30, 66]$~GeV.  
In this Higgs portal scenario, the gamma-ray line signal originating from $S\to Z\gamma$ is highly suppressed compared to the continuum signal, because of the smallness of its branching ratio for the value of $m_S$ preferred by the GC excess data (see Figs.~\ref{fig:BrS} and \ref{fig:gc-mxsv-resolution-s}).
As shown in Fig.~\ref{fig:gc-mxsv-resolution-s}, the observed spectral line width is very sensitive to the boost of $S$. 
For further comparison with the effect due to the energy resolution of the instrument, we consider a higher energy resolution of $\xi=$2\%, which would be achievable in the DAMPE  \cite{Bernardini:2017han} and GAMMA-400  \cite{Topchiev:2017xfp} experiments, and show the results in Fig.~\ref{fig:gc-mxsv-resolution0.02}.  One can thus expect that for the case $m_S\gtrsim 0.99\, m_X$, the prominent gamma-ray line signal, generated from $S \to \gamma \gamma$, is distinguishable against the continuum spectrum.
Once Fermi-LAT can accumulate as much as 15 years of data \cite{Charles:2016pgz}, the gamma-ray signal with energy $\gtrsim 40$~GeV (corresponding to a larger $\text{Br}(S \to \gamma \gamma$)) predicted in the Higgs portal scenario is very likely to be directly examined in the near future.

  We find that the best fit is $m_X\simeq m_S \simeq 86$~GeV, featuring a $p$-value of 0.42 (see Table~\ref{tab:GC}). In other words, the corresponding gamma-ray line peaks at 43~GeV.  It is very interesting to note that Liang {\it et al.} recently found a line-like structure at $\sim 43$~GeV with the significance $\sim 3.0\sigma$ after analyzing 85 month Pass 8 Fermi-LAT data (P8R2\_ULTRACLEAN\_V6) in the directions of 16 Galaxy clusters which are expected to have large $J$ factors \cite{Liang:2016pvm}.  Further extensive analyses of this line signal should be crucial for testing this scenario and  identifying the nature of dark matter.

\subsection{Constraints from other measurements}\label{sec:constraints}

 \begin{figure}[t!]
  \begin{center}
   \includegraphics[width=0.43\textwidth]{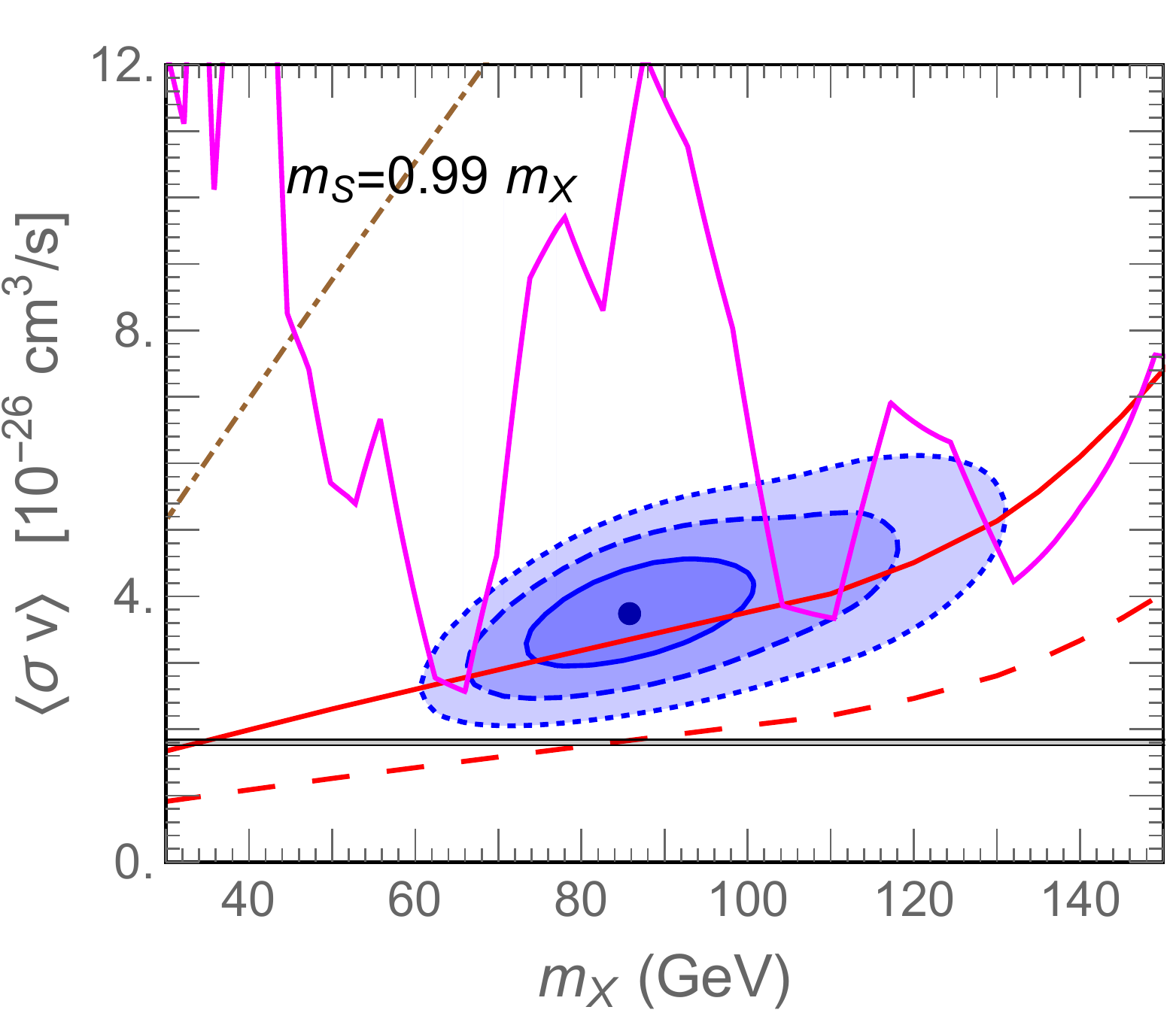} \hskip0.2cm
   \includegraphics[width=0.43\textwidth]{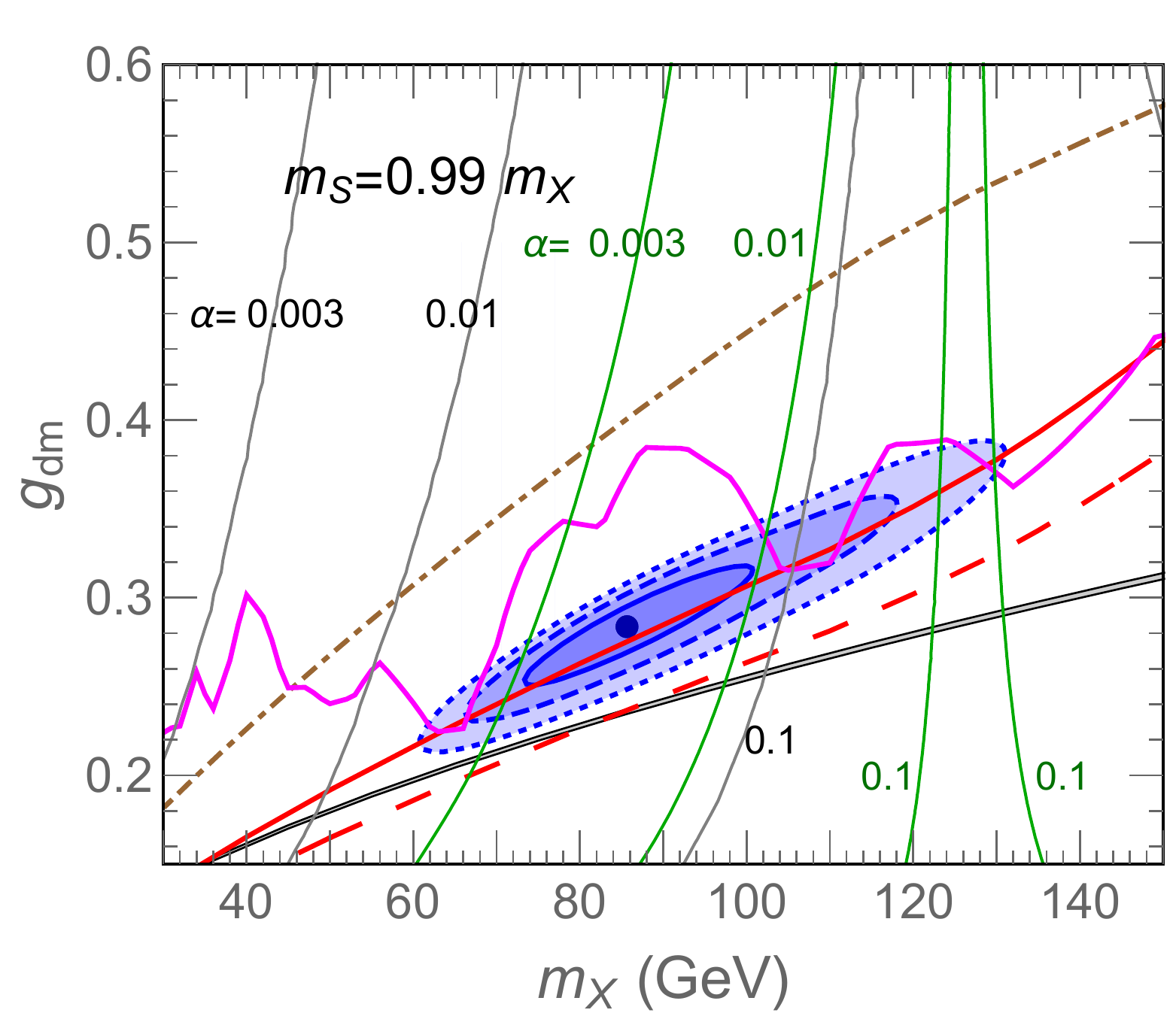}
\\ \vskip0.cm
   \includegraphics[width=0.43\textwidth]{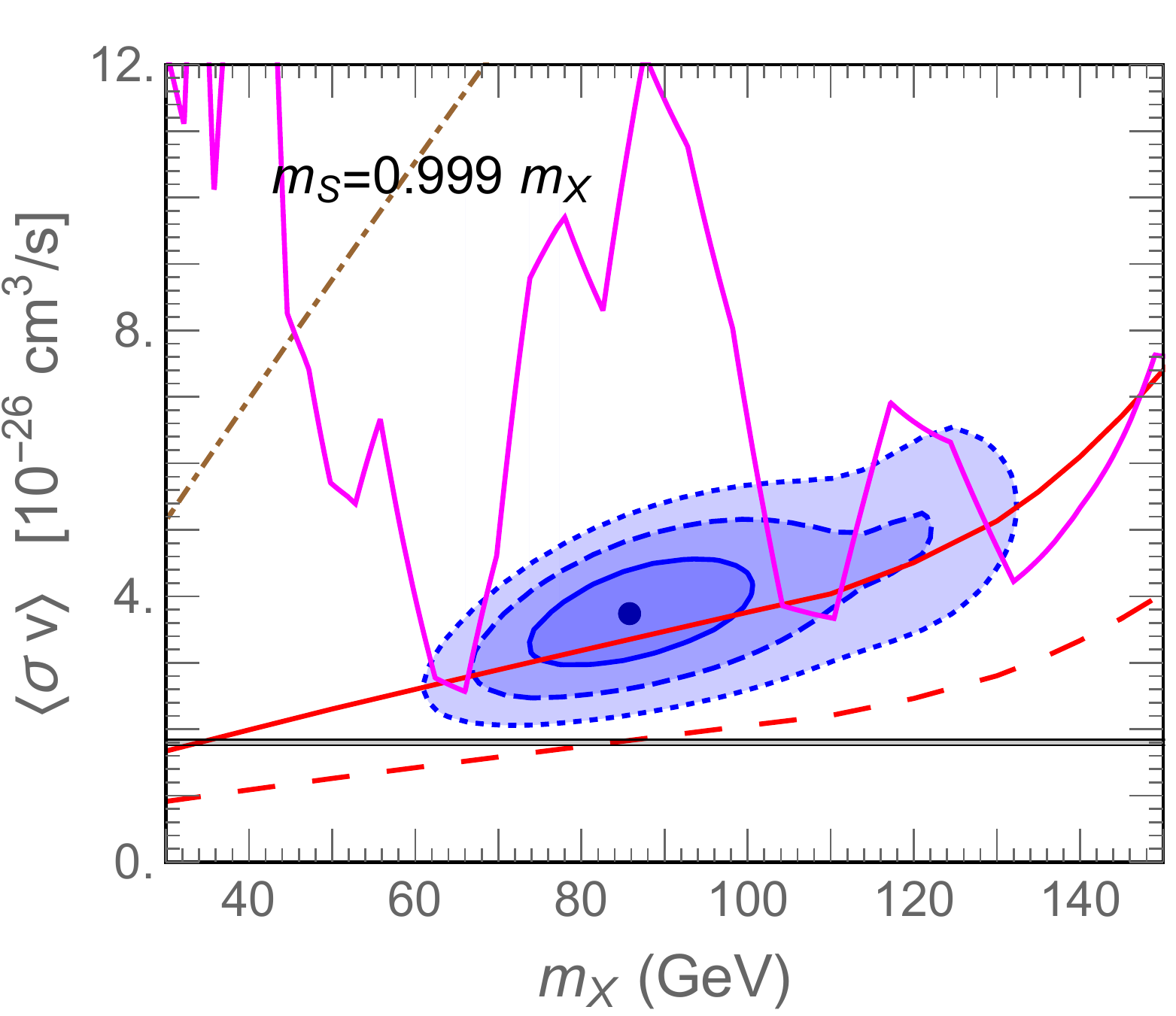} \hskip0.2cm
   \includegraphics[width=0.43\textwidth]{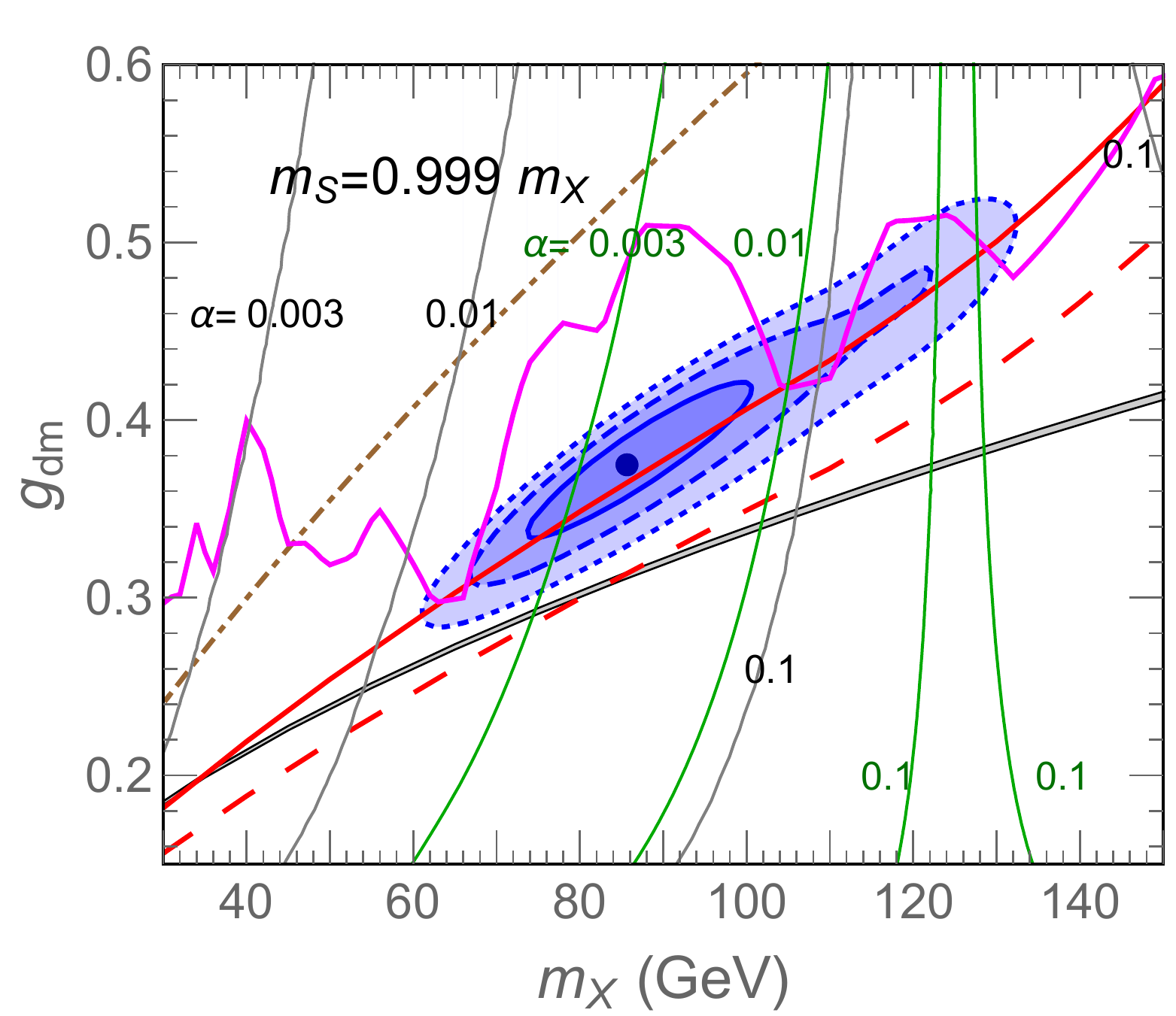} 
\caption{Left panel:  GC allowed regions for a generic Higgs portal DM model, using $\rho_\odot = \text{0.4 GeV/cm}^3$ and $\gamma=1.2$ in the gNFW profile. The dot denotes the best-fit, and blue contours correspond to $p$-values $=$ 0.3 (solid), 0.15 (dashed), and 0.05 (dotted).   For $\alpha \gtrsim 2\times 10^{-6}$, the correct relic density in the WIMP scenario is accounted for by the narrow gray range, while for $\alpha  \lesssim 2\times 10^{-6}$,  a larger annihilation cross section could be needed. The 95\%  C.L. bound from the Fermi-LAT NFWc gamma-ray line search  within the ROI: $R_{\rm GC}=3^\circ$ is depicted 
by the solid magenta line. The current Fermi-LAT dSph limit at 95\% C.L. and its projected sensitivity are depicted as the solid  and long-dashed red lines, respectively, while the Planck  CMB 95\% C.L. bound is sketched as the dot-dashed brown line.  
Right panel: Same notations as the left panel but in the $(m_X, g_{\rm dm})$ plane for the secluded vector DM model.  
The 95\% C.L. bounds from XENON1T and LZ projected sensitivity are indicated by the gray and green lines, which, with $\alpha$ values denoted,
correspond to the use of $\rho_\odot =$ 0.4 $\text{GeV/cm}^3$;  in the region $m_X < m_h$ (or $m_X > m_h$), the RHS (or LHS) of each line  is allowed. }
\label{fig:gc-constraints}
\end{center}
\end{figure}

 \begin{figure}[t!]
  \begin{center}
   \includegraphics[width=0.435\textwidth]{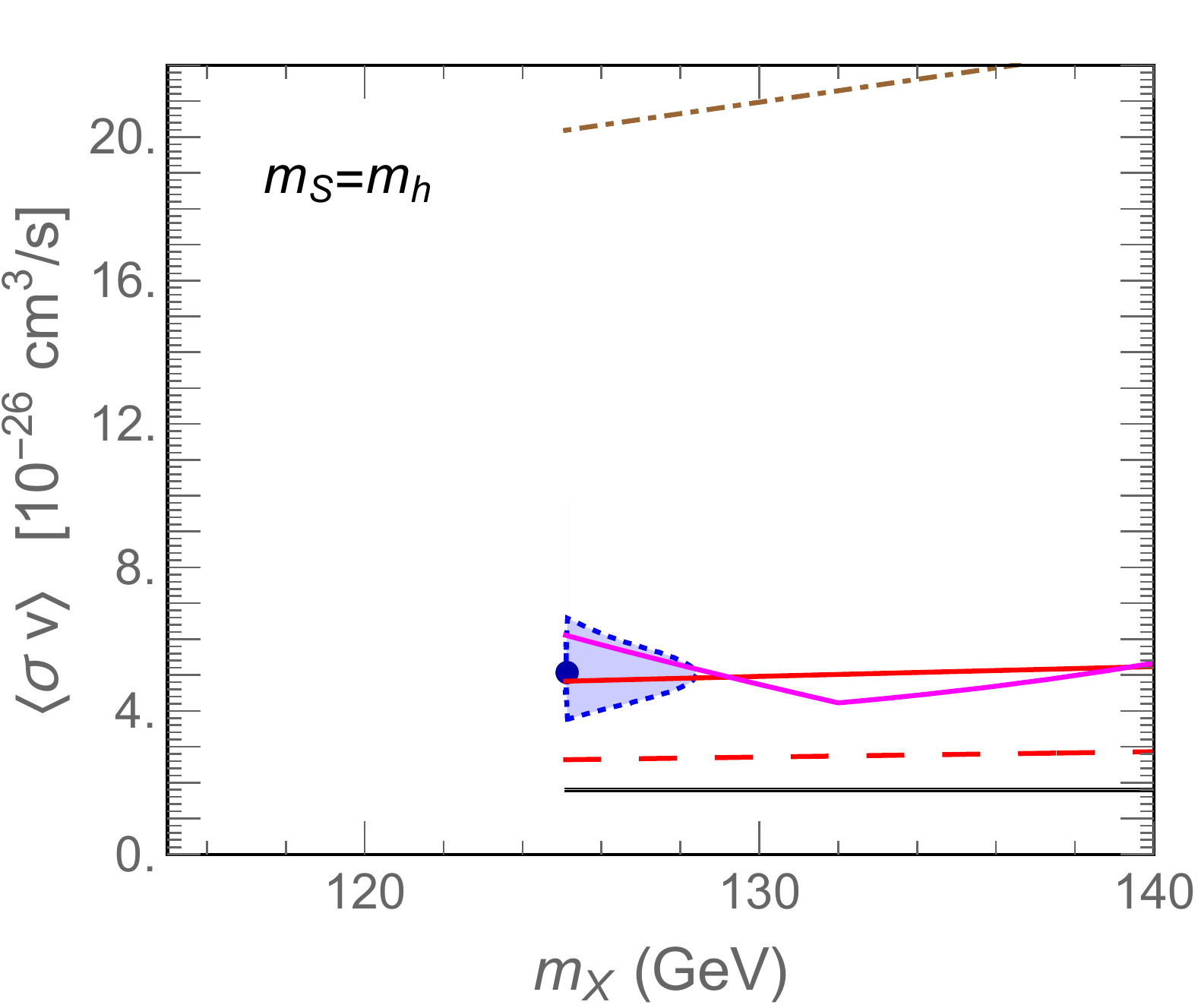} \hskip0.09cm
   \includegraphics[width=0.435\textwidth]{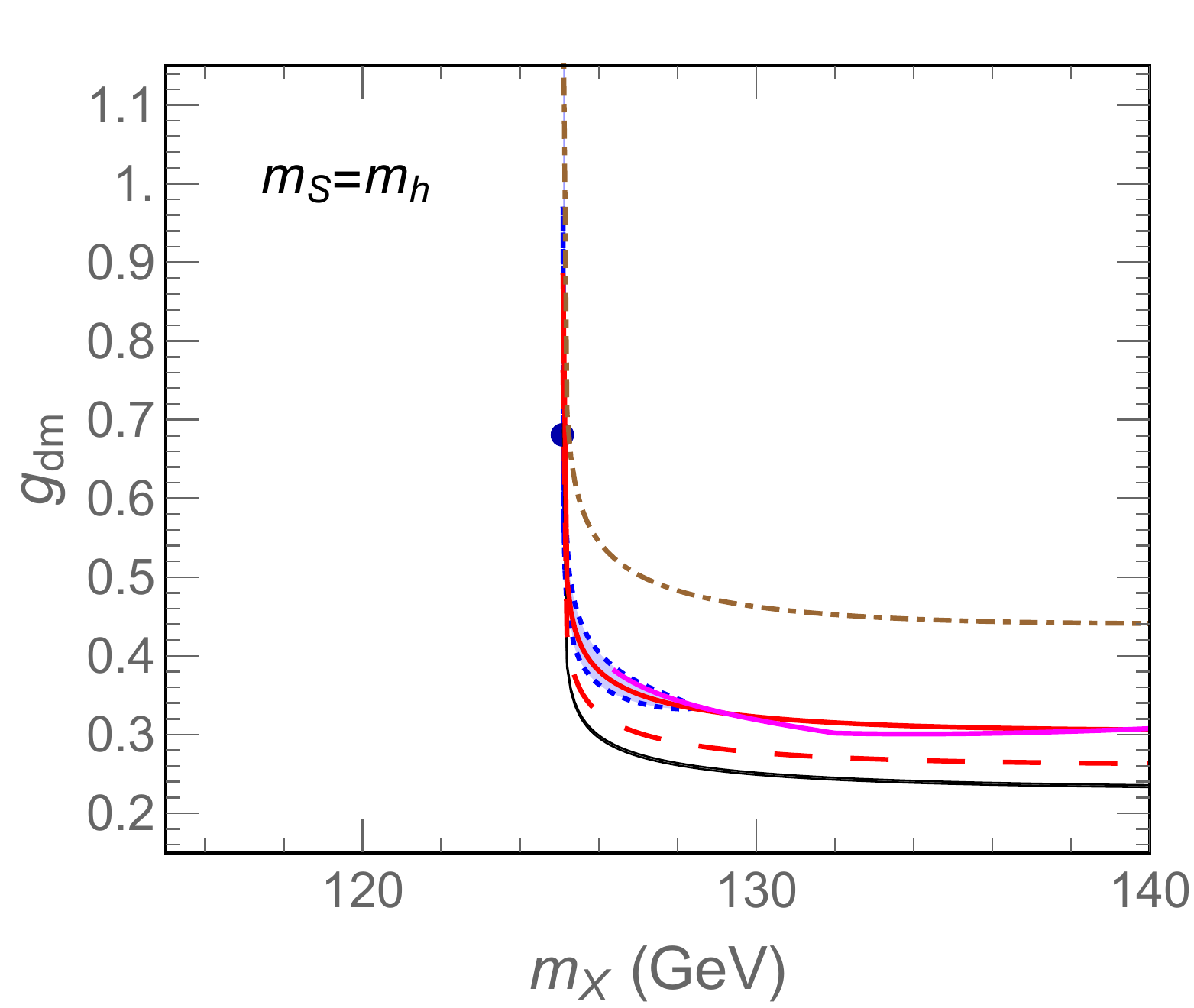} 
\caption{Same as Fig.~\ref{fig:gc-constraints} but for $m_S=m_h$
}
\label{fig:gc-constraints-2}
\end{center}
\end{figure}

Here we present the parameter constraints from various measurements.  For the analysis shown in Figs.~\ref{fig:gc-constraints} and \ref{fig:gc-constraints-2}, we adopt the fiducial value $\rho_\odot  = 0.4$ ~GeV/cm$^3$,  and fix $\gamma$ to be 1.2 used in the CCW gamma-ray flux.  The dependence of the results on the different values of  $\rho_\odot$ will be further given in Appendix~\ref{app:DMdensity}.

In the left panel of Figs.~\ref{fig:gc-constraints} and \ref{fig:gc-constraints-2}, we show the GC region favored by the Femi gamma-ray excess data, and constraints from other measurements  in the ($m_X$, $\langle \sigma v\rangle$) plane. This result is valid for a generic Higgs portal DM model, no matter what the fundamental properties of dark matter are. 
On the other hand, in the right panel of Figs.~\ref{fig:gc-constraints} and \ref{fig:gc-constraints-2}, we consider the secluded vector dark matter model as presented in Sec.~\ref{sec:model}, and thus translate the results of the left panel to the  $(m_X, g_{\rm dm})$ plane, where some regions, dependent on the value of $\alpha$, can be further constrained by the direct detection. 
As shown in the right panel of Fig.~\ref{fig:gc-constraints}, a smaller mass difference of $X$ and $S$ requires a larger $g_{\rm dm}$ to account for the GC data  due to the fact that the phase space for $XX\to SS$ vanishes in the limit $m_S \to m_X$.

The detailed constraints from various measurements will be discussed as follows.

\underline{Fermi-LAT gamma-ray line search:}  The Fermi-LAT collaboration has recently placed constraints on the gamma-line signals \cite{Ackermann:2013uma,Ackermann:2015lka}. The resultant limit depends on the mass and density profile of DM. We consider the Fermi R3 (ROI) fit from which the constraint, compared with other ROI results, is more restrictive. The R3 is defined to a very small circular regions of radius $R_{\rm GC}=3^\circ$ centered on the GC, and optimized for the contracted NFW\,\footnote{The contracted NFW is called the generalized NFW in this paper.} (NFWc) profile with $\gamma=1.3$.
 Since this choice of ROI strongly depends on the value of $\gamma$ (see the discussion in Appendix B of Ref.~\cite{Ackermann:2013uma}),  and since Fermi R3 line limit and CCW  data correspond to different ROIs, we  do not rescale the inner slope of the halo profile of the former one to $\gamma=1.2$ to match each other. We remark that if the Fermi line data did not depend on $\gamma$, such a rescaling would weaken the constraint on the annihilation cross section  by a factor of two.

In Figs.~\ref{fig:gc-constraints} and \ref{fig:gc-constraints-2},  the 95\% confidence level (C.L.) bound from the updated Fermi-LAT R3 (NFWc) gamma-ray line search (5.8 years of Pass 8 data) \cite{Ackermann:2015lka} is depicted by the solid magenta line corresponding to the use of  $\rho_\odot = \text{0.4 GeV/cm}^3$. 
One should note that the Fermi gave the gamma-ray line limit for direct DM annihilation to the photon pair, while in our case four photons are produced per annihilation. Therefore, in our case the measured gamma-ray line energy is $m_X/2$, and the limit for $\langle \sigma v\rangle \times \text{Br} (S\to \gamma\gamma)$ is equivalent to the value of $2 \langle\sigma v\rangle_{\gamma\gamma}$ given in Ref.~\cite{Ackermann:2015lka}.

\underline{Fermi-LAT observations of dwarf spheroidal galaxies:}
We perform a combined likelihood analysis of  28 kinetically confirmed and 17 candidate dSphs with 6 years of the Fermi-LAT data\footnote{The individual likelihood functions for given dSphs are available from the website:  ``http://www-glast.stanford.edu/pub\_data/1203/ ".} (passing the P8R2 SOURCE event class selections), where gamma-ray energies are in the range from 500~MeV to 500~GeV \cite{Fermi-LAT:2016uux}. 
 As in Ref.~\cite{Fermi-LAT:2016uux}, we use the spectroscopically determined J-factors with errors for the confirmed dSphs, and adopt  predicted values from the distance scaling relationship with a nominal uncertainty of 0.6 dex for the newly discovered candidates.  
 We refer readers to Ref.~\cite{Yang:2018fje}  for the detailed description of the likelihood analysis that we have used here. 
 
 The solid and long-dashed red lines shown in Fig.~\ref{fig:gc-constraints} represent the current dSph limit at 95\% C.L. and Fermi-LAT projected sensitivity, respectively.  Here we have assumed 60 dSphs ($\equiv N_{\rm dSph}$) observed and the 15 years of data ($\equiv N_{\rm data}$) collected for the projected sensitivity, which approximately rescales with $\sqrt{N_{\rm dSph} } \times \sqrt{ N_{\rm data}}$ \cite{Anderson:2015rox}.

As shown in Figs.~\ref{fig:gc-constraints} and \ref{fig:gc-constraints-2}, the parameter space  is much more restricted by the current dSphs constraint, compared with other measurements. 
The scenario that the hidden sector dark matter interacts with the SM through the Higgs portal can be further tested by the dSphs projection.

\underline{Planck cosmic microwave background:}
The CMB provides a probe into the DM annihilation at the epoch of recombination, and thus offers a complementary constraint compared with experiments of the gamma-ray observations.
Planck sets a bound on the annihilation parameter, $p_{\rm ann}$, from TT, TE, EE+lowP (temperature and polarization) data combinations \cite{Ade:2015xua},
\begin{align}
p_{\rm ann} \equiv f_{\rm eff} \frac{\langle \sigma v \rangle_{\rm CMB}}{m_X} < 4.1 \times 10^{-28} \  \text{cm}^3 \text{s}^{-1}  \text{GeV}^{-1} \,, 
\end{align}
where $\langle \sigma v \rangle_{\rm CMB}$  at the epoch of recombination $\simeq \langle \sigma v\rangle$ at the present day for s-wave DM annihilation (as the secluded vector dark model that we study in this paper), and  the efficiency factor $f_{\rm eff}$ is the fraction of the energy that is injected into the intergalactic medium from DM annihilations at redshift z.  The  efficiency factor depends on the spectra of $e^\pm$ pairs and photons produced following DM annihilations, 
\begin{align}
f_{\rm eff} =\frac{1}{2 m_X} \int_0^{m_X}  E dE \bigg[ 2 f_{\rm eff}^{e^-}(E)  \bigg( \frac{d N_{e^-}}{dE}\bigg)_X + f_{\rm eff}^\gamma (E) \bigg( \frac{dN_{\gamma}} {dE}\bigg)_X \bigg] \,,
\end{align}
where we use $f_{\rm eff}^{\gamma, e^-}(E)$ curve results suited for the ``3 keV" baseline prescription obtained by Slatyer~\cite{Slatyer:2015jla}, and
$(dN_{e^- /\gamma} / dE)_X$ is the electron/photon spectrum produced per DM annihilation in the DM rest frame.  
The calculation for $(d N_{e^-}/ dE)_X$, which originates from two-body and three-body $S$ decays following the DM annihilation $XX\to SS$, is completely the same as that for $(dN_{\gamma} / dE)_X$ described in Sec.~\ref{sec:gamma-spectrum}, but using PPPC4DMID to generate the electron spectrum instead of the photon spectrum.

The Planck  CMB 95\% C.L. bound is sketched as the dot-dashed brown line in Figs.~\ref{fig:gc-constraints} and \ref{fig:gc-constraints-2}. The current Planck CMB limit seems to be considerably weaker than the Fermi-LAT dSphs limit.

\underline{Correct relic density:} The thermodynamic evolution of the hidden sector (vector) dark matter interacting with the SM through the Higgs portal has been studied in Ref.~\cite{Yang:2019bvg}. Here we will present the main properties, and refer readers to Ref.~\cite{Yang:2019bvg} for the detailed results. The value of $\alpha$ is relevant to the coupling strengths of $S$ to the SM particles, and thus determine the decoupling temperature below which the hidden sector is kinetically decoupled from the SM bath. 

For $\alpha \gtrsim 2\times 10^{-6}$, the correct relic density is set by the $XX \leftrightarrow SS$ interaction, so that the DM particles can be in chemical and thermal equilibrium with $S$ particles and with the SM bath (through $S$) before freeze-out. While this result is consistent with the conventional WIMP scenario,  the annihilation cross section corresponding to the narrow gray range in Figs.~\ref{fig:gc-constraints}  and \ref{fig:gc-constraints-2} can account for the correct relic density. 
As for $\alpha$ less than $2\times 10^{-6}$,  the dark sector has been kinetically decoupled from the thermal reservoir, before it becomes nonrelativistic. For this case, 
 the DM annihilation cross section and coupling contant $g_{\rm dm}$,  providing a correct relic density, could be boosted to the upper side of the gray range in the left panel and right panel of Figs.~\ref{fig:gc-constraints} and \ref{fig:gc-constraints-2}, respectively.

\underline{XENON1T result and LUX-ZEPLIN (LZ) projected sensitivity:} Considering a specific DM model, one can set limits on (coupling) parameters from the direct detection experiments. For the secluded vector DM model shown in Sec.~\ref{sec:model}, the elastic scattering cross section of $X$ off a nucleon ($N$), independent of the nuclear spin, is referred to as the ``spin-independent cross section", which via the $t$-channel interactions with exchange of $S$ and $h$ is given by
\begin{align}
\sigma_N =  \frac{\mu_{XN}^2 m_N^2 f_N^2 g_{\rm dm}^2}{4\pi} \frac{\sin^2 2\alpha}{v_H^2} \left( \frac{1}{m_S^2} - \frac{1}{m_h^2} \right)^2 \,,
\end{align}
where $\mu_{XN} = m_X m_N/ (m_X + m_N)$ is the reduced mass of $X$ and $N$, and $f_N=  \sum_q \langle N | \bar q q |N \rangle m_q/m_N \simeq 0.3$  \cite{Alarcon:2011zs,Cline:2013gha}.

In the right panel of Fig.~\ref{fig:gc-constraints}, using  $\rho_\odot = \text{0.4 GeV/cm}^3$,
the 95\% C.L. bounds from XENON1T  \cite{Aprile:2017iyp} and LUX-ZEPLIN (LZ)  projected sensitivity  \cite{Akerib:2018lyp}
are respectively indicated by the gray and green solid contours, where the corresponding $\alpha$ values are denoted, and, in the region $m_X < m_h$ (or $m_X > m_h$), the RHS (or left hand side ($\equiv$ LHS)) of each line is allowed. 
We observe that XENON1T and LZ projected sensitivity are able to constrain the $m_X-g_{\rm dm}$ parameter space preferred by the GC gamma-ray excess data for $\alpha\gtrsim 0.01$ and $\gtrsim 0.002$, respectively. 

 In Fig.~\ref{fig:gc-constraints-2}, we do not consider direct detection limits on the case of $m_S=m_h$, for which there is no constraint for this perfect degenerate case.  However, for a generic case of $|m_S -m_h|< 4$~GeV and $|\alpha| \gtrsim 0.17 \,  (0.02)$,  the region $g_{\rm dm} <1$ evades the XENON1T searches (LZ projected sensitivity) for $\rho_\odot \in [0.25,0.6]\, \text{GeV/cm}^3$.

\underline{Big bang nucleosynthesis (BBN):}  The BBN measurement can set a lower bound on the mixing angle, $\alpha$.
When the scalar mediator $S$ decays out-of-equilibrium into SM particles, the Universe becomes radiation-dominated again and experiences the reheating owing to large entropy injection. (See the discussions in Ref.~\cite{Yang:2019bvg}.)
We can constrain $\alpha$ through the observation limit on the reheating temperature.
 If the reheating temperature $T_{\rm RH}$ was on the order of the neutrino decoupling temperature, then the neutrinos would not be well thermalized ~\cite{Hasegawa:2019jsa}; if so, the relative rate of light element abundances would be changed, too.
From the $Y_p +D/H$ analysis (with the helium nucleon fraction $Y_p \equiv 4n_{\rm He}/ n_b$), the authors of Ref.~\cite{Hasegawa:2019jsa} have obtained a lower bound
 at 95\% C.L. on the reheating temperature, $T_{\rm RH} \gtrsim 4.1$~MeV.
Further using the relation~\cite{Hasegawa:2019jsa}
\begin{align}
T_{\rm RH} \simeq  0.7 \left( \frac{\Gamma_S}{{\rm sec}^{-1}}\right)^{1/2} {\rm MeV},
\label{bbn-time}
\end{align}
we can get the $S$ width, $\Gamma_S \gtrsim (0.03~\text{sec})^{-1}$, i.e. $\alpha \gtrsim 1 \times10^{-10}$  for the present study.

\section{Discussions}\label{sec:discussions}

\subsection{Mixing angle $\alpha$ constraint in the hidden Higgs portal dark matter model:
from the thermodynamic point of view}\label{sec:alpha}

If $\alpha$ is extremely small, the hidden sector can be decoupled from the SM bath at the very high temperature $T\gg m_{X,S}$. Like the hot dark matter case,
 after decoupling, the relativistic hidden sector almost maintains the same temperature as the bath, and its comoving number density is also approximately unchanged.  Here we have neglected temperature variation of the SM bath due to the change of its relativistic degrees of freedom, when its temperature drops below $m_t$ or $m_h$.  

Therefore, for the case that the hidden sector was in thermal equilibrium with the bath in the earlier stage but later on was decoupled from the bath even at the very high temperature, once the decoupling has occurred, the relativistic hidden sector evolves with a temperature which is almost the same as the temperature of the bath. Moreover, for this case, when the hidden sector becomes nonrelativisitc, it gets hotter  than the bath during the cannibal epoch \cite{Yang:2019bvg}. In Ref.~\cite{Yang:2019bvg}, we have presented a comprehensive study on thermodynamic evolution of the hidden sector for this secluded vector DM model.  Here we would like to discuss the minimum value of $\alpha$ for which the hidden sector was once in thermal equilibrium with the bath when $T\gtrsim m_S$. This part did not mention in Ref.~\cite{Yang:2019bvg}.

In the following discussion, we assume that, before decoupling, number changing interactions among the dark sector particles are still active and guarantee their thermal equilibrium with zero chemical potential.
We separately discuss the requirement of interactions, including (i) $SS\leftrightarrow \text{SM~SM}$, (ii) $S \leftrightarrow \text{SM~SM}$, and (iii) $S\, \text{SM} \leftrightarrow  S\, \text{SM}$, that can account for the thermal equilibrium between $S$ and the SM bath at a temperature $T$ which is larger than $m_S$.

If the hidden sector is in equilibrium with the thermal bath through the interaction $SS\leftrightarrow \text{SM~SM}$ at  a temperature $T\gtrsim m_S$, we need to have 
 $n_S^{\rm eq} \langle \sigma v \rangle_{SS \to \text{SM~SM}} \gtrsim H$, which describes the $S$ production rate from the inverse annihilation is larger than the cosmic dilution rate. 
 Here $n_S^{\rm eq}$ is the equilibrium number density (with zero chemical potential) and $H$ is the Hubble rate.
 In the limits of large energy and small $\alpha$, because $\langle \sigma v \rangle_{SS \to \text{SM~SM}} \propto \alpha^2/s^2$ which is suppressed in the high temperature due to the fact that $\sqrt{s} \propto T$, therefore we can simply take $T\approx m_S$ to obtain the lower bound of the mixing angle, $\alpha \gtrsim 10^{-4}$ for this interaction. (See Appendix B of Ref.~\cite{Yang:2019bvg} for the exact form of the $SS\to \text{SM~SM}$ amplitude.)
 
 If thermal equilibrium between $S$ and the SM bath is due to $S \leftrightarrow \text{SM~SM}$ and hold at $T\gtrsim m_S$, we have the inverse decay rate $ \Gamma_S \gtrsim H$. Since $H\propto T^2$, we can simply take $T\approx m_S$ to get $\alpha \gtrsim 10^{-5}$ (see also the result shown in the right panel of Fig.~1 in Ref.~\cite{Yang:2019bvg}).
 
As for the elastic scattering $S\, \text{SM} \leftrightarrow  S\, \text{SM}$, we adopt the definition of temperature for the relativistic $S$,
\begin{align}
T_S =\frac{g_S}{n_S(T_S)} \int \frac{d^3 p_S}{(2\pi)^3} \frac{{\bf p}_S^2}{3 E_S}  f_S (T_S) \,, \label{eq:temp-1}
\end{align} 
where the distribution function $f_S \simeq \exp[-(E_S-\mu_S)/T_S]$ with $\mu_S$ the chemical potential, $E_S$ is the distribution function, $g_S=1$ is the internal degrees of freedom, $n_S$ is the number density, and ${\bf p}_S$ are  energy and 3-momentum of $S$, respectively.  See  further discussions for the definition of temperature in Appendix~\ref{sec:temp-S}. Eq.~(\ref{eq:temp-1}) is a good approximation for the temperature definition even at the very high temperature, $T_S \gg m_S$. Solving the Boltzmann moment equation,  we obtain the temperature evolution of $S$ for $T_S\gg m_S$,
\begin{align}
 \frac{d T_S}{dt} + H T_S 
 = \frac{1}{ n_S(T_S)}  \left[- \left( \frac{d n_S(T_S) }{dt} +3 H n_S (T_S) \right) T_S 
+ \{  S\, \text{SM} \leftrightarrow S\, \text{SM} \}_{\rm coll}  +\dots \right]\,,  \label{eq:boltz-t}
\end{align}
where the elastic collision term is described by a semi-relativistic Fokker-Planck equation \cite{Binder:2016pnr},
\begin{align}
 \{  S\, \text{SM} \leftrightarrow S\, \text{SM} \}_{\rm coll} 
 & \simeq 
   \int   \frac{d^3p_S}{(2\pi)^3 }   \frac{ {\bf p}_S^2}{3 E_S} \frac{\partial}{ \partial {\bf p}_S} \cdot 
    \left[  \gamma(T) \,    \left( {\bf p}_S f_S(T_S) + E_S T \frac{\partial f_S(T_S)}{\partial{\bf p}_S} \right)  \right]  \,.
\end{align}
Here the momentum relaxation rate is given by 
 \begin{align}
\gamma (T) =
\frac{1}{6 E_S \, T \, \Big[1- \frac{{\bf p}_S^2}{3E_S^2} \Big] } \sum_f  \int  \frac{d^3 k}{(2\pi)^3} f_f (T) (1 - f_f (T)) \frac{|{\bf k}|}{\sqrt{{\bf k}^2 +m_f^2}} 
\int_{-4 {\bf k}^2}^0 dt (-t) \frac{d \sigma_{S f \to S f }}{dt} \,, \label{eq:gamma}
\end{align}
where $f$ is the relevant relativistic SM species, and $f_f(T)$ is its distribution function at temperature $T$, ${\bf k}$ is the 3-momentum of $f$, $t$ is the momentum transfer squared between $S$ and $f$. Note that this formula is a good approximation for a relativistic $S$ under the condition $E_S \gg \sqrt{t}$.
Taking the limits $E_S\gg m_S$ and $T_{(S)}\gg m_S$, we then obtain
\begin{align}
\{ S\, \text{SM} \leftrightarrow S\, \text{SM} \}_{\rm coll} 
 \simeq -   \bar\gamma(T_S)  \,  n_S(T_S) \,  (T_S -T) ,
\end{align}
where $\bar\gamma \equiv (2/3)\, \gamma(T) \, E_S/T_S$,  which is (assuming that $t \ll m_S^2$)
\begin{align}
\bar\gamma  \simeq  \sum_{f }  
&\frac{40 N_c^f m_S}{\pi^3} \bigg( \frac{g_{SSS}\, g_{S ff}}{m_S} + \frac{m_S}{m_h}  \frac{g_{hSS} \, g_{h ff}}{m_h} \bigg)^2 \nonumber\\
& \times \bigg[
    \zeta(6) (1-2^{-5})\Big( \frac{T}{m_S}  \Big)^6 + \frac{\zeta(4)}{1620} (1-2^{-3})\Big( \frac{m_f}{m_S}\Big)^2 \Big( \frac{T}{m_S} \Big)^4
  \bigg] 
  \frac{m_S}{T_S} \,.
  \label{eq:bargamma}
\end{align}
Note that in Ref.~\cite{Yang:2019bvg}, we consider the case satisfying $T \lessapprox m_S$.
If the elastic scattering $S\, \text{SM} \leftrightarrow  S\, \text{SM}$ can maintain $S$ and the SM bath in thermal equilibrium at $T\gg m_S$, the energy gained by $S$  through the elastic scattering is larger in magnitude than the Hubble cooling rate,
\begin{align}
 \bar\gamma \,  n_S \,  (T_S)\,  T \gtrsim  \,  H(T) \, n_S (T_S) \,  T_S \,.
\end{align}
Because $H \propto T^2$ and $\bar\gamma \propto \alpha^2 T^5$ (from Eq.~(\ref{eq:bargamma})), we thus have $\alpha \propto T_{\rm el}^{-3/2}$ with $T_{\rm el}$ being the decoupling temperature for the elastic scattering interaction. A smaller mixing angle $\alpha$ will result in a higher $T_{\rm el}$.
Using the result shown in Fig.~5 of  Ref.~\cite{Yang:2019bvg}, from that we have $\alpha\sim 10^{-6}$ when $T_{\rm el} \sim m_S$, therefore for the case that the hidden sector is kinetically decoupled from the bath at $T_{\rm el} (\gg m_S)$, the corresponding mixing angle reads
\begin{align}
\alpha \approx 10^{-6} \Big( \frac{T_{\rm el}}{m_S} \Big)^{-3/2}.
\end{align}
As will be discussed below, the vacuum will become unstable when $T\gtrsim 10^{10}$~GeV for the secluded vector DM model. If $T_{\rm el}$ is below this value, we shall need $\alpha \gtrsim 10^{-18}$.

\subsection{Theoretical vacuum stability for the secluded vector dark matter model}\label{sec:th-constraints}

Before concluding this paper, we study the scale-dependence of vacuum stability for the secluded vector dark matter model. The vacuum is required to be stable at the tree-level potential, i.e., the potential should be bounded from below and satisfies,
\begin{align}
\lambda_H>0, \quad
\lambda_S>0,  \quad
( \lambda_{HS} + \sqrt{\lambda_H \lambda_S} >0 \quad  {\rm if}\  \lambda_{HS}<0)  \,.
\label{eq:vacuum-stability}
\end{align}
Meanwhile, requiring $\det M_{\rm Higgs}^2 >0$, we also have $\lambda_{HS}^2 - \lambda_H \lambda_S<0 $. From this and Eq.~(\ref{eq:vacuum-stability}) we thus get  $-|\lambda_{HS}| + \sqrt{\lambda_H \lambda_S} >0$.

\begin{figure}[t!]
\begin{center}
\includegraphics[width=0.42\textwidth]{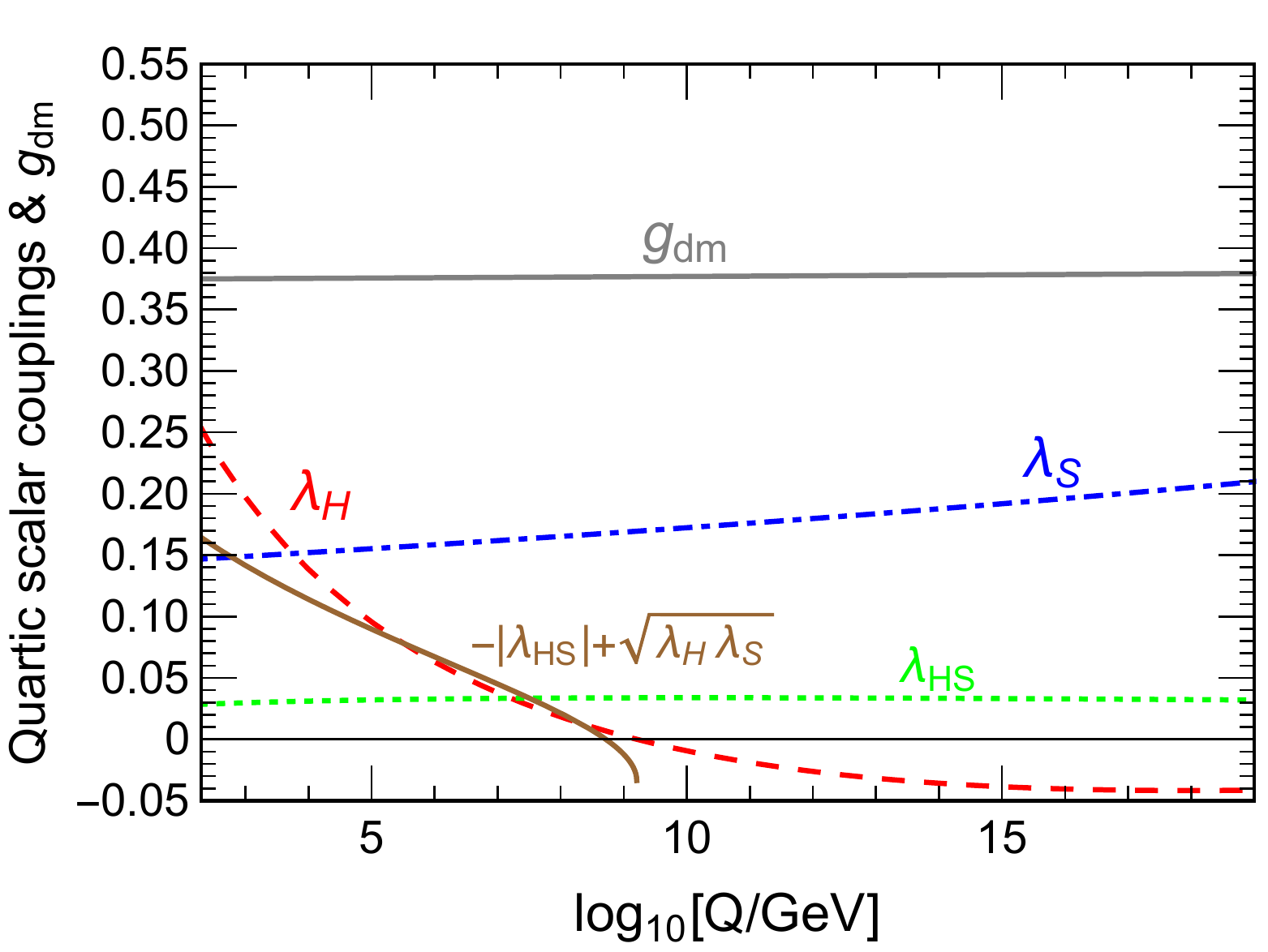} \hskip0.3cm
\includegraphics[width=0.42\textwidth]{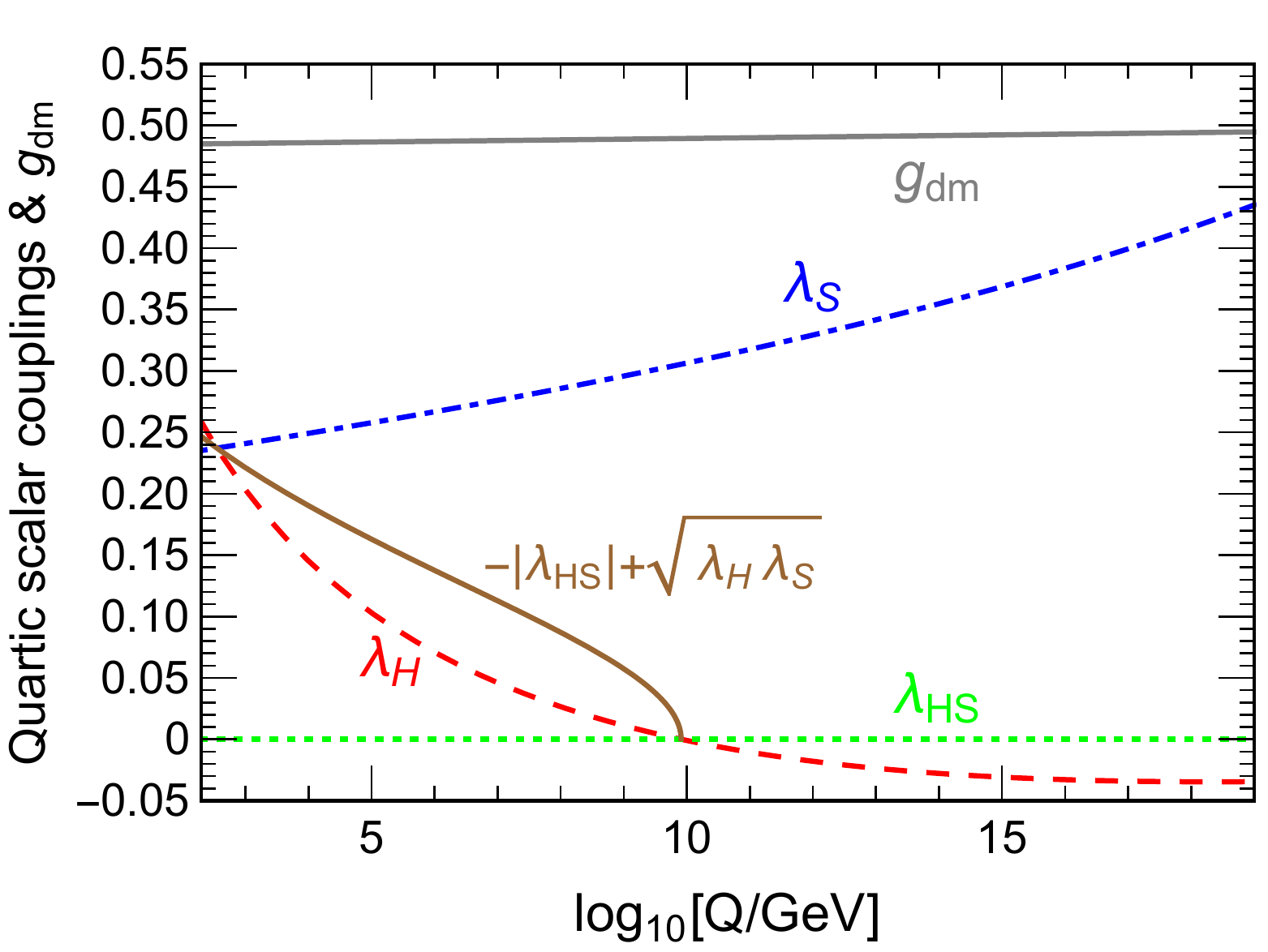}\\
\includegraphics[width=0.42\textwidth]{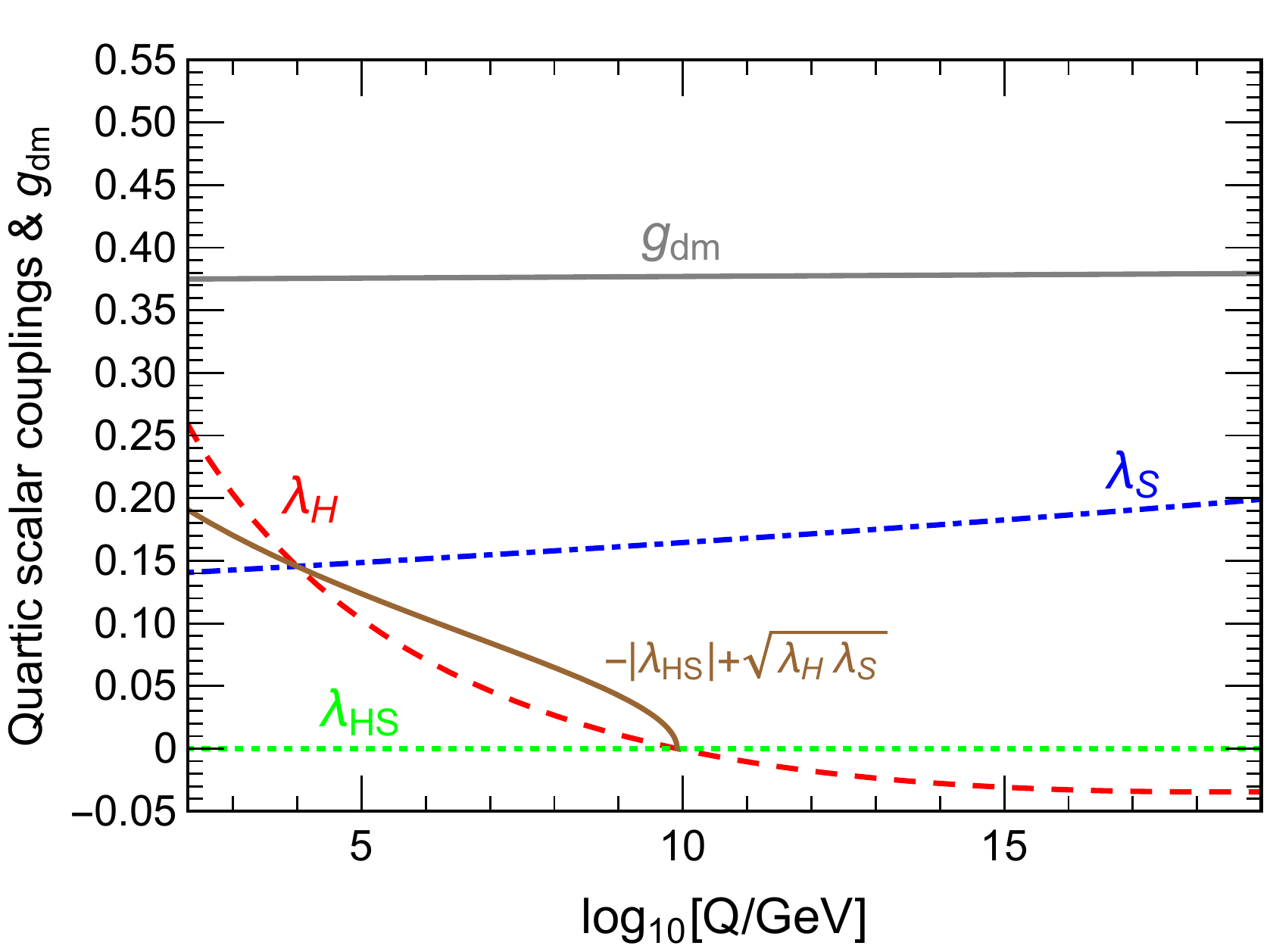} \hskip0.3cm
\includegraphics[width=0.42\textwidth]{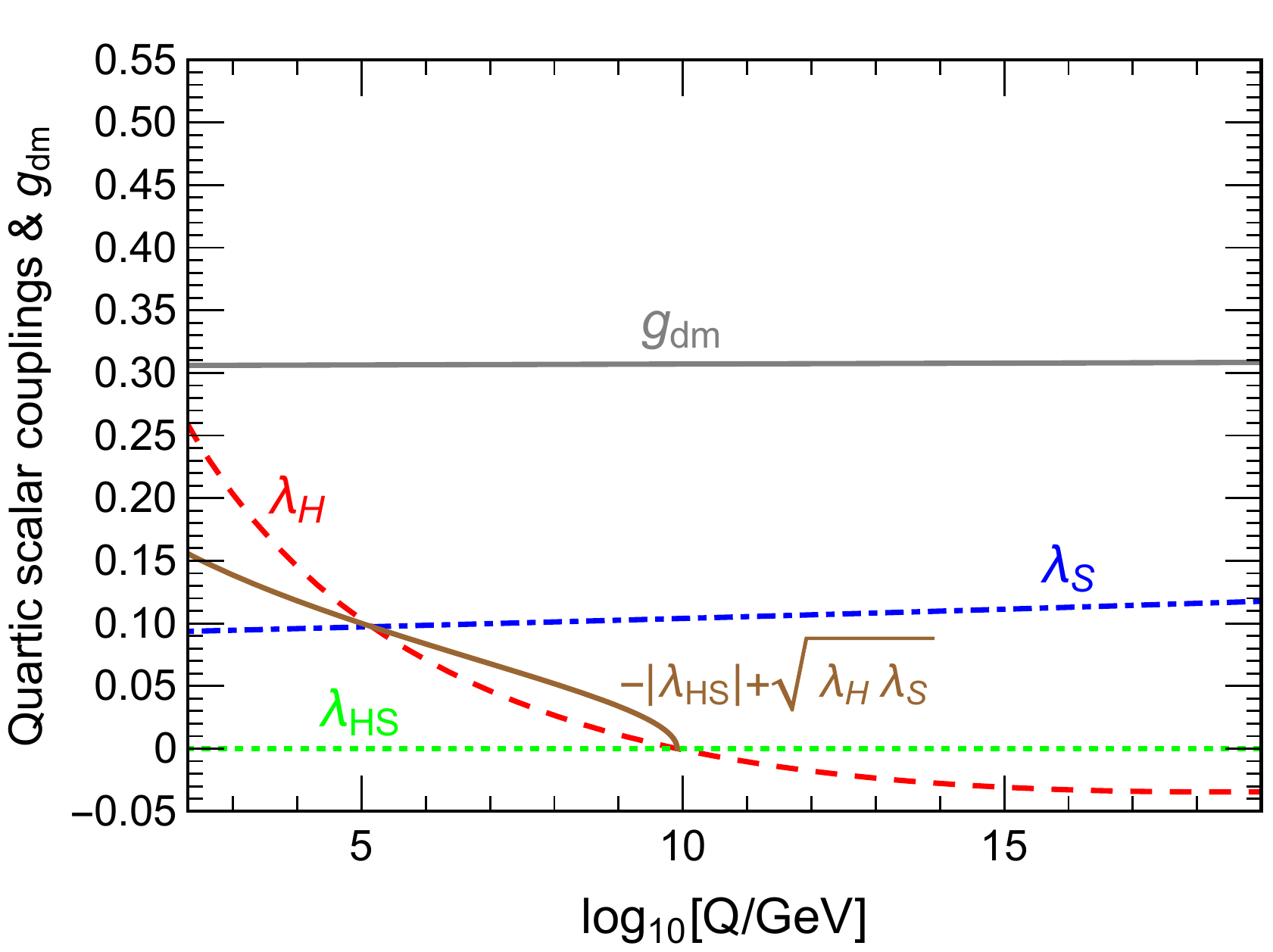}\\
\caption{ Running of quartic scalar couplings and $g_{\rm dm}$, as functions of the renormalization scale $Q$. Here at scale $Q=m_t$ (the top quark mass), we use $m_X=m_S=86$~GeV, and $(g_{\rm dm}, \alpha) =$ (0.375,0.2), (0.375, $\sim 0$), (0.485, $\sim 0$), and (0.306, $\sim 0$) for the upper left, lower left, upper right, and lower right panels, respectively.
 In the limit of  $m_S \to m_X$ and $\alpha\to 0$, the scale-dependence of the quartic scalar couplings, insensitive to the values of $m_X$ and $m_S$, depends only on the initial value of $g_{\rm dm}$, for which, as the reference value, we have used the central values of the GC excess fits for the case $m_S=0.999\, m_X$ shown in Fig.~\ref{fig:gc-constraints-app}.
}
\label{fig:reg-quartic-couplings}
\end{center}
\end{figure}

It was pointed out in Ref.~\cite{Baek:2012se}, where one-loop $\beta$-functions were considered, that the top quark can drive $\lambda_H$ to become negative at a certain higher scale, such that the electroweak vacuum is no longer the global minimum. To examine the vacuum stability, we study the renormalization group equations (RGEs) of the quartic scalar couplings, for which using SARAH \cite{Staub:2008uz,Staub:2010jh,Staub:2013tta,Staub:2015kfa} the $\beta$ functions are calculated up to the two-loop level, and collected in Appendix~\ref{app:rges}.  We find that, in the limit of  $m_S \to m_X$ and $\alpha\to 0$, the scale-dependence of the quartic scalar couplings, highly insensitive to the values of $m_X$ and $m_S$, depends only on the initial value of $g_{\rm dm}$.  
As seen in Fig.~\ref{fig:reg-quartic-couplings}, for the $m_X\approx m_S$ case with a sizable mixing angle $\alpha$, the stability condition is violated, i.e., $\lambda_H$ becomes negative, at the scale $Q$ less than $10^{10}$~GeV, but the violating scale will approach to   $10^{10}$~GeV in vanishing $\alpha$ limit.  In other words, the present secluded vector DM model is an effective theory suitable for the scale below $10^{10}$~GeV. Above this temperature, the Universe might experience the reheating and could be dominated by more massive particles during the oscillation epoch.

\section{Conclusions}\label{sec:conclusions}

The gamma-ray line signal generated from the DM annihilation could be a clear signature which is distinguishable from astrophysical backgrounds and reveals the particle nature of DM. 
We are motivated by the recent studies that direct DM annihilation to two SM-like Higgses, produced close to rest,  is capable of accounting for the GC gamma-ray excess data but with a little lower $p$-value $\lesssim 0.13$, and also motived by the fact that the quality of fit can be significantly improved if DM mostly annihilates to a lighter Higgs pair, which soften the gamma-ray spectrum to have a better fit to the observation peaked at $1-3$~GeV.

 We therefore consider a Higgs portal DM model where the hidden scalar mediates the  interaction of DM with the SM due to its mixing with the SM Higgs.
 In this model, the DM is secluded in the hidden sector and can annihilate directly to a pair of  lighter scalar mediators, each of which, nearly degenerate with DM in mass,  subsequently decays into the SM particles.

For the case of $m_X \simeq m_S$,  we have obtained that the parameter region $m_X\in[60, 132]~\text{GeV}$ can provide a good spectral fit to the Fermi GC gamma-ray excess data ($p$-value $\geq 0.05$), showing the energy of the gamma-ray line $\in [30, 66]$~GeV.  
The best fit to the data yields $m_X\simeq m_S \simeq 86$~GeV,  featuring a $p$-value of 0.42, so that the corresponding gamma-ray line arising from $S \to \gamma\gamma$ peaks at 43~GeV.  
The observed spectral line width, depending on the energy resolution of the detector (see Fig.~\ref{fig:gc-mxsv-resolution-s} vs. \ref{fig:gc-mxsv-resolution0.02}), is sensitive to the Lorentz-boost from the mediator rest frame to the DM center of mass frame, that directly correlates with the mass difference of $m_X$ and $m_S$.
In the Higgs portal model, we expect that, for the secluded DM case $0.99 \lesssim m_S/m_X <1 $,  a prominent gamma-ray line arising from $S \to \gamma \gamma$ can be distinguished from the continuum spectrum, while the line signal originating from $S\to Z\gamma$ is highly suppressed.

The fitted value of the low-velocity DM annihilation cross section depends on the DM distribution. Adopting $\gamma=1.2$ and $\rho_\odot=0.4$~GeV/cm$^3$,
 a good fit to the GC excess emission gives $\langle \sigma v \rangle \in [ 2.0, 6.8]\times 10^{-26}\, \text{cm}^3/s$.  Using a smaller (or larger) $\gamma$ and/or $\rho_\odot$, the value of $\langle \sigma v\rangle$ can be further raised (or lowered).
 We have derived constraints on the annihilation cross section from the Fermi-LAT gamma-ray line search, Fermi-LAT dSphs gamma-ray observations, and Planck CMB measurement. These detections can offer complementary probes. Currently,  the dSphs constraint on the parameter space favored by the GC excess emission is more restrictive than that derived from  other measurements.  Considering the renormalizable secluded vector dark matter model,  we have shown  the results  in the  $m_X-g_{\rm dm}$ parameter space, where some regions favored by the GC excess emission can be excluded by XENON1T  for $\alpha \gtrsim 0.01$ and further by the LZ projected sensitivity for $\alpha \gtrsim 0.002$.

The constraint from the Planck BBN measurement requires  the $S$ width $\Gamma_S \gtrsim (0.03~\text{sec})^{-1}$, from which we can put a lower bound on the mixing angle $\alpha \gtrsim 1 \times10^{-10}$.
On the other hand, for a case with a small mixing angle $\alpha \lesssim 2 \times 10^{-6}$, the hidden sector has been kinetically decoupled from the bath before it becomes nonrelativistic \cite{Yang:2019bvg}. As  such,  the correct relic density is described by a  DM annihilation cross section which could be significantly boosted above the conventionally WIMP value \cite{Yang:2019bvg}. In this paper, we have discussed the case with an extremely small value of $\alpha$, for which  the relativistic hidden sector can be decoupled from the SM bath at very high temperatures $T\gg m_{X,S}$, and, after decoupling, almost maintains the same temperature as the bath until $T\sim m_{S,X}$.
Assuming that the number changing interactions among the dark sector particles can guarantee their thermal equilibrium with zero chemical potential before decoupling, 
we have shown that the mixing angle $\alpha$ and elastic decoupling temperature $T_{\rm el}$ satisfy the relation: $\alpha \approx 10^{-6} (T_{\rm el}/ m_S)^{-3/2}$.
We have also shown that in this present scenario the vacuum of the secluded vector DM model will become unstable when $T\gtrsim 10^{-10}$. Thus, if $T_{\rm el}\gtrsim 10^{-10}$ is required for the secluded vector DM model, we obtain $\alpha \gtrsim 10^{-18}$.

For this Higgs portal (vector) DM model, the dSphs projected sensitivity can further probe the region of the annihilation cross section, which will be able to approach the conventional WIMP value. Furthermore,  extensive analyses for the line signal with the energy in the range of 30$-$66~GeV should be crucially important in testing this Higgs portal scenario and identifying the nature of dark matter in the future.

\acknowledgments \vspace*{-1ex}
 This work was supported in part by the Ministry of Science and Technology, Taiwan, under Grant No. 108-2112-M-033-002.

\appendix

\section{The partial decay widths of the hidden scalar  $S$}\label{app:s-width}

For the $S$ decay with $m_S \lesssim 200$~GeV,  the widths of the ${\bar f} {f}$ and $V V^{(*)}$ modes (with $f \equiv$ quark, charged lepton, $V\equiv$ on-shell $W, Z$, and $V^* \equiv$ off-shell $W, Z$) are given by
\begin{align}
 \Gamma(S \rightarrow \bar{f} f) 
  = &(1+ \Delta_{ff}) N_c^f  \frac{m_S}{8\pi} g_{Sff}^2  \left( 1- \frac{4 m_f^2}{m_S^2} \right)^{3/2} \theta(m_S-2m_f) \,,  \label{eq:S-width-fermion} \\
\Gamma (S \rightarrow VV^*) 
  = & \frac{3 }{128 \pi^3 v_H^2 } m_S \;  g_{SVV}^2  \;  \delta_V' R_T(x) \; \theta(m_S-m_V)  \theta(2m_V-m_S) 
  \nonumber \\
 & + \frac{m_{S}^3}{128 \pi m_V^4} \,  g_{SVV}^2 \, \delta_V \,   \sqrt{1-4x} \, (1-4x +12x^2)   \theta(m_S - 2m_V)  \,,
\label{HVV-3-2body}
\end{align}
and the widths of the modes generated by loop induced decays into the $gg, \gamma \gamma$, and $Z \gamma$ are given by
\begin{align}
\Gamma(S \rightarrow g g) 
& =(1 + \Delta_{gg})  \frac{\alpha_s^2 m_S^3}{128  \pi^3 }
      \left|\sum_{q\equiv {\rm quark}}   \frac{g_{Sqq}}{m_q} A_{1/2} \left( \tau_q \right)   \right|^2 \; , 
\label{app:Sgg}  \\
\Gamma\, (S\to \gamma\gamma)  
&= (1+\Delta_{\gamma\gamma}) \frac{ \alpha^{2}\, m_{S}^{3}}{256\,\pi^{3}} 
     \left|  \sum_{f \equiv q, \ell}  N_{c}^f Q_f^2 \frac{g_{Sff}}{m_f}  A_{1/2}(\tau_f) + \frac{g_{SWW} }{2 m_W^2} A_1(\tau_W) \right|^2 \,, 
\label{app:Sgaga} \\
\Gamma (S\to Z\gamma ) 
&= \frac{ m_W^2\, \alpha \,m_{S}^{3}} 
{128\,\pi^{4} v_{\rm SM}^2} \left( 1-\frac{m_Z^2}{m_S^2} \right)^3 \left|
 \sum_{f \equiv q, \ell}  N_{c}^f  \frac{Q_f \hat{v}_f}{\cos\theta_W} \frac{g_{Sff}}{m_f}   \bar{A}_{1/2} (\tau_f,\lambda_f) + 
 \frac{g_{SWW} }{ 2m_W^2} \bar{A}_1(\tau_W,\lambda_W) \right|^2  \,,
\label{app:hzga}
\end{align}
where  $\Delta_{\ell\ell}=0,  \Delta_{qq}=5.67\alpha_s(\mu)/\pi$, $\Delta_{gg} \simeq (215/12)\alpha_s(\mu)/\pi$, and $\Delta_{\gamma\gamma} \simeq 0$ for $m_S<350$~GeV  are the NLO QCD correction factors \cite{Djouadi:2005gj}, $N_c^{q(\ell)} \equiv 3\, (1)$ for the quark (lepton),  $Q_f$ is the charge of the fermion $f$, $g_{Sff}=-s_\alpha m_f/v_{\rm SM} $,  $g_{SVV}=- 2 s_\alpha m_V^2 /v_{\rm SM} $, $\delta_W=2, \delta_Z=1$, $\delta'_W=1$, $\delta_Z' = \frac{7}{12} - \frac{10}{9}
\sin^2\theta_W+ \frac{40}{27}\sin^4\theta_W$, $\hat{v}_f=2I_f^3-4 Q_f \sin\theta_W^2$ with $\sin\theta_W=0.23$ and $I_f^3$ being the left-handed weak isospin of the fermion, 
\begin{align}
R_T(x) 
 =  \frac{3(1-8x+20x^2)}{(4x-1)^{1/2}} \arccos \left( \frac{3x-1}
   {2x^{3/2}} \right) -\frac{1-x}{2x} (2-13x+47x^2)  - \frac{3}{2}(1-6x+4x^2) \ln x  \,,
\label{app:hvvRT}
\end{align}
with $x\equiv m_V^2/m_S^2$  \cite{Keung:1984hn,Djouadi:2005gi}, and
 the form factors  induced by spin--$\frac{1}{2}$ (top-)quark-loop ($A_{1/2}$ and ${\bar A}_{1/2}$) and by spin--1 $W$-loop  ($A_{1}$ and ${\bar A}_{1}$) are given by  \cite{Djouadi:2005gj,Djouadi:2005gi}
\begin{align}
A_{1/2}(\tau_i ) & =  2 \tau_i  [1+(1-\tau_i  )f(\tau_i )]   \,, 
\label{app:Af}  \\   
A_1(\tau_i ) & =  - [2  +3 \tau_i +3 \tau_i  (2 -\tau_i )f(\tau_i )] \,,
\label{app:Aw} \\
\bar{A}_{1/2} (\tau_i ,\lambda_i  ) 
& =  \left[I_1(\tau_i ,\lambda_i  ) - I_2(\tau_i ,\lambda_i  )
\right]  \,,
\\
\bar{A}_1(\tau_i ,\lambda_i  ) 
& =  c_W \left\{ 4\left(3-\frac{s_W^2}{c_W^2} \right)
I_2(\tau_i ,\lambda_i  ) + \left[ \left(1+\frac{2}{\tau_i }\right) \frac{s_W^2}{c_W^2}
- \left(5+\frac{2}{\tau_i } \right) \right] I_1(\tau_i ,\lambda_i  ) \right\}   \,,
\label{app:hzgaform}
\end{align}
with 
 {\allowdisplaybreaks
\begin{align}
I_1(\tau_i ,\lambda_i  ) 
& =  \frac{\tau_i \lambda_i  }{2(\tau_i -\lambda_i  )} + \frac{\tau_i^2\lambda_i^2}{2(\tau_i -\lambda_i  )^2} \left[ f(\tau_i )-f(\lambda_i  ) 
\right] + \frac{\tau_i^2\lambda_i  }{(\tau_i -\lambda_i  )^2} \left[ g(\tau_i ) - 
g(\lambda_i  ) \right] \;,
 \\
I_2(\tau_i ,\lambda_i  ) 
& =  - \frac{\tau_i \lambda_i  }{2(\tau_i -\lambda_i  )}
 \left[ f(\tau_i )- f(\lambda_i ) \right] \,,  \\
g(\tau_i )  
&=  
\left\{ 
   \begin{array}{lr} 
       \sqrt{\tau_i  -1} \arcsin \sqrt{\tau_i^{-1}}
        \, , \ \ & \text{for}\quad  \tau_i  \geq 1 
      \\ 
      \frac{\sqrt{1-\tau_i } }{2} \left[ \ln \frac{1+\sqrt{1-\tau_i }}{1-\sqrt{1-\tau_i }} - i \pi\right] 
        \, , & \text{for}\quad \tau_i  < 1 
   \end{array} 
 \right. \;,
\label{app:gtau} \\
f(\tau_i ) & = g^2(\tau_i ) /(\tau_i  -1)  \,,
\\
\tau_i  &=\frac{4 m_i^2}{m_S^2}  \,,   \qquad \lambda_i  =\frac{4 m_i^2}{m_Z^2} \,.
\end{align}
}
Here, we take the renormalization scale $\mu=m_S/2$.

\section{The annihilation cross section for $XX \to SS$} \label{app:XX2SS}

\begin{figure}[t!]
\begin{center}
\includegraphics[width=0.8\textwidth]{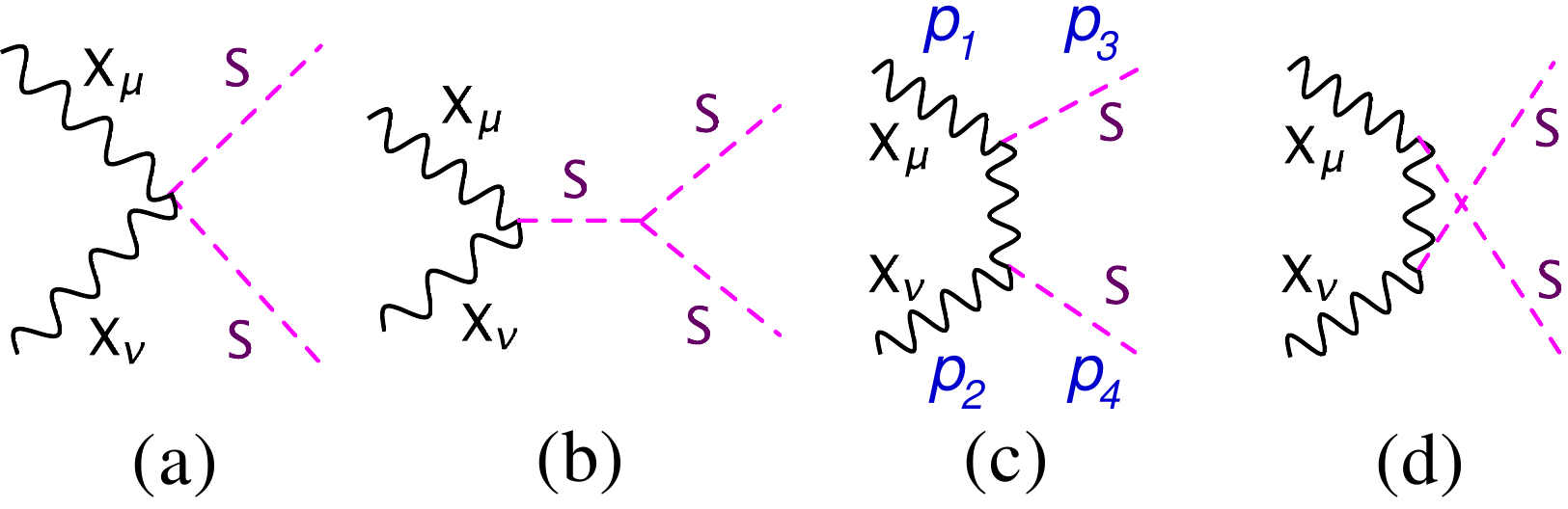}
\caption{The dominant contributions to the DM annihilation cross section, where (a), (b), (c), and (d) are diagrams for the 4-vertex, $s$-, $t$-, $u$-channels, respectively. Here $p_i$ are the momenta of the particles.}
\label{fig:vdm_ann}
\end{center}
\end{figure}

As shown in Fig.~\ref{fig:vdm_ann}, the  cross section  for the $XX\to SS$ in the laboratory frame, where one of the incoming particles is at rest with $v_{\rm lab}$ being the relative velocity measured,  is given by \cite{Yang:2019bvg}
\begin{align}
\sigma v_{\text{lab}}
&= (\sigma v_{\text{lab}})_{\text{4v,s}}  + (\sigma v_{\text{lab}})_{\text{t,u}}  + (\sigma v_{\text{lab}})_{\text{int}}  \,,
\end{align}
where
  {\allowdisplaybreaks
\begin{align}
(\sigma v_{\text{lab}})_{\text{4v,s}} 
&=
\left(3+ \frac{s ( s-4 m_X^2 )}{4 m_X^4} \right) 
\frac{c_\alpha^2 g_X^2 \sqrt{s-4 m_S^2} }{72 \pi  \left( (s-m_S^2 )^2 +  \Gamma_S^2 m_S^2 \right) \sqrt{s} (s-2 m_X^2 )}
\nonumber\\
& \quad \times \left[
\left(c_\alpha g_X (s- m_S^2) - g_{SSS} m_X \right)^2 + c_\alpha^2 \Gamma_S^2 g_X^2 m_S^2 \right] \,, 
\label{eq:4vs}
\\
 (\sigma v_{\text{lab}})_{\text{t,u}} 
 &=
\frac{c_\alpha^4 g_X^4 \sqrt{s-4 m_S^2} }{288 \pi m_X^4  \sqrt{s}  (s-2 m_X^2)}  \nonumber\\
&
\times \Bigg[4 m_S^4+4 s m_S^2+s^2  
+\frac{2 (m_S^8-8 m_X^2 m_S^6+24 m_X^4 m_S^4-32 m_X^6 m_S^2+48 m_X^8) }{m_S^4-4 m_X^2 m_S^2+m_X^2 s}
\nonumber\\
& \quad 
- \frac{4 \big(3 m_S^8-8 m_X^2 m_S^6  + (4 m_X^2 m_S^2- m_S^4) (8 m_X^4+s^2) -2 m_X^2 (24 m_X^6-2 s^2 m_X^2+s^3) \big) }{( s-2 m_S^2) \sqrt{s-4 m_S^2} \sqrt{s-4 m_X^2}} \nonumber\\
& \qquad
\times \ln \left(\frac{s-2 m_S^2 + \sqrt{s-4 m_S^2} \sqrt{s-4 m_X^2}}{s-2 m_S^2 - \sqrt{s-4 m_S^2} \sqrt{s-4 m_X^2}}\right)
\Bigg] \,,  
\label{eq:tu}
\\
(\sigma v_{\text{lab}})_{\text{int}}
&=c_\alpha^3 g_X^3 
\frac{
c_\alpha  g_X \left( (s-m_S^2 )^2 +  \Gamma_S^2 m_S^2 \right) -  g_{SSS} m_X (s-m_S^2) }{144 \pi m_X^4 \left( \left(s- m_S^2 \right)^2+ \Gamma_S^2 m_S^2\right) \left(s- 2 m_X^2\right) \sqrt{s} \sqrt{s-4 m_X^2}}
\nonumber\\
&
 \times \Bigg[
\sqrt{s-4 m_S^2} \sqrt{s-4 m_X^2}  \big(  s (6 m_X^2 -s) - 2 (2 m_X^2+s ) m_S^2 \big) \nonumber\\
& \quad -
2 \left( (2 m_X^2+s) m_S^4-4 m_X^2  (2 m_X^2+s ) m_S^2+2 m_X^2  (12 m_X^4-2 s m_X^2+s^2 ) \right) \nonumber\\
& \qquad
\times \ln \left(\frac{s-2 m_S^2 + \sqrt{s-4 m_S^2} \sqrt{s-4 m_X^2}}{s-2 m_S^2 - \sqrt{s-4 m_S^2} \sqrt{s-4 m_X^2}}\right)
\Bigg] \,,
\label{eq:int}
 \end{align}
with  the center-of-mass energy of $\sqrt{s}\simeq 2m_X$  for the low-velocity DM annihilation, and  the triple $SSS$ coupling being
\begin{align}
g_{SSS}=  -\frac{ 3 c_\alpha^3  m_S^2}{v_S} + \frac{3 s_\alpha^3 m_S^2}{v_{H}} \,, \label{eq:gsss}
\end{align}
}
and $v_S = m_X/g_{\rm dm}$. Here $ (\sigma v_{\text{lab}})_{\text{4v,s}}$ is the cross section resulting from the 4-vertex and $s$-channel diagrams,   $(\sigma v_{\text{lab}})_{\text{t,u}}$ is from the $t$- and $u$-channels, and $(\sigma v_{\text{lab}})_{\text{int}}$  is  from the interference between (4-vertex, $s$) and ($t$, $u$). An interesting property is the DM annihilation amplitudes,
\begin{align}
& \text{Fig.~\ref{fig:vdm_ann}(a), (b), (c), (d)}  \nonumber \\
& =
 2 i  c_\alpha^2 g_{\rm dm}^2 \epsilon_{1,\mu}  \epsilon_2^{\mu} \,, \, 
 -2 i  \frac{c_\alpha g_{\rm dm} m_X g_{SSS}}{ s-m_S^2  +  i \Gamma_S m_S }  \epsilon_{1,\mu}  \epsilon_2^{\mu} \,, \,
 4 i \frac{c_\alpha^2 g_{\rm dm}^2 m_X^2}{t-m_X^2}  \epsilon_{1,\mu}  {\bf P}_{\mu\nu}\epsilon_2^{\nu} \,,  \,
 4 i \frac{c_\alpha^2 g_{\rm dm}^2 m_X^2}{u-m_X^2} \epsilon_{1,\mu}  {\bf Q}_{\mu\nu}\epsilon_2^{\nu}   \nonumber \\
 & =
( i 2  g_{\rm dm}^2 , \, 
  i 2 g_{\rm dm}^2  , \, 
  -4 i  g_{\rm dm}^2 , \, 
  -4 i  g_{\rm dm}^2 ) \epsilon_{1,\mu}  \epsilon_2^{\mu}, \, \hskip1cm  \text{in the limit}\, s=4m_X^2, m_S \to m_X, \alpha \to 0,
 \label{xxss-amplitude-ratio}
\end{align}
where $\epsilon_{i, \mu}$ is the polarization vector of the initial DM particle, and
\begin{align}
{\bf P}_{\mu\nu} & \equiv g_{\mu\nu} - \frac{ (p_1 -p_3)_\mu (p_1 -p_3)_\nu}{m_X^2} \,, \nonumber\\
{\bf Q}_{\mu\nu} & \equiv g_{\mu\nu} - \frac{ (p_1 -p_4)_\mu (p_1 -p_4)_\nu}{m_X^2} \,.
\end{align}
 In Eq.~(\ref{xxss-amplitude-ratio}) we have used $\Gamma_S\approx 0$ in the $\alpha \to 0$  limit, and ${\bf P}_{\mu\nu} \approx {\bf Q}_{\mu\nu}\approx g_{\mu\nu}$,  $t=(p_1-p_3)^2\approx 0$ and $u=(p_1-p_4)^2\approx 0$ in the limits of $s\to 4m_X^2$ and $m_X\to m_S$. Therefore, in this limit, the total amplitude can be well approximated as $ - 4 i g_{\rm dm}^2 \epsilon_{1,\mu}  \epsilon_2^{\mu}$.
In other words, for a nearly degenerate case of $X$ and $S$ with a small mixing angle $\alpha$, we have $\sigma v_{\text{lab}} \approx (\sigma v_{\text{lab}})_{\text{4v,s}} $, which is numerically confirmed.
Note that we have neglected the diagram $XX\to h^* \to SS$, which, corresponding to Fig.~\ref{fig:vdm_ann}(b) but with the propagator replaced by $h$, is further suppressed by $\sin^2\alpha$ in the amplitude level, because the coupling of the $X-X-h$ vertex is  ``$i s_\alpha g_{\rm dm} m_X$", while the coupling of the $h-S-S$ vertex is 
\begin{align}
g_{hSS} 
&=
-  \frac{c_\alpha^2 s_\alpha (2 m_S^2 -m_h^2)}{v_S} - \frac{c_\alpha s_\alpha^2 (2 m_S^2 -m_h^2)}{v_{H}}  \\
& \simeq -  \frac{ s_\alpha  g_{\rm dm} (2 m_S^2 -m_h^2)}{m_X}, \, \text{in the small $\alpha$ limit}.
\label{eq:ghss}
\end{align}
For most of the GC favored regions in the Higgs portal model, we find that $m_X <m_h (=125.18~{\rm GeV})$, i.e. $XX\to hh$ is kinematically forbidden.  Nevertheless, 
as shown in Fig.~\ref{fig:gc-mxsv-resolution-s}, if $S$ and $h$ are degenerate in mass, only a very small GC region, corresponding to $p$-value $\lesssim 0.09$ and $m_X \in [m_h, 128~\text{GeV}]$, is allowed; in this region, the DM annihilation is still dominated by $XX \to SS$, while the $XX\to hh$ amplitude, for which the 4-vertex, $u$-, and $t$-channels  $\propto \sin^2\alpha$  and $s$-channel $\propto \sin\alpha$,  is relatively suppressed by $\sin\alpha$.

 \begin{figure}[t!]
  \begin{center}
   \includegraphics[width=0.43\textwidth]{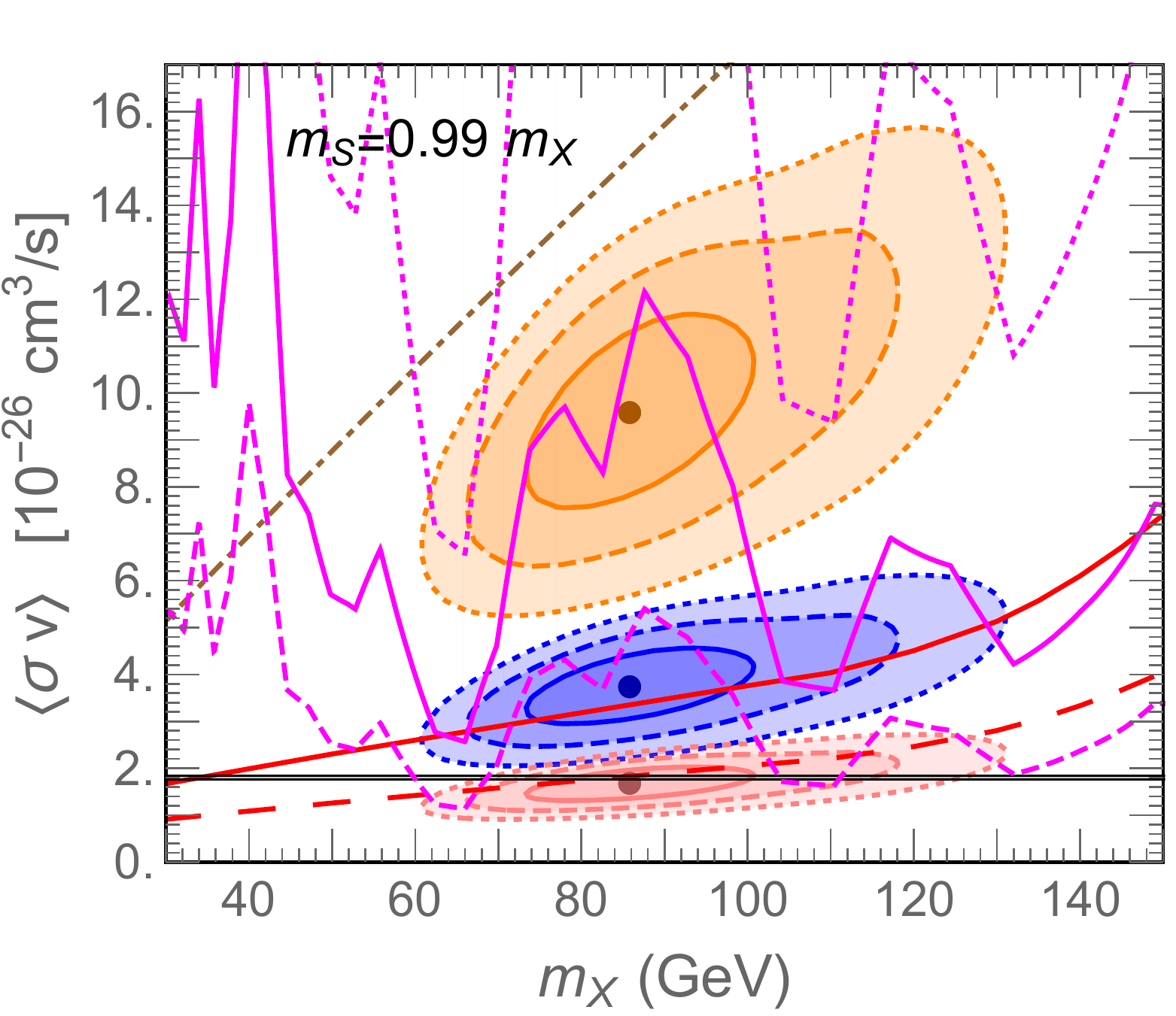} \hskip0.2cm
   \includegraphics[width=0.43\textwidth]{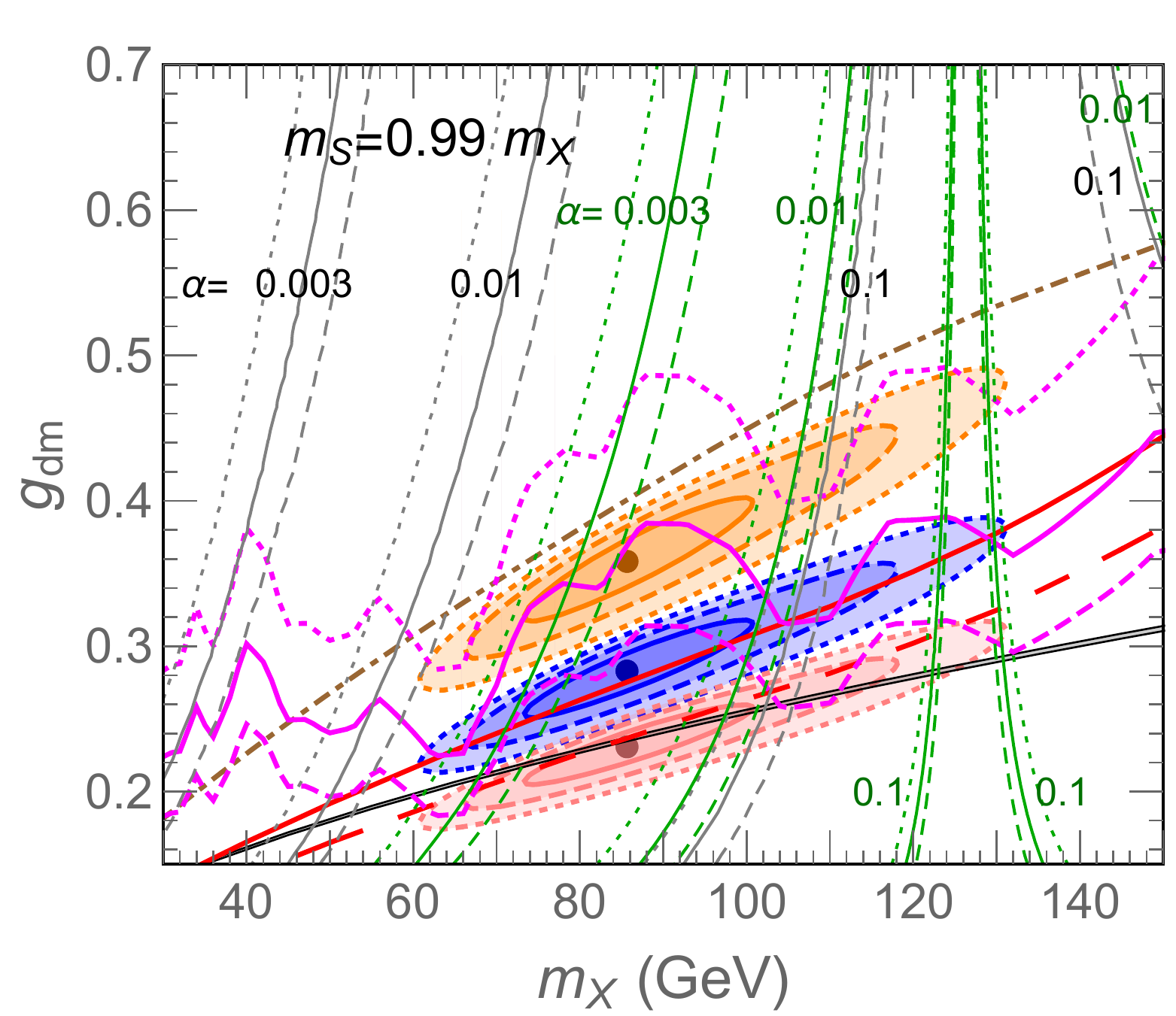}
\\ \vskip0.cm
   \includegraphics[width=0.43\textwidth]{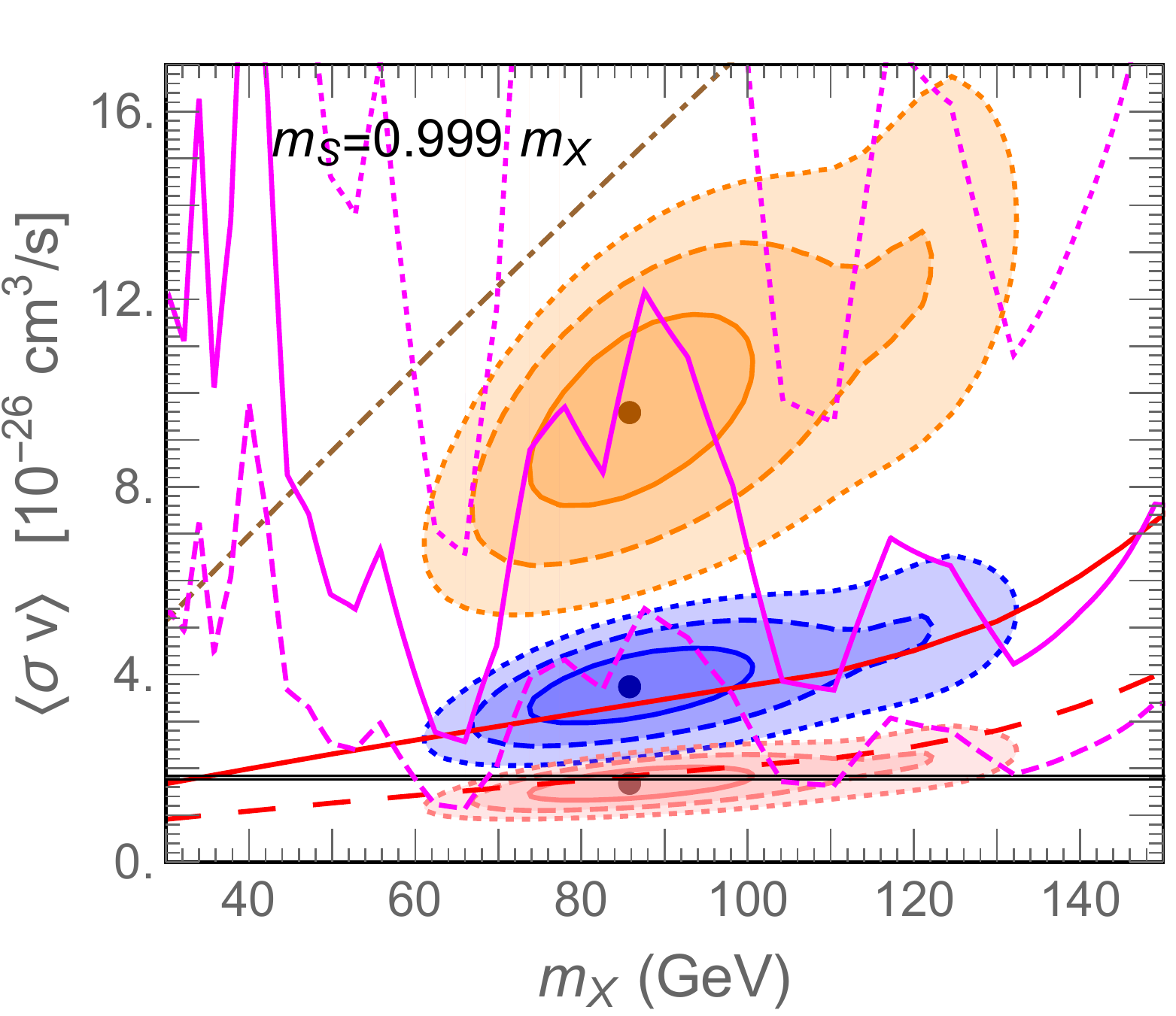} \hskip0.2cm
   \includegraphics[width=0.43\textwidth]{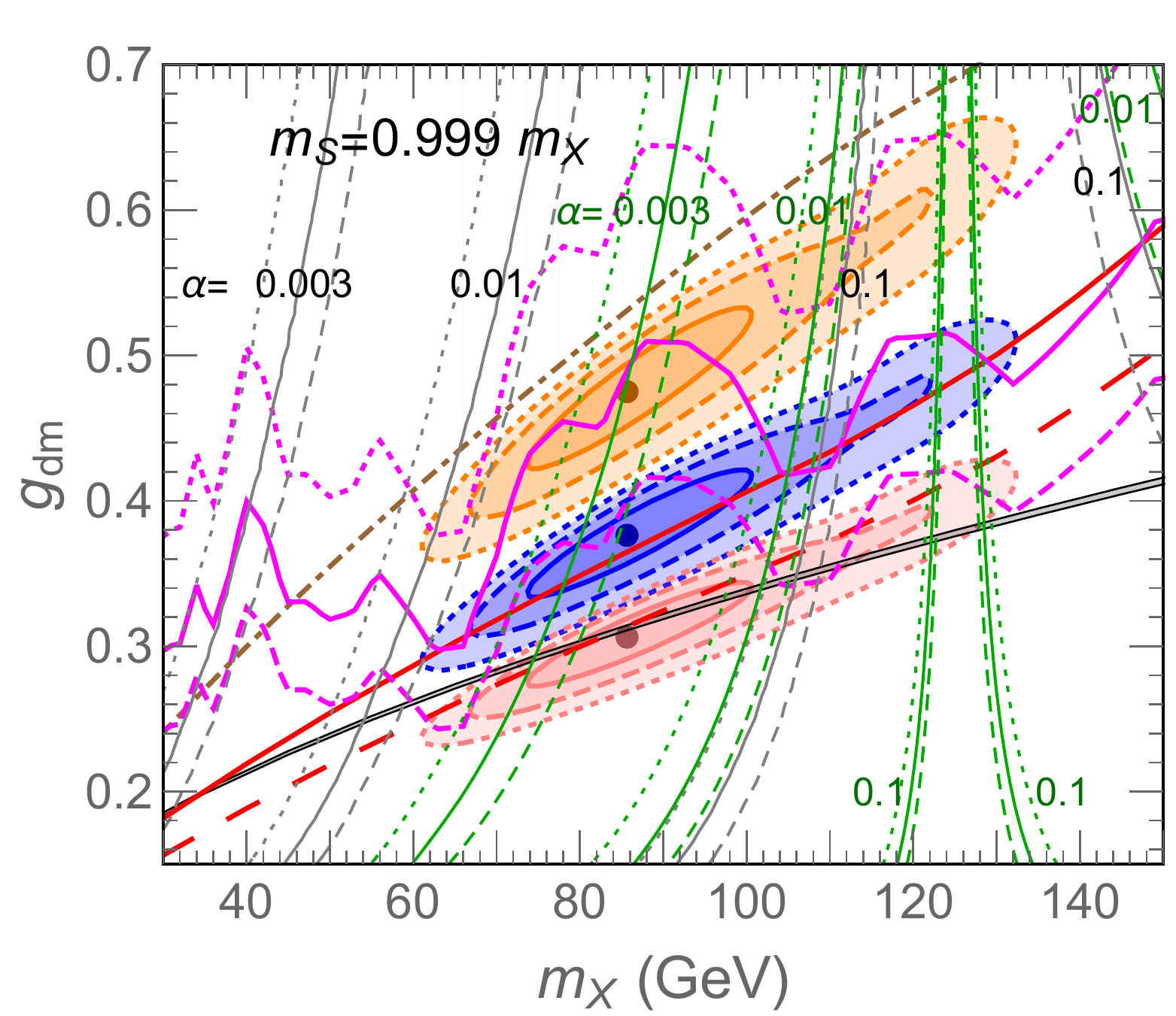} 
\caption{Same as Fig.~\ref{fig:gc-constraints} but  (i) for the GC allowed regions with the  blue, orange, and pink contours refer to $\rho_\odot =$ 0.4, 0.25, and $\text{0.6 GeV/cm}^3$, respectively, (ii) for the 95\%  C.L. bounds from the Fermi-LAT NFWc gamma-ray line search  within the ROI: $R_{\rm GC}=3^\circ$ are depicted 
for three values: $\rho_\odot =$ 0.4 (solid), 0.25 (dotted), and 0.6 (dashed) $\text{GeV/cm}^3$, and  (iii) for 
the 95\% C.L. bound from XENON1T and LZ projected sensitivity, given in the right panel, are indicated by the gray and green lines, which, with $\alpha$ values denoted,
are further given for three cases of $\rho_\odot =$ 0.4 (solid), 0.25 (dotted), and 0.6 (dashed) $\text{GeV/cm}^3$. }
\label{fig:gc-constraints-app}
\end{center}
\end{figure}

 \begin{figure}[t!]
  \begin{center}
   \includegraphics[width=0.435\textwidth]{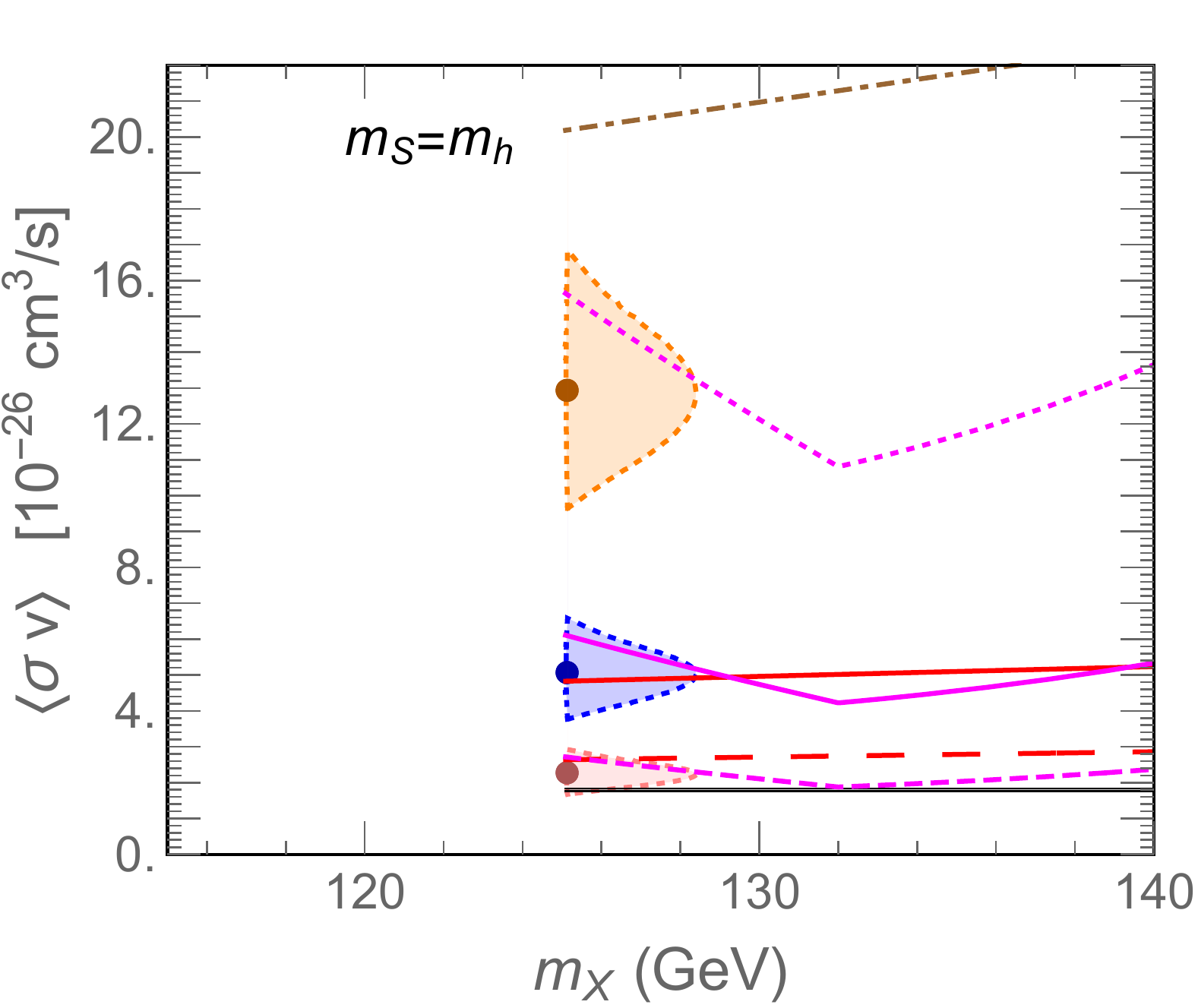} \hskip0.09cm
   \includegraphics[width=0.435\textwidth]{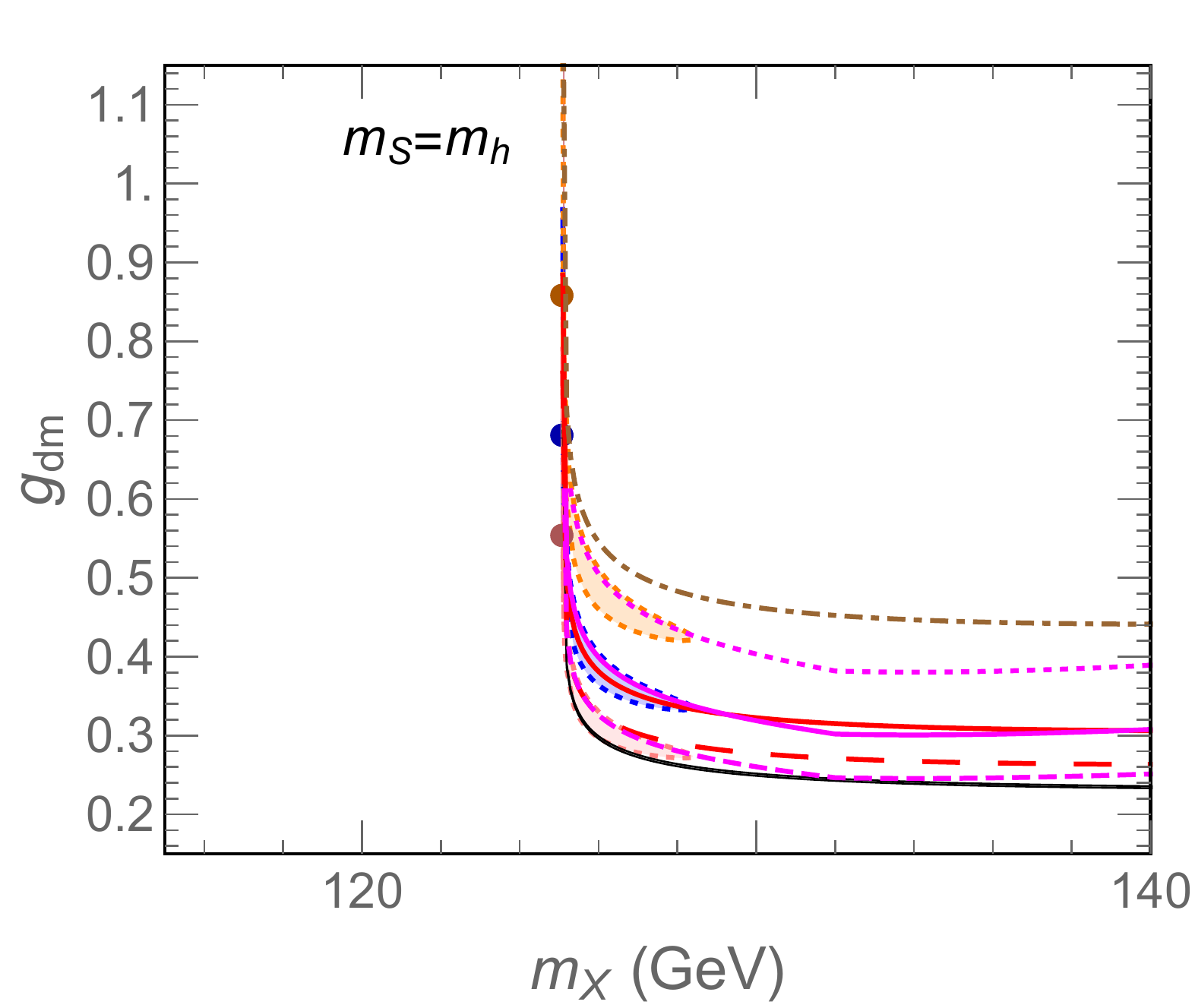} 
\caption{Same as Fig.~\ref{fig:gc-constraints-app} but for $m_S=m_h$
}
\label{fig:gc-constraints-2-app}
\end{center}
\end{figure}

\section{The dependence of the allowed parameter space on the value of $\rho_\odot$}\label{app:DMdensity}

The uncertainties of the dark matter distribution near the Galactic center and its local density are still large. 
 CCW have analyzed the GC inner slope for 60 Galactic diffuse emission models, and found  $\gamma=1.2 \pm 0.1$ preferred within a ROI:  $| \ell| \leq 20^\circ$ and $2^\circ \leq |b| \leq 20^\circ$.  To further illustrate the dependence on variation of  $\rho_\odot \in$ [0.25, 0.6] GeV/cm$^3$  for the GC excess favored region compared with other constraints, in Figs.~\ref{fig:gc-constraints-app} and \ref{fig:gc-constraints-2-app}, we employ three values of $\rho_\odot  = 0.25, 0.4$ and 0.6~GeV/cm$^3$. On the other hand, if using a smaller (or larger) $\gamma$, we can na\"ively expect  from the change of the $J$-factor that the value of $\langle \sigma v\rangle$ is further raised (or lowered) for the GC favored region. This appendix is a complement to Sec.~\ref{sec:constraints}.

The bounds of the Femi gamma-ray line search \cite{Ackermann:2015lka}  and direct detections  \cite{Aprile:2017iyp,Akerib:2018lyp} also depend on the value of $\rho_\odot$, which are depicted  in Figs.~\ref{fig:gc-constraints-app} and \ref{fig:gc-constraints-2-app} by the solid, dotted, and dashed magenta lines, corresponding to the use of  $\rho_\odot =$ 0.4, 0.25, and $\text{0.6 GeV/cm}^3$, respectively. 
For the right panel of Fig.~\ref{fig:gc-constraints-2-app}, the direct detection does not set the bound on this perfect degenerate case.  Note again that for $\rho_\odot \in [0.25,0.6]\, \text{GeV/cm}^3$, the parameter space,  where $g_{\rm dm} <1$, $|m_S -m_h|< 4$~GeV and $|\alpha| \gtrsim 0.17 \,  (0.02)$, can evade the bound from the XENON1T measurement (LZ projected sensitivity).

\section{The definition of temperature for $S$}\label{sec:temp-S}

In the high temperature limit $E_S\gg m_S$, the temperature of $S$ satisfies the relation,
\begin{align}
T_S &=\frac{g_S}{n_S(T_S)} \int \frac{d^3 p_S}{(2\pi)^3} \frac{{\bf p}_S^2}{3 E_S}  f_S (T_S) 
 +\frac{g_S}{n_S(T_S)} \int \frac{d^3 p_S}{(2\pi)^3} \frac{{\bf p}_S^2}{3 E_S}  f_S^2 (T_S)   \,,
\label{app:temp-1}
\end{align} 
where $f_S=[\exp((E-\mu_S)/T_S) -1]^{-1}$ with $\mu_S$ the chemical potential,  $g_S=1$ is the internal degrees of freedom, and $n_S$ is the number density.
Assuming that $S$ is in chemical equilibrium with zero chemical potential, 
we can approximate the RHS of Eq.~(\ref{app:temp-1}) as
\begin{align}
T_S 
\Bigg(  
      \frac{ \int_0^\infty \frac{x^3}{e^x -1} dx}{3   \int_0^\infty \frac{x^2}{e^x -1} dx} 
  +  \frac{ \int_0^\infty \frac{x^3}{ (e^x -1)^2} dx}{3   \int_0^\infty \frac{x^2}{e^x -1} dx} 
 \Bigg)
 &=  T_S \Bigg( \frac{\pi^4}{90 \zeta(3)}+ \frac{-\pi^4/15 +6\zeta(3) }{6 \zeta(3)  } \Bigg) \nonumber\\
 &\simeq T_S (0.90 +0.10)  
 \,,
\label{app:temp-1p}
\end{align} 
which shows that the term $\propto f_S^2$ on the RHS of Eq.~(\ref{app:temp-1}) gives about 10\% correction in amount.
On the other hand, in high temperatures,  we can expect that the average number of particles in each state of the phase space  is much less than 1, i.e.,  $1 + f_S \simeq 1$, and thus approximate the $S$ distribution as
\begin{align}
f_S =  e^{-(E_S-\mu_S)/T_S} (1\pm f_S) \simeq e^{-(E_{S}-\mu_S)/T_S}  \,.
\end{align}
Using the approximate distribution,   we obtain
\begin{align}
\frac{g_S}{n_S(T_S)} \int \frac{d^3 p_S}{(2\pi)^3} \frac{{\bf p}_S^2}{3 E_S}  f_S (T_S) =T_S \,,
\label{app:temp-2}
\end{align} 
which is a good approximation for the temperature of $S$. Here the approximation of the thermal equilibrium number density (with $\mu_S=0$) is less than the exact value by a factor of 17\%.
We thus use Eq.~(\ref{app:temp-2}) as the benchmark to derive the Boltzmann moment equation of the hidden scalar temperature, which is suitable at high temperatures ($\gg m_S$).

\section{RGEs up to two-loop order}\label{app:rges}

The renormalization group equations up to two-loop order for $g_{\rm dm}$ and scalar quartic couplings are described by
\begin{align}
\frac{d\lambda}{dt} = \frac{1}{16\pi^2} \beta_{\lambda}^{(1)} + \frac{1}{(16\pi^2)^2} \beta_{\lambda}^{(2)} \,,
\end{align}
where $\lambda \equiv g_{\rm dm}, \lambda_H, \lambda_{HS}, \lambda_S$, and $t \equiv \ln Q$, with $Q$ being the renormalization scale. Here by using SARAH \cite{Staub:2008uz,Staub:2010jh,Staub:2013tta,Staub:2015kfa}, the $\beta$-functions are given by
 {\allowdisplaybreaks
 \begin{align} 
 \beta_{g_{\rm dm}}^{(1)}  = &  
\frac{1}{3} g_{\rm dm}^{3} \,,
\\ 
\beta_{g_{\rm dm}}^{(2)}  = &  4 g_{\rm dm}^{5}   \,,
\\ 
\beta_{\lambda_{H}}^{(1)}  =  &
 \frac{27}{100} g_{1}^{4} +\frac{9}{10} g_{1}^{2} g_{2}^{2} +\frac{9}{4} g_{2}^{4} +2 \lambda_{HS}^{2} -\frac{9}{5} g_{1}^{2} \lambda_{H} 
 -9 g_{2}^{2} \lambda_{H} +12 \lambda_{H}^{2} +12 \lambda_H y_t^2 -12 y_t^4 \,,
 \\ 
\beta_{\lambda_{H}}^{(2)}  = & 
-\frac{3411}{1000} g_{1}^{6} -\frac{1677}{200} g_{1}^{4} g_{2}^{2} -\frac{289}{40} g_{1}^{2} g_{2}^{4}+\frac{305}{8} g_{2}^{6} 
 +16 g_{\rm dm}^{2} \lambda_{HS}^{2} -8 \lambda_{HS}^{3} +\frac{1887}{200} g_{1}^{4} \lambda_{H}  \nonumber \\ 
  & +\frac{117}{20} g_{1}^{2} g_{2}^{2} \lambda_{H} 
  -\frac{73}{8} g_{2}^{4} \lambda_{H}  -10 \lambda_{HS}^{2} \lambda_{H} +\frac{54}{5} g_{1}^{2} \lambda_{H}^{2} +54 g_{2}^{2} \lambda_{H}^{2} -78 \lambda_{H}^{3} 
 \nonumber \\ 
 & -\frac{171}{50} g_{1}^{4} y_t^2 +\frac{63}{5} g_{1}^{2} g_{2}^{2}  y_t^2 -\frac{9}{2} g_{2}^{4}  y_t^2 +\frac{17}{2} g_{1}^{2} \lambda_{H}  y_t^2 
  +\frac{45}{2} g_{2}^{2} \lambda_{H}  y_t^2 +80 g_{3}^{2} \lambda_{H}  y_t^2 -72 \lambda_{H}^{2}  y_t^2
 \nonumber \\ 
 &-\frac{16}{5} g_{1}^{2}  y_t^4 -64 g_{3}^{2}  y_t^4 -3 \lambda_{H}  y_t^4  +60  y_t^6 \,,
\\ 
\beta_{\lambda_{HS}}^{(1)}  =  &
\frac{1}{10} \lambda_{HS} \Big( 40 \lambda_{HS}  + 40 \lambda_S  -45 g_{2}^{2}  -60 g_{\rm dm}^{2}  + 60 \lambda_{H}   + 60  y_t^2  -9 g_{1}^{2} \Big) \,,
\\ 
\beta_{\lambda_{HS}}^{(2)}  =  & 
 \frac{1671}{400} g_{1}^{4} \lambda_{HS} +\frac{9}{8} g_{1}^{2} g_{2}^{2} \lambda_{HS}  -\frac{145}{16} g_{2}^{4} \lambda_{HS} 
 +\frac{86}{3} g_{\rm dm}^{4} \lambda_{HS} +\frac{3}{5} g_{1}^{2} \lambda_{HS}^{2}  +3 g_{2}^{2} \lambda_{HS}^{2} +4 g_{\rm dm}^{2} \lambda_{HS}^{2} 
 \nonumber \\ 
 & -11 \lambda_{HS}^{3} +32 g_{\rm dm}^{2} \lambda_{HS} \lambda_S -24 \lambda_{HS}^{2} \lambda_S -10 \lambda_{HS} \lambda_{S}^{2} +\frac{36}{5} g_{1}^{2} \lambda_{HS} \lambda 
 +36 g_{2}^{2} \lambda_{HS} \lambda_{H} -36 \lambda_{HS}^{2} \lambda_{H} 
\nonumber \\ 
 &-15 \lambda_{HS} \lambda_{H}^{2}  
  +\frac{17}{4} g_{1}^{2} \lambda_{HS}  y_t^2  +\frac{45}{4} g_{2}^{2} \lambda_{HS}  y_t^2 
 +40 g_{3}^{2} \lambda_{HS}  y_t^2 -12 \lambda_{HS}^{2}  y_t^2 -36 \lambda_{HS} \lambda_{H}  y_t^2 
 \nonumber\\
 & -\frac{27}{2} \lambda_{HS} y_t^4 \,,
\\ 
\beta_{\lambda_S}^{(1)}  = & 
2 \Big(2 \lambda_{HS}^{2}  -6 g_{\rm dm}^{2} \lambda_S  + 6 g_{\rm dm}^{4}  + 5 \lambda_S^2 \Big) \,,
\\ 
\beta_{\lambda_S}^{(2)}  = & 
 -\frac{416}{3} g_{\rm dm}^{6} +\frac{24}{5} g_{1}^{2} \lambda_{HS}^{2} +24 g_{2}^{2} \lambda_{HS}^{2} -16 \lambda_{HS}^{3} 
+\frac{316}{3} g_{\rm dm}^{4} \lambda_S  -20 \lambda_{HS}^{2} \lambda_S +56 g_{\rm dm}^{2} \lambda_{S}^{2}  -60 \lambda_{S}^{3} \nonumber \\ 
 &  -24 \lambda_{HS}^{2}  y_t^2 \,,
\end{align}
}
where $y_t$ is the Yukawa coupling, and $g_3, g_2$ and $g_1$ are respectively the SU(3)$_C$, SU(2)$_L$ and U(1)$_Y$ gauge couplings, with $g_1= \sqrt{5/3} g_Y$ written in SU(5) normalization.  The RGEs for the relevant SM parameters, $g_i$ and $y_t$, also taken into account up to two loops by using SARAH, are not shown here for saving space and can be referred to Ref.~\cite{Buttazzo:2013uya}.

\end{document}